\documentclass[apsrev4-1,twocolumn,prd,showpacs,amssymb,floatfix,nofootinbib,superscriptaddress,aps]{revtex4-1}
\usepackage[utf8]{inputenc}
\usepackage{epsf,color,colordvi}
\usepackage{graphicx}
\usepackage{amsmath,amsthm,amssymb,bm}
\usepackage{times}
\usepackage{subfig}
\usepackage{placeins}
\usepackage{braket}
\usepackage{multirow}

\usepackage[normalem]{ulem}

\definecolor{darkgreen}{cmyk}{1,0,1,0.4}

\definecolor{pink}{cmyk}{0.4,1,0.3,0}

\def\com2#1{\textcolor{red}{\it{#1}}}

\usepackage[inline]{enumitem}

\def\g {\gamma}

\def\bar {\overline}

\def\nn{\nonumber}
\def\bstt{$B_s\to\tau\tau$}
\def\bKmt{$B\to K\mu\tau$}
\def\bKnn{$B\to K\nu\bar{\nu}$}
\def\bKstnn{$B\to K^*\nu\bar{\nu}$}

\def\st {\sin\theta}
\def\ct {\cos\theta}

\begin{document}

\title{An in-depth analysis of $b\to c(s)$ semileptonic observables with possible $\mu - \tau$ mixing}

\author{Srimoy Bhattacharya}
\email{bhattacharyasrimoy@gmail.com}
\affiliation{The Institute of Mathematical Sciences,\\ Taramani, Chennai 600113, \\   India}
\author{Aritra Biswas}
\email{iluvnpur@gmail.com}
\affiliation{School of Physical Sciences, Indian Association for the Cultivation of Science,\\
2A \& 2B Raja S.C. Mullick Road, Jadavpur,\\ Kolkata 700 032, India}
\author{Zaineb Calcuttawala}
\email{zaineb.calcuttawala@gmail.com}
\affiliation{Department of Physics, University of Calcutta\\
92 Acharya Prafulla Chandra Road, Kolkata 700009, India}
\author{Sunando Kumar Patra}
\email{sunando.patra@gmail.com}
\affiliation{Department of Physics, Indian Institute of Technology\\
Kanpur-208016, India}

\begin{abstract}
In a couple of recent publications \cite{Choudhury:2017qyt,Choudhury:2017ijp}, the authors attempted to achieve simultaneous explanation of the persistent flavor anomalies in $b\to s$ and $b\to c$ semileptonic decays with a minimal scheme by using only three unknown new parameters. The analysis was obtained with a handful of precise observables. Motivated by their proposal, in this paper we reanalyze the models proposed in the aforementioned papers with a total of 170 observables from those channels including newly available measurements, correlated theoretical results and constraints. We validate our results by searching for the most influential points and outliers. By analyzing the parameter spaces and their relationship with the constraints, we gain new insight and statistical significance in those models. We also provide a new and precise calculation of $R(J/\Psi)$, obtained during the analysis.
\end{abstract}


\date{\today} 

\maketitle

\section{Introduction}

Observables in heavy flavor physics have long served as instrumental probes in the search for beyond Standard Model (BSM) new physics (NP). Over the last few years in particular, deviations have been observed between the SM estimates and the experimental values for the ratios involving $R(D^{(*)})=\left(\Gamma(B\to D^{(*)}\tau\nu)\right) / \left(\Gamma(B\to D^{(*)}\ell\nu)\right)$~\cite{Lees:2012xj,Lees:2013uzd,Huschle:2015rga,Sato:2016svk,Aaij:2015yra,Hirose:2016wfn,Hirose:2017dxl,Aaij:2017uff,Aaij:2017deq} and $R(K^{(*)})=\frac{\Gamma(B\to K^{(*)}\mu\mu)}{\Gamma(B\to K^{(*)}ee)}$~\cite{Aaij:2014ora,Aaij:2017vbb} measured by BaBar, Belle and LHCb ($\ell = \mu$ or $e$). At the quark level, these involve semileptonic charged current $b\to c\ell\nu$ and neutral current $b\to s\ell\ell$ transitions respectively. Modes with hadronic initial and final states are subject to large corrections from strong interactions, that are not yet satisfactorily understood. Semileptonic modes, on the other hand, allow a much more agreeable control since the dominant uncertainties are due to the form factors and CKM elements. These get canceled to a large extent in the ratios mentioned above and the observables are hence regarded as relatively clean probes for NP. The pattern also suggests that such NP, if present, will violate lepton-flavor universality (LFU), a feature that is one of the principle characteristics of the SM. While the charged current ratios have the potential to probe and quantify the apparent non-universal coupling of the leptons to the $W$ boson for the third generation, their neutral current counterparts do the same for the corresponding couplings to the Z boson for the second generation. An observed trend is that the experimental values for the $R(D^{(*)})$ ratios lie above while the $R(K^{(*)})$ values lie below their corresponding SM predictions. Currently, the $R(D)$ and $R(D^*)$ ratios exceed the global arithmetic average for the corresponding SM values due to HFLAV by $2.3\sigma$ and $3.0\sigma$ respectively. The global deviation for these two anomalies taken together is about $3.78\sigma$. $R(K)$ is $2.6\sigma$ less than SM while the low ($0.045<q^2<1.1$ GeV$^2$) and high-$q^2$ ($1.1<q^2<6.0$ GeV$^2$) bins for $R(K^*)$ lag behind by $2.1-2.3\sigma$ and $2.4-2.5\sigma$ respectively. These ratios have hence attracted much attention from the community\footnote{Theoretically cleaner probes for LFU violation should in principle involve the branching ratios (BRs) involving the purely leptonic decay modes like $B_s\to l^+l^-$ and $B_c\to l\nu$ $(l=e,\mu,\tau)$, since the only theoretical uncertainty due to the decay constants will cancel in the ratio. However, PDG~\cite{pdg} reports no measurement for the BRs of the leptonic $B_c$ modes, while only upper limits are available for $\mathcal{B}(B_s\to e^+e^-, \tau^+\tau^-)$. The only measurement is $\mathcal{B}(B_s\to\mu^+\mu^-)$ with $\sim 20\%$ uncertainty, capable of accommodating LFU violating NP. Hopefully, with more precise and complete data on these modes in the future, this issue on LFUV can be settled for good.}. From the theoretical side, there have been many efforts to explain these both from model-independent points of view and from within the framework of various existing models
\cite{Bobeth:2001sq,Altmannshofer:2008dz,Hiller:2014yaa,Crivellin:2012ye,Crivellin:2015lwa,Beaujean:2015gba,Bhattacharya:2015ida,Becirevic:2016yqi,Das:2016vkr,Choudhury:2016ulr,Bhattacharya:2016mcc,Bardhan:2016uhr,Bhattacharya:2016zcw,Das:2017kfo,Bardhan:2017xcc,Biswas:2017vhc,Capdevila:2017bsm,Altmannshofer:2017yso,DAmico:2017mtc,Hiller:2017bzc,Geng:2017svp,Ciuchini:2017mik,Celis:2017doq,Becirevic:2017jtw,Cai:2017wry,Kamenik:2017tnu,Sala:2017ihs,DiChiara:2017cjq,Ghosh:2017ber,Alok:2017jaf,Bonilla:2017lsq,Guo:2017gxp,Chen:2017usq,Baek:2017sew,Iguro:2017ysu,Cline:2017aed,Alok:2017qsi,Descotes-Genon:2017ptp,Bhattacharya:2018kig,Biswas:2018jun,Babu:2018vrl,Huang:2018nnq,Chen:2018hqy,deBoer:2018ipi,Greljo:2018ogz,Abdullah:2018ets,Martinez:2018ynq,Aloni:2018ipm,Colangelo:2018cnj,Kim:2018oih,Iguro:2018vqb,Blanke:2018yud,Mandal:2018kau,Bansal:2018nwp,Asadi:2018sym,Iguro:2018fni,Hu:2018veh,Carena:2018cow,Hu:2018lmk,Robinson:2018gza,Feruglio:2018fxo,Chkareuli:2019xqi,Baek:2019qte,Hutauruk:2019crc,Geng:2018xzd,Fornal:2018dqn,Stangl:2018kty,Capdevila:2018jhy,Maji:2018gvz,Faisel:2018bvs,Rocha-Moran:2018jzu,Allanach:2018odd,Ciuchini:2018anp,Singirala:2018mio,Allanach:2018lvl,Angelescu:2018tyl,Faber:2018qon,Duan:2018akc,Rajeev:2018txm,Kumbhakar:2018uty,Guadagnoli:2018ojc,Hati:2018fzc,King:2018fcg,Baek:2018aru,Arbey:2018ics,Earl:2018snx,Sahoo:2018ffv}. Only some of them attempt to explain the anomalies together \cite{Altmannshofer:2017poe,Crivellin:2017zlb,Blanke:2018sro,Aebischer:2018acj,Biswas:2018iak,Heeck:2018ntp,Li:2018lxi,Datta:2018xty,Li:2018rax,Crivellin:2018yvo,Trifinopoulos:2018rna,Kumar:2018kmr,Becirevic:2018rnv,Becirevic:2018afm,Marzo:2019ldg}.

Subsequently, lepton flavor universality violating (LFUV) hints have also been observed in other exclusive channels with the same underlying quark transitions. The $R_{J/\psi}$ ratio~\cite{Aaij:2017tyk}, involving the $b\to cl\nu$ transition is one such example which is about $2\sigma$ above its corresponding SM prediction\footnote{This ratio suffers from  substantial uncertainties due to the uncertainties in the theoretical predictions, which heavily depend on the corresponding parametrization for the $B_c\to J/\psi$ form factors.}. Similarly, the $q^2=[1,6]$ GeV$^2$ bin for the $B_s\to\phi\mu\mu$~\cite{Aaij:2015esa} decay is an example for another exclusive transition with the underlying quark structure $b\to s\mu\mu$ which lags behind the SM. However, leptonic processes such as $B_s\to\mu\mu$~\cite{Tanabashi:2018oca} and radiative processes such as $B\to X_s\gamma$ show no such deviation from their corresponding SM results, although they entail the same transition at the quark level. This is also true for the mass difference $\Delta M_s$ and the mixing phase $\phi_s$~\cite{Aaij:2014zsa} for the $B_s$ system. Hence, the observed pattern for the apparent LFUV is a complicated one.

Recently, an interesting proposal was put forward~\cite{Choudhury:2017qyt,Choudhury:2017ijp} where the authors devised a class of `models' in order to accommodate the `clean' LFUV observables entailing both $b\to c\ell\nu$ and $b\to s\ell\ell$ transitions, while also making predictions for a few modes as signatures for the validity of their proposed models. Starting from an effective Lagrangian at low energies, they tried to explain these observables in a phenomenologically motivated minimalistic NP scenario, i.e., with the help of a minimal set of NP operators, with small values for the Wilson coefficients (WCs) corresponding to a small enough NP scale of a few TeV. They found that a reasonable fit includes at least two and at most three (with possible symmetry relations between the corresponding WCs) new current-current operators. These operators were constructed out of flavor eigenstates, augmented by change of basis for the charged leptonic fields. They showed that this class of models explain the above mentioned observables with only three model parameters.

In the scope of our current article, we consider this class of minimal models, without changing their simplistic appeal, and analyze them in the light of a large number of $b\to c,s$ observables. After recreating the earlier analysis, and examining the detailed constrained parameter spaces under scrutiny of a bunch of statistical tools, we motivate a more rigorous fit, involving a total of 170 observables, with greater degrees of freedom and greater chances of finding outliers. In the process, we calculate the theoretical results for observables independently, to get coherent measures of covariances, which also gives us the opportunity to obtain a precise theoretical estimate of $R(J/\Psi)$. 

The paper is organized as follows: In section~\ref{model}, we provide a short summary of the model(s) discussed by the aforementioned articles. Next, in section \ref{sec:motivation}, we motivate the need for reanalyzing these models. Section \ref{sec:method} contains a short but detailed description of the data, method, and the tools used in our analysis. In section \ref{sec:results}, we describe and analyze our results, and summarize our conclusions in section \ref{sec:conclusions}.

\section{Theory and Motivation}
\subsection{The models: A short recap}\label{model}
This section will be a quick and short recap of the `models' put forward by the aforementioned refs.~\cite{Choudhury:2017qyt,Choudhury:2017ijp}. We will try to brush through the arguments behind the operator structures that the authors put forward
, closely following~\cite{Choudhury:2017ijp} in what follows. A detailed discussion about the models is provided in appendix \ref{app:models}.

With their original motivation to attribute the charge and neutral anomalies to the same source with a minimal number of NP parameters, the authors consider operators consisting of vector and axial vector currents alone, with the corresponding hamiltonian(s) consisting of at most two four-fermi operators at the scale $m_b$ \footnote{The scalar and/or tensor operators, although can be potential sources for the charge current anomalies, are not particularly useful for their neutral current counterparts. The explanation of $R(K)$ with scalar (or pseudoscalar currents) is incompatible with $B_s\to\mu\mu$ branching ratio. The role of tensor operators in the case of the neutral current anomalies is discussed in~\cite{Bardhan:2017xcc}.}. We list the effective Hamiltonians and the Wilson coefficients for several transitions used in this analysis in appendix \ref{app:effHamil}.

While such a NP Lagrangian has to respect the SM gauge symmetries, LFU cannot be a symmetry of the corresponding Hamiltonian. The authors observe that to achieve this, it is sufficient for the corresponding fields to belong to the second and third generation of fermions only. The scale of this NP has to be higher than the electroweak scale due to collider constraints. They write such operators in terms of weak eigenstates involving second and third generation quark fields, but only third generation leptons. The mass eigenstates for the $\mu$ and $\tau$ leptons are generated via appropriate rotations of the charged lepton states.

Considering effective theory, a ``model" in this discussion implies a combination of (at most) two four-fermion operators at $m_b$ scale. The NP scale is considered to be 1 TeV for our analysis.
In order to explain the anomalies in both the charged current sector $b \to c$ and the neutral current sector $b \to s$ simultaneously, the NP operators will involve the second and third generation quark fields and only the third generation lepton field, which can be appropriately rotated from the flavor basis to the mass basis to give rise to the mass eigenstates of the $\mu$ and $\tau$ leptons.

\begin{align}
\label{eq:rot}
\tau = \ct \, \tau' + \st \, \mu' \ ,
\qquad \quad 
\nu_\tau = \ct \, \nu_\tau' + \st \, \nu_\mu'\,.
\end{align}

Following the short hand notation introduced in ref. \cite{Choudhury:2017ijp}:
\begin{eqnarray}
(x,y) &\equiv& \bar x_L \g^\mu y_L \ \ \ \  \forall\ \  x,y\,\nonumber\\
\{x,y\} &\equiv& \bar x_R \g^\mu y_R \ \ \ \  \forall\ \  x,y\,, 
\end{eqnarray} 
we arrive at the following `model':
\begin{align}
\label{eq:op1}
{\rm Model~ I:}\quad  \mathcal{O}_{\rm I} =& \sqrt{3} \, A_1 \, (\bar Q_{2L} \g^\mu L_{3L})_3 \, 
(\bar L_{3L} \g_\mu Q_{3L})_3 \nonumber\\
& -2  A_2 \, (\bar Q_{2L} \g^\mu L_{3L})_1 \, (\bar L_{3L} \g_\mu Q_{3L})_1 .
\end{align}
In the above, $A_{1,2}$\footnote{The model parameters $A_i$ $\equiv$ $A_i/\Lambda^2$ where $\Lambda$ is chosen to be 1 TeV.} are the WC's with dimension of (mass)$^{-2}$. $Q_i$ and $L_i$ stand for the
$i$-th generation (and weak-eigenstate) $SU(2)_L$ quark and lepton
doublet fields respectively. The subscripts `3' and `1' represent the
$SU(2)_L$ triplet and singlet structure of the currents respectively\footnote{The factor of $\sqrt{3}$ has been introduced explicitly to account for	the Clebsch-Gordan coefficients.}. As mentioned earlier, only the second and
third generation quark doublets and the third generation 
lepton doublet alone are involved in Eq.~\ref{eq:op1}. Terms with the potential to explain the $b\to s \mu \mu$ anomalies are generated when one considers the simplest of field rotations for the
left-handed leptons from the unprimed (flavor) to the primed (mass)
basis, as mentioned in Eq.~\ref{eq:rot}. In a similar fashion, following set of `model's can be defined as well:
\begin{align}
{\rm Model~ II:}\quad \mathcal{O}_{\rm II} =& -\sqrt{3} \, A_1 \, (\bar Q_{2L} \g^\mu Q_{3L})_3 \,(\bar L_{3L} \g_\mu L_{3L})_3\nonumber\\ 
& + \sqrt{3}  A_2 \, (\bar Q_{2L} \g^\mu L_{3L})_3 \, (\bar L_{3L} \g_\mu Q_{3L})_3.\label{eq:op2}\\
{\rm Model~ III:}\quad  \mathcal{O}_{\rm III} =& 
-\sqrt{3} \, A_1 \, (\bar Q_{2L} \g^\mu Q_{3L})_3 \, (\bar L_{3L} \g_\mu L_{3L})_3 \nonumber\\
&+2 \, A_3 \, (\bar Q_{2L} \g^\mu Q_{3L})_1 \, (\bar L_{3L} \g_\mu L_{3L})_1.\label{eq:op3}
\end{align}

Introducing right-handed leptonic currents preceded by a new WC $A_5$\footnote{The factor of $\sqrt{2}$ is, as in the previous case, a Clebsch-Gordon factor which parametrizes the strength of the right-handed tauonic current.}, we get 
\begin{align}
{\rm Model~ IV:}\quad  \mathcal{O}_{\rm IV}  = &  \sqrt{3} \, A_1 \, \left[
- (\bar Q_{2L} \g^\mu Q_{3L})_3 \, (\bar L_{3L} \g^\mu L_{3L})_3 \right.\nonumber\\
&\left. + \frac{1}{2} \, (\bar Q_{2L} \g^\mu L_{3L})_3 \, (\bar L_{3L} \g^\mu Q_{3L})_3
\right] \nonumber\\
&+  \sqrt{2} \, A_5 \,  (\bar Q_{2L} \g^\mu Q_{3L})_1 \, (\bar \tau_R \g^\mu \tau_R).\label{eq:op4}
\end{align}
As is shown in appendix \ref{app:models}, $A_5\simeq 3A_1/4$ suppresses any NP contribution to $B_s\to\tau\tau$ for this model. This also generates a tiny contribution to $B\to K^{(*)}\mu\mu$ comparable to the SM contribution without the leptonic mixing tuned to unnaturally low values\footnote{The introduction of right handed leptonic current also opens the possibility of introducing independent mixing in the right-handed sector. However the authors do not explore the corresponding phenomenology in order to satisfy their requirement of ``minimality''.}. 
Similarly, we can define:
\begin{align}
\label{eq:op5}
{\rm Model~ V:}\quad   \mathcal{O}_{\rm V}  = & -\sqrt{3} \, A_1 \, (\bar Q_{2L} \g^\mu Q_{3L})_3 \, (\bar L_{3L} \g^\mu L_{3L})_3 \nonumber\\ 
& +  A_1 \, (\bar Q_{2L} \g^\mu Q_{3L})_1 \, (\bar L_{3L} \g^\mu L_{3L})_1 
\nonumber\\
& +  \sqrt{2} \, A_5 \, (\bar Q_{2L} \g^\mu Q_{3L})_1 \, (\bar \tau_R \g^\mu \tau_R).
\end{align}
Looking at the operator corresponding to Model V in terms of the component fields (Eq. \ref{eq:Va_comp}), $A_5\simeq A_1$ is preferred by $B_s\to\tau\tau$.

\begin{figure*}
	\centering
	\subfloat[Mod I: $A_1$ vs. $A_2$ (C0)]{\includegraphics[height=3.7cm]{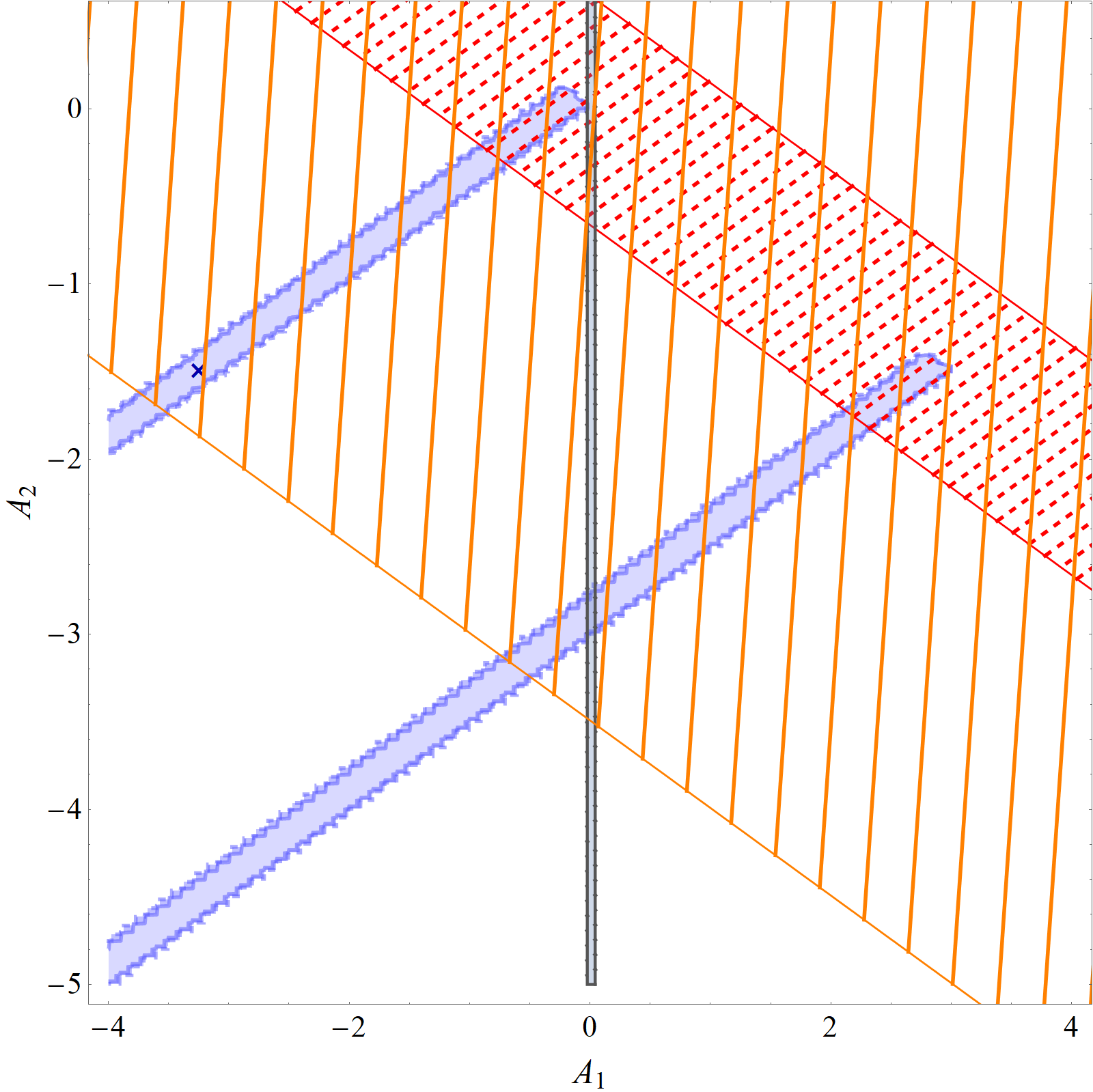}\label{fig:fewM1a1a2}}~~
	\subfloat[Mod I: $A_1$ vs. $A_2$ (C1)]{\includegraphics[height=3.7cm]{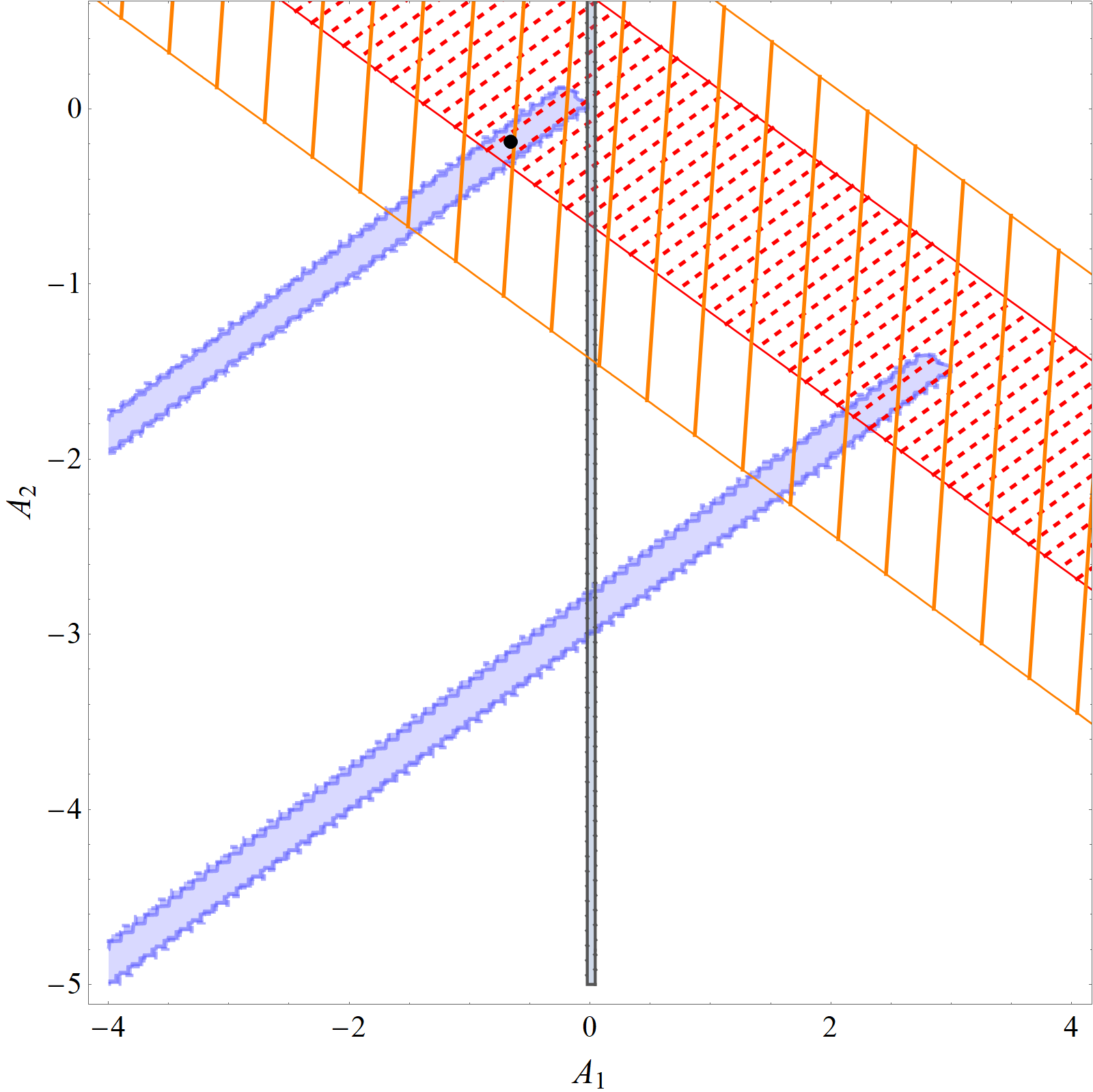}\label{fig:fewM1a1a2C}}~~
	\subfloat[Mod I: $A_2$ vs. $\sin\theta$ (C0)]{\includegraphics[height=3.7cm]{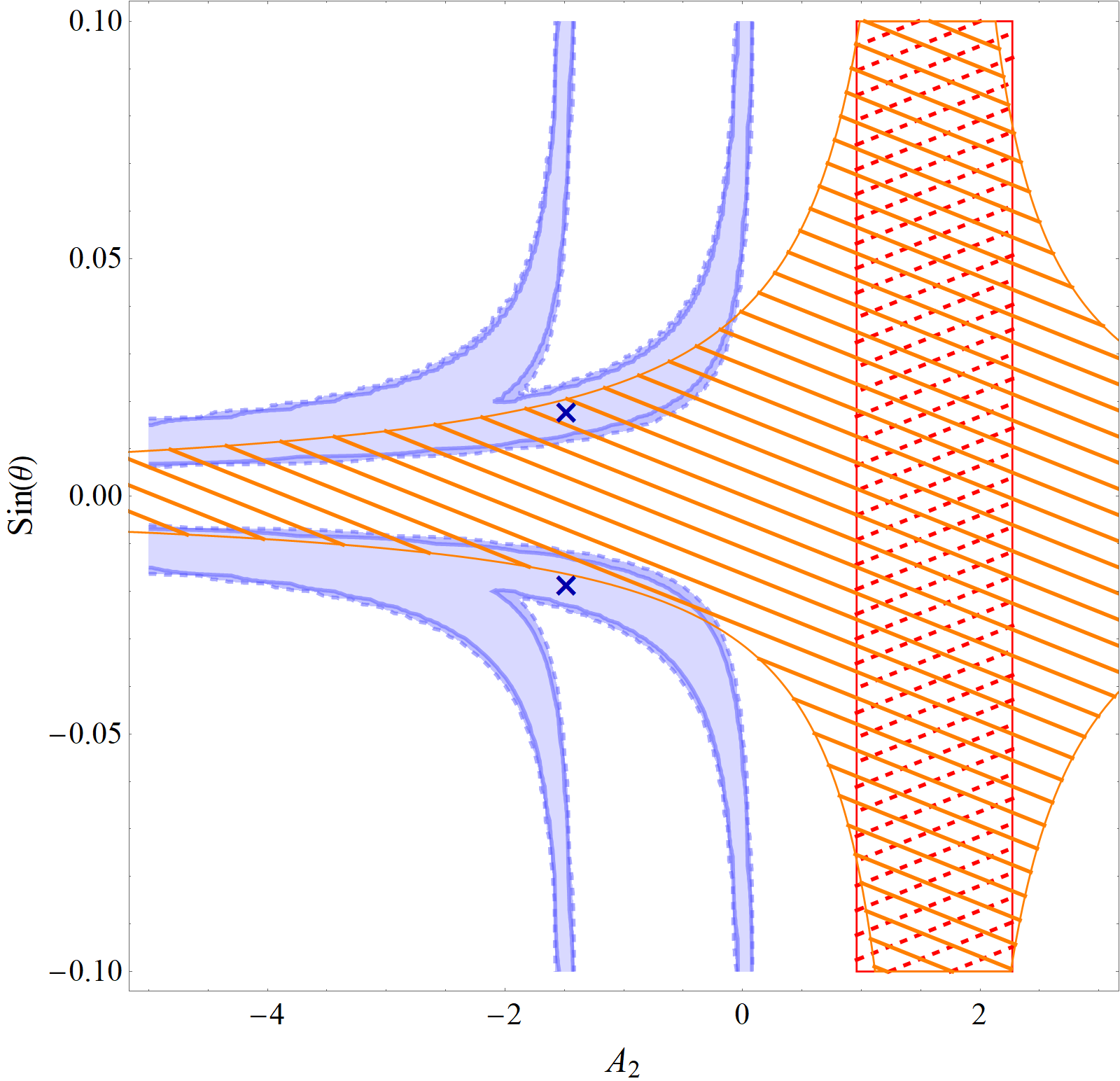}\label{fig:fewM1a2st}}~~
	\subfloat[Mod I: $A_2$ vs. $\sin\theta$ (C1)]{\includegraphics[height=3.7cm]{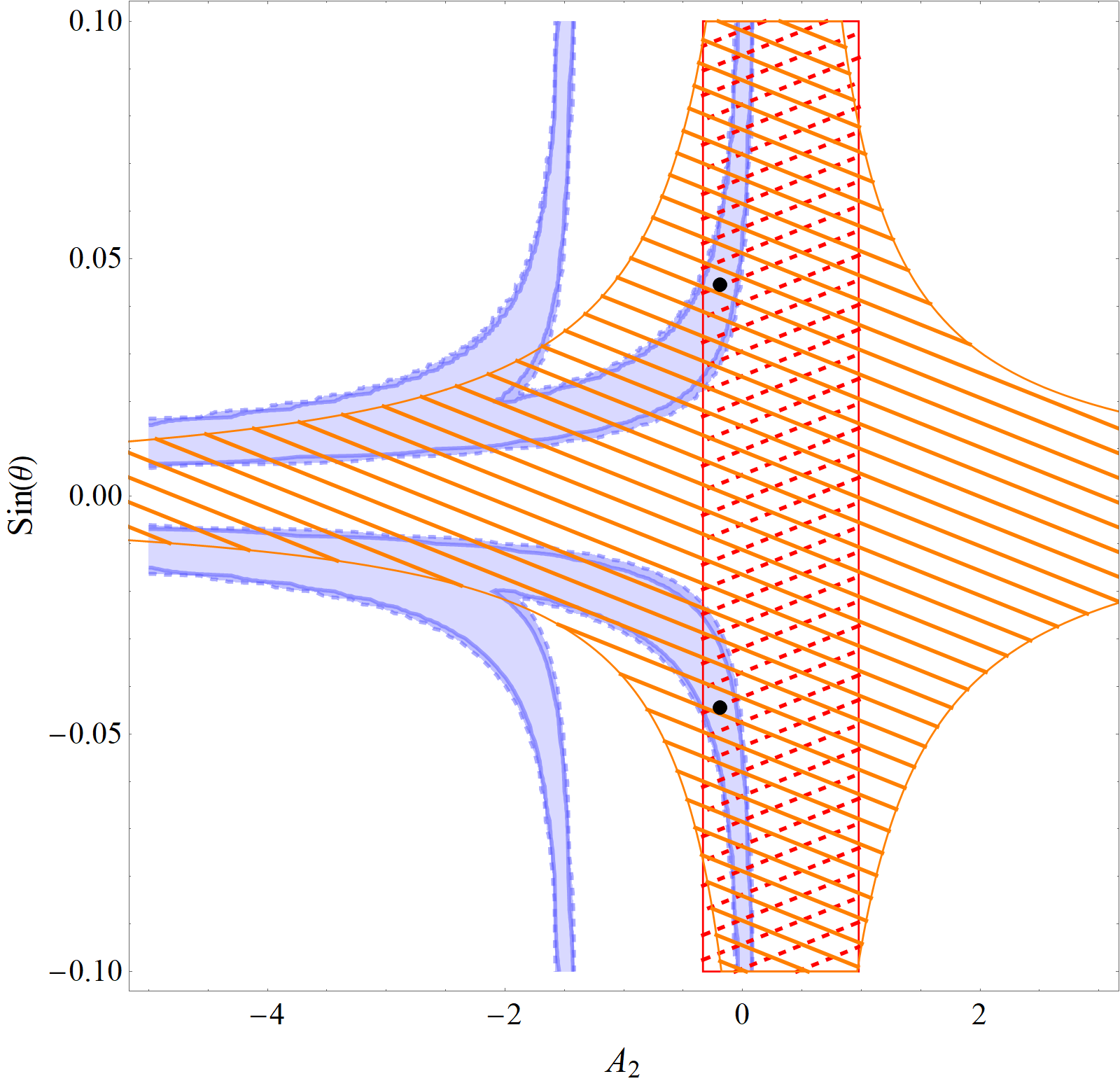}\label{fig:fewM1a2stC}}\\
	\subfloat[Mod II: $A_1$ vs. $A_2$ (C0)]{\includegraphics[height=3.7cm]{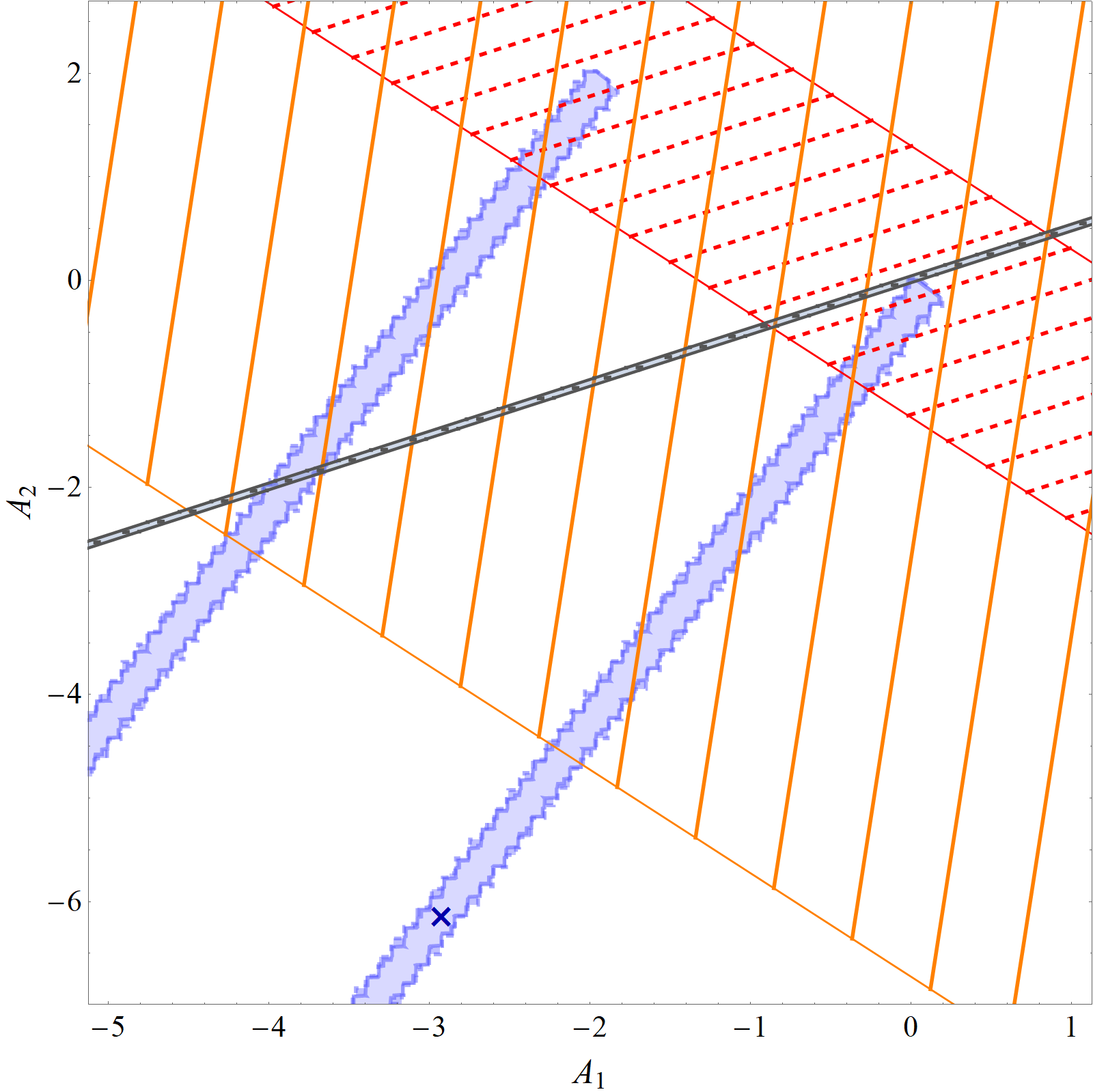}\label{fig:fewM2a1a2}}~~
	\subfloat[Mod II: $A_1$ vs. $A_2$ (C1)]{\includegraphics[height=3.7cm]{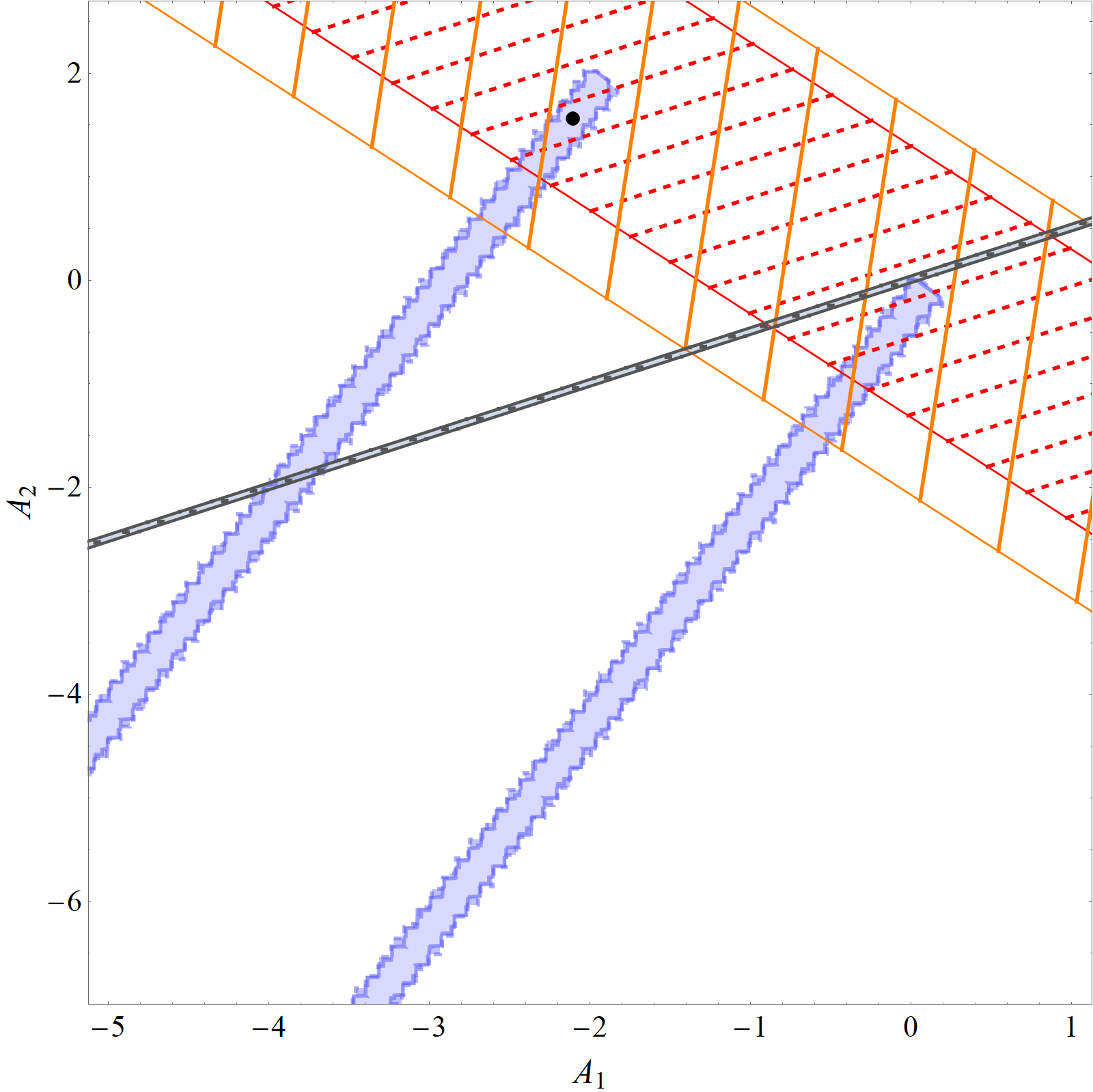}\label{fig:fewM2a1a2C}}~~
	\subfloat[Mod II: $A_2$ vs. $\sin\theta$ (C0)]{\includegraphics[height=3.7cm]{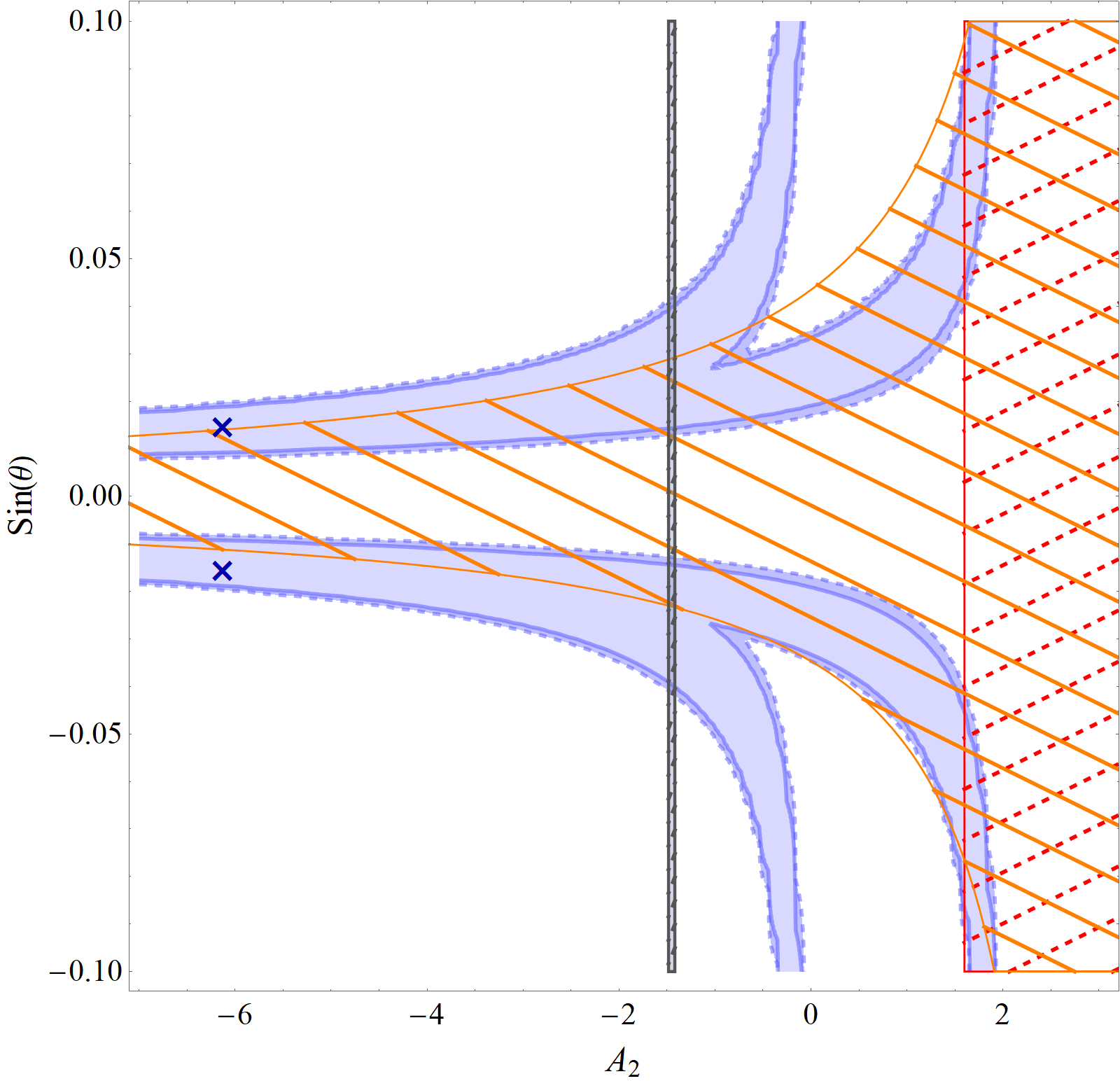}\label{fig:fewM2a2st}}~~
	\subfloat[Mod II: $A_2$ vs. $\sin\theta$ (C1)]{\includegraphics[height=3.7cm]{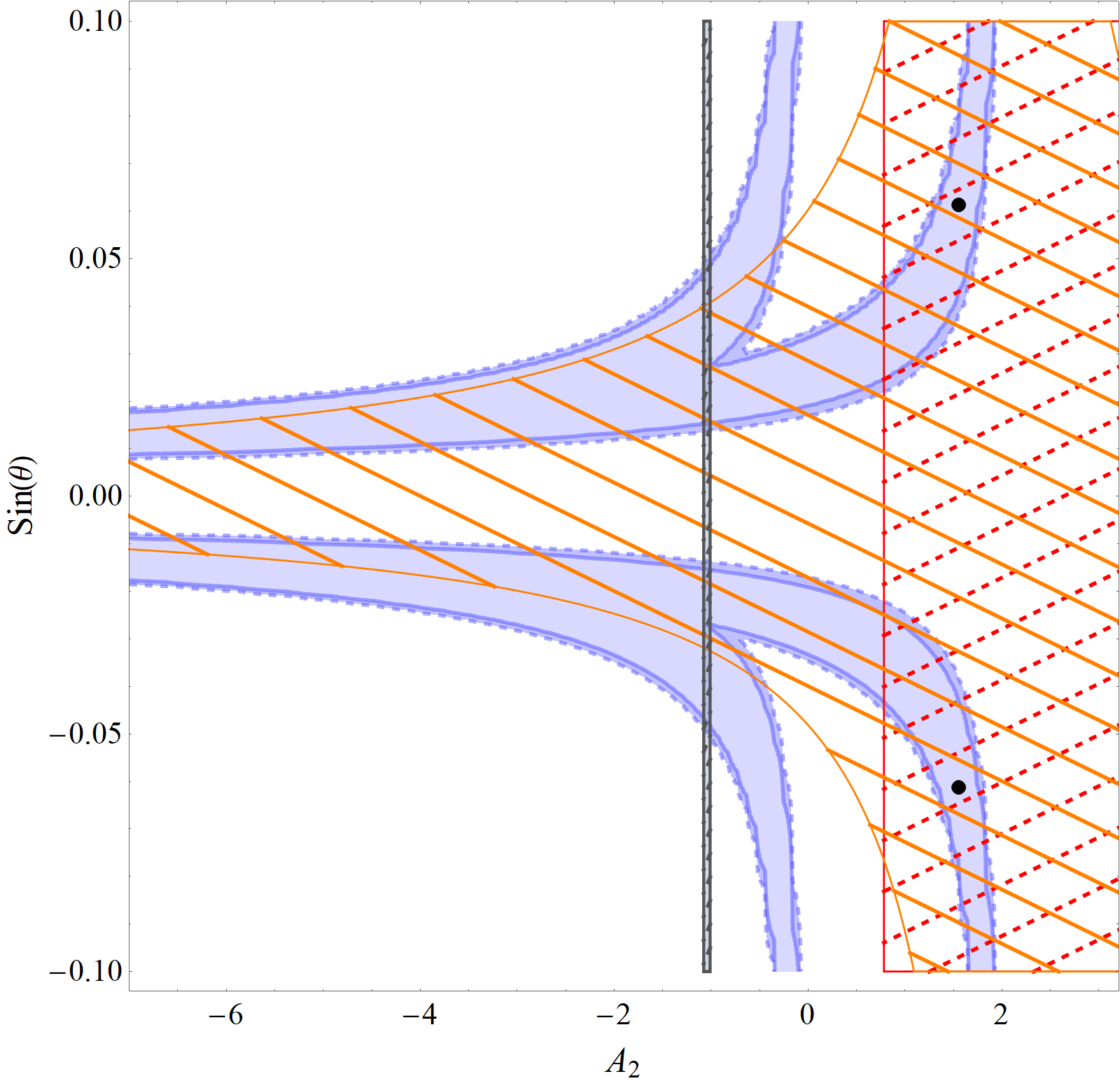}\label{fig:fewM2a2stC}}\\
	\subfloat[Mod III: $A_1$ vs. $A_3$ (C0)]{\includegraphics[height=3.7cm]{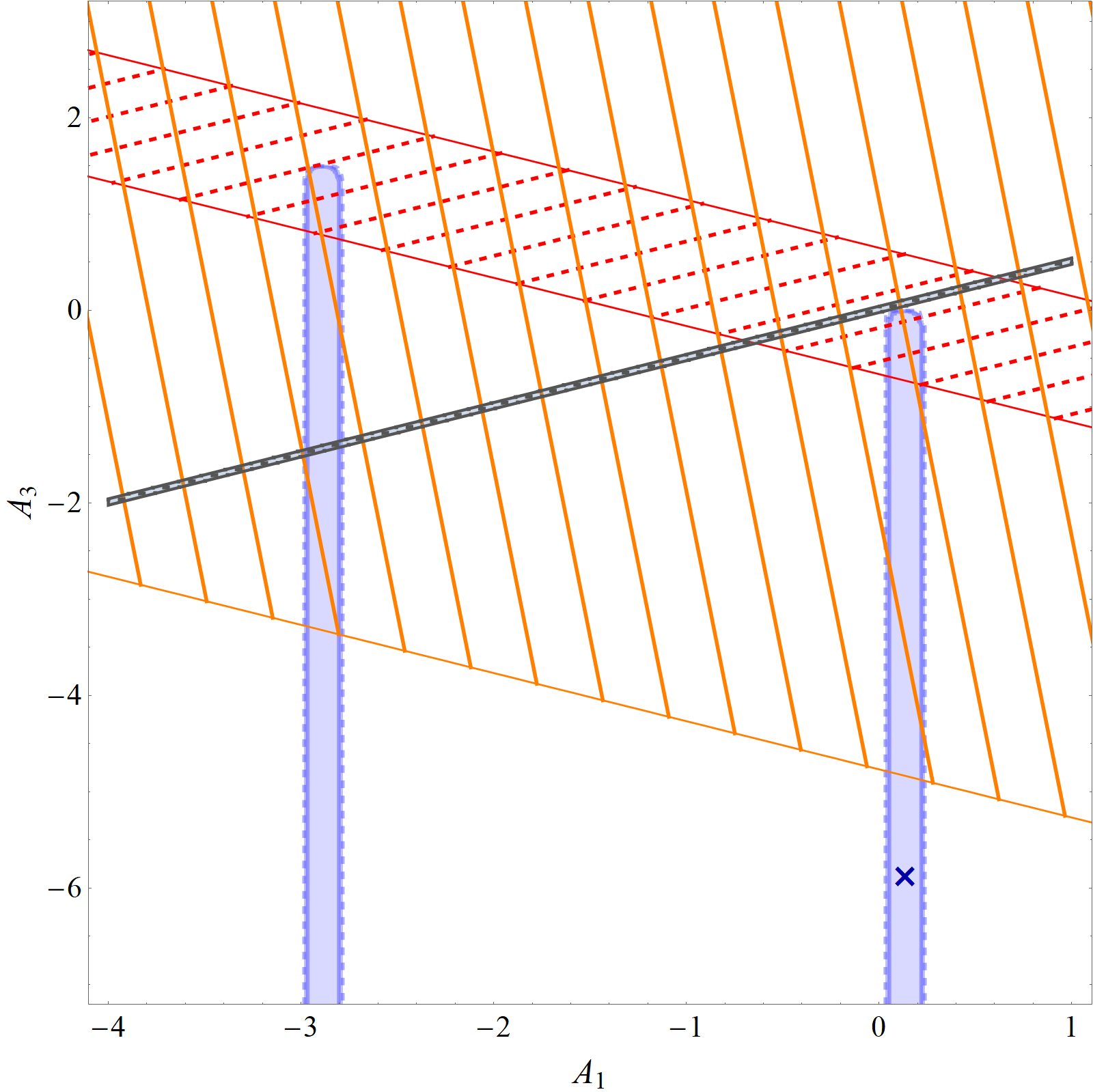}\label{fig:fewM3a1a3}}~~
	\subfloat[Mod III: $A_1$ vs. $A_3$ (C1)]{\includegraphics[height=3.7cm]{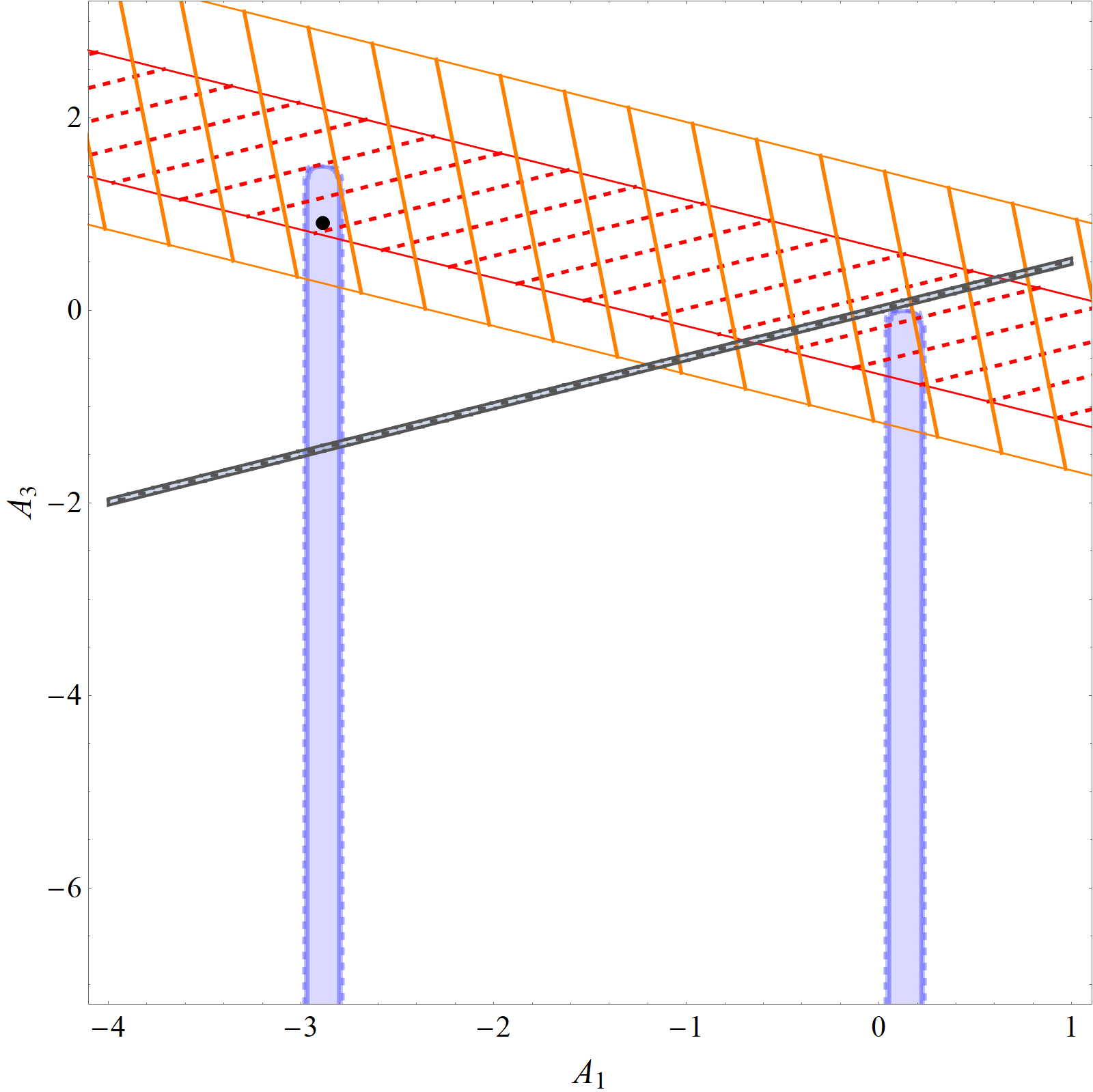}\label{fig:fewM3a1a3C}}~~
	\subfloat[Mod III: $A_3$ vs. $\sin\theta$ (C0)]{\includegraphics[height=3.7cm]{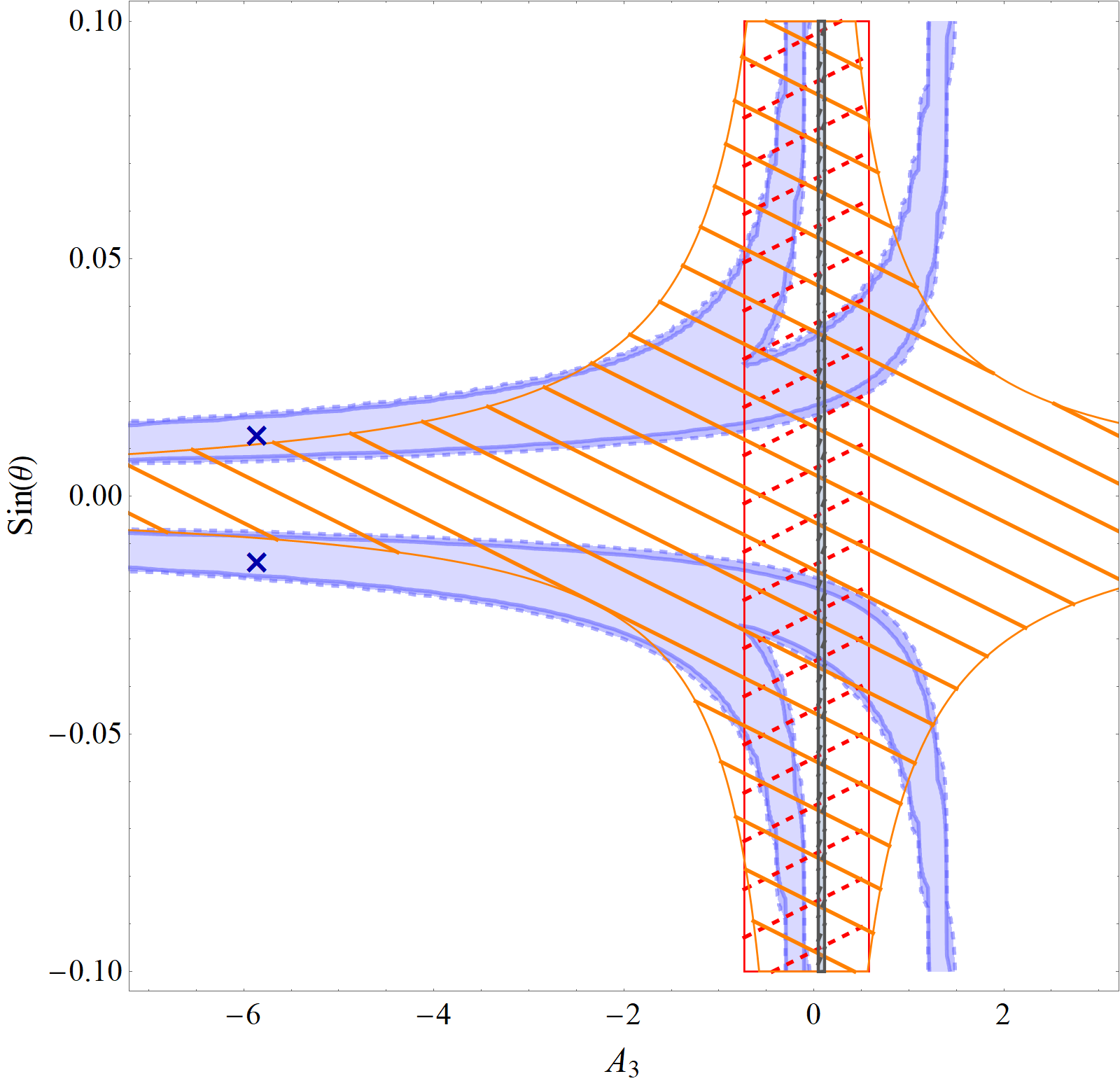}\label{fig:fewM3a3st}}~~
	\subfloat[Mod III: $A_3$ vs. $\sin\theta$ (C1)]{\includegraphics[height=3.7cm]{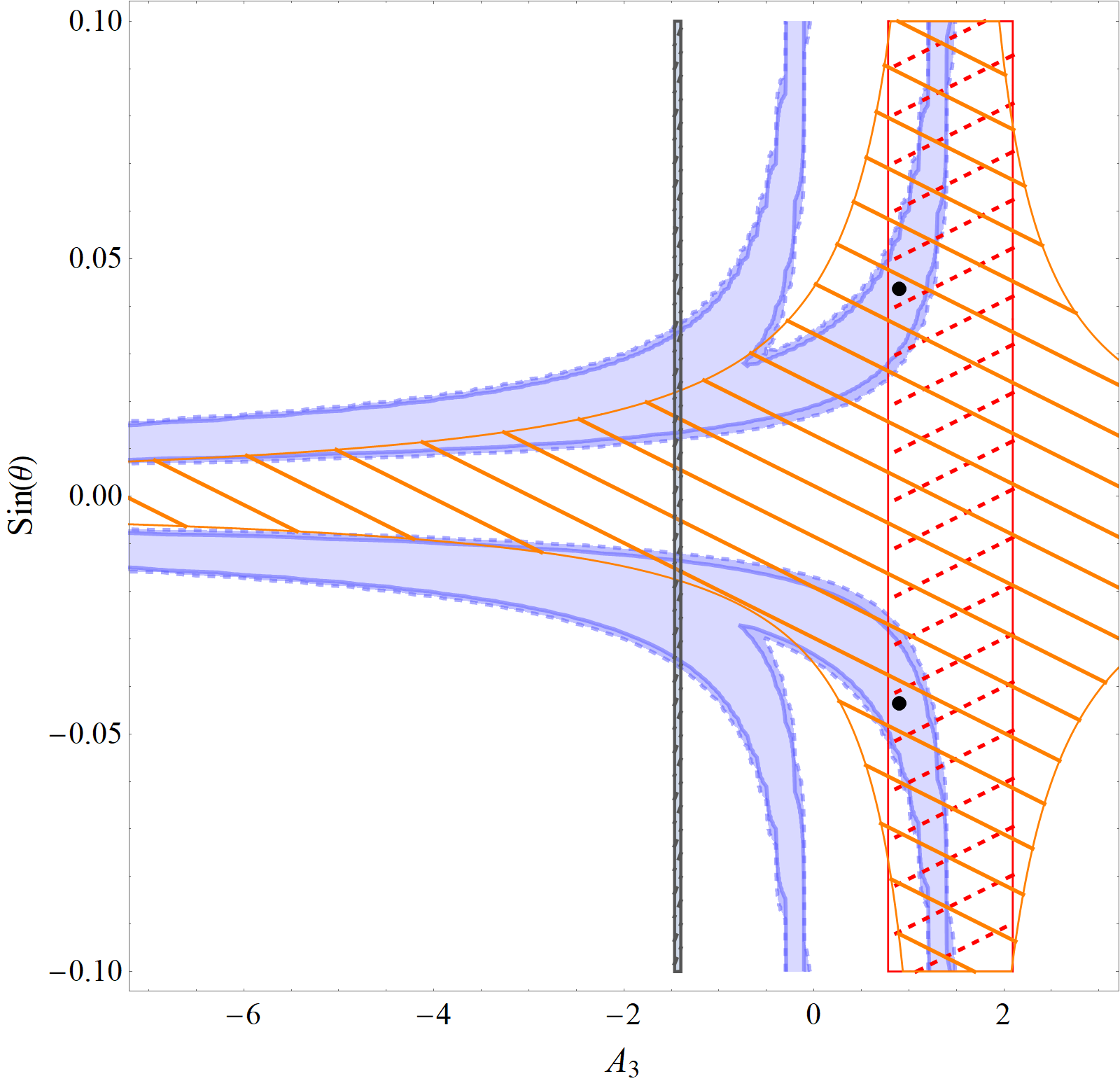}\label{fig:fewM3a3stC}}\\
	\subfloat[Mod IV: $A_1$ vs. $A_5$ (C0)]{\includegraphics[height=3.7cm]{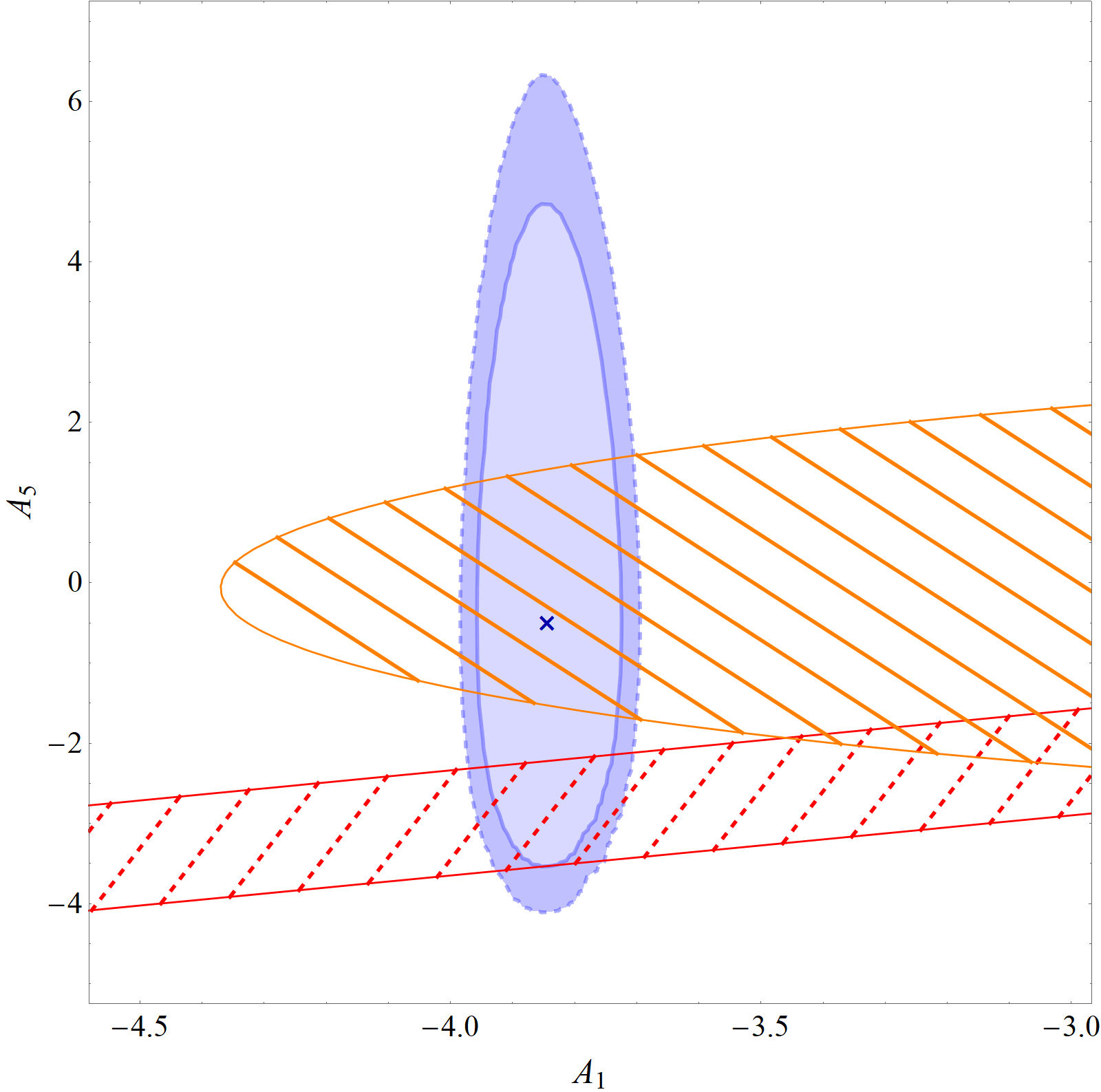}\label{fig:fewM4a1a5}}~~
	\subfloat[Mod IV: $A_1$ vs. $A_5$ (C1)]{\includegraphics[height=3.7cm]{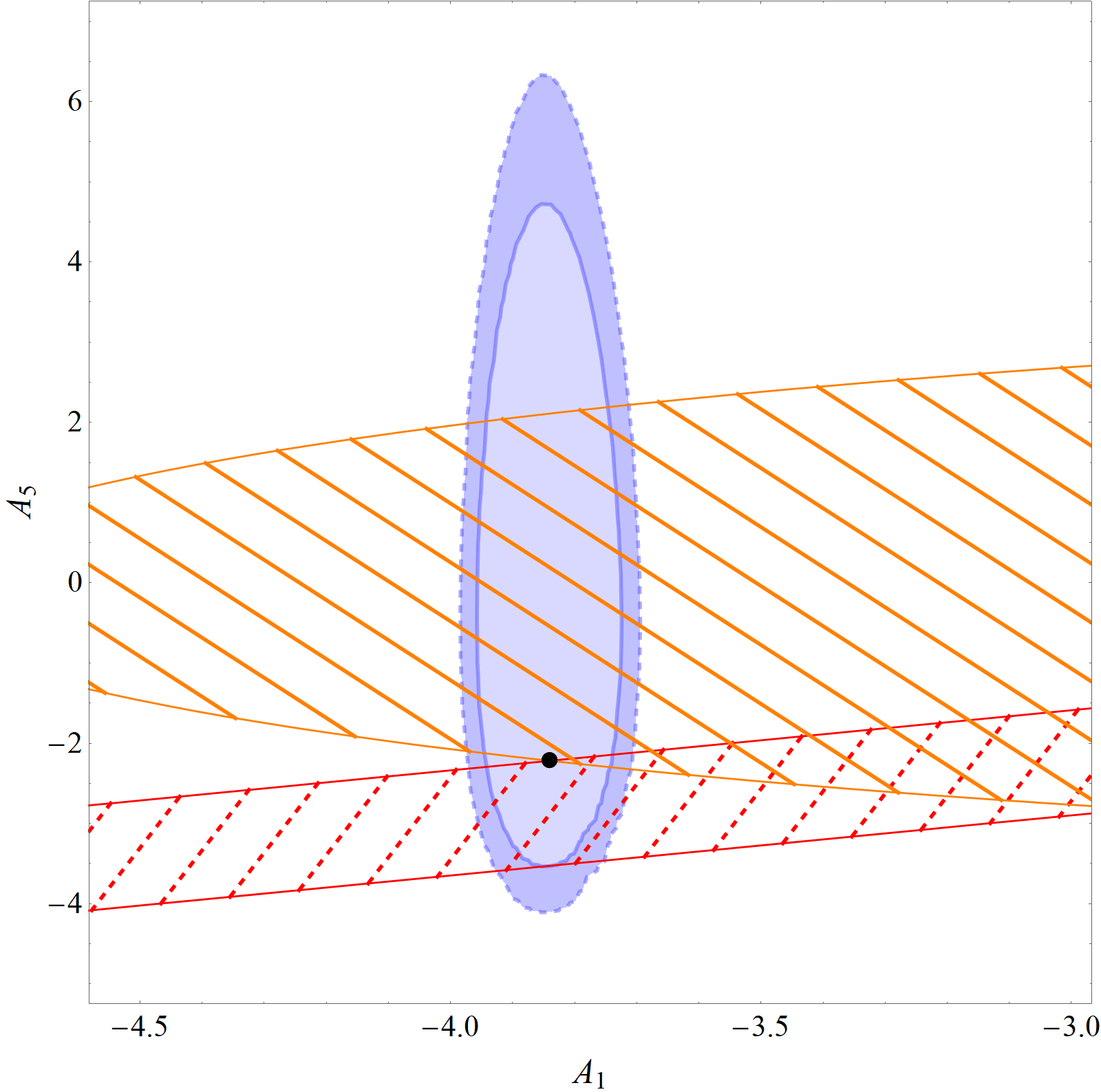}\label{fig:fewM4a1a5C}}~~
	\subfloat[Mod IV: $A_5$ vs. $\sin\theta$ (C1)]{\includegraphics[height=3.7cm]{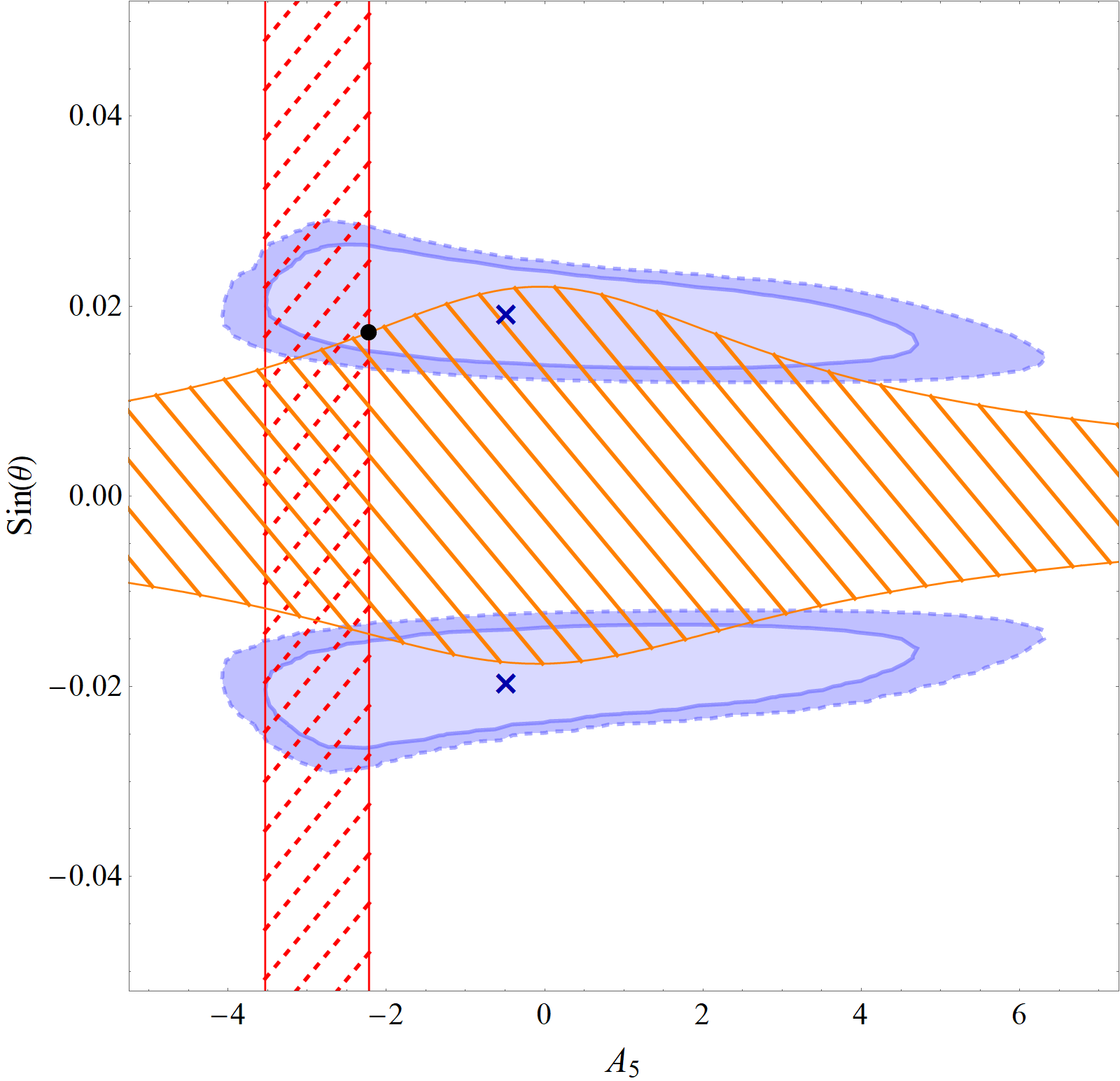}\label{fig:fewM4a5stC}}~~
	\subfloat[Mod V: $A_1$ vs. $A_5$ (C0)]{\includegraphics[height=3.7cm]{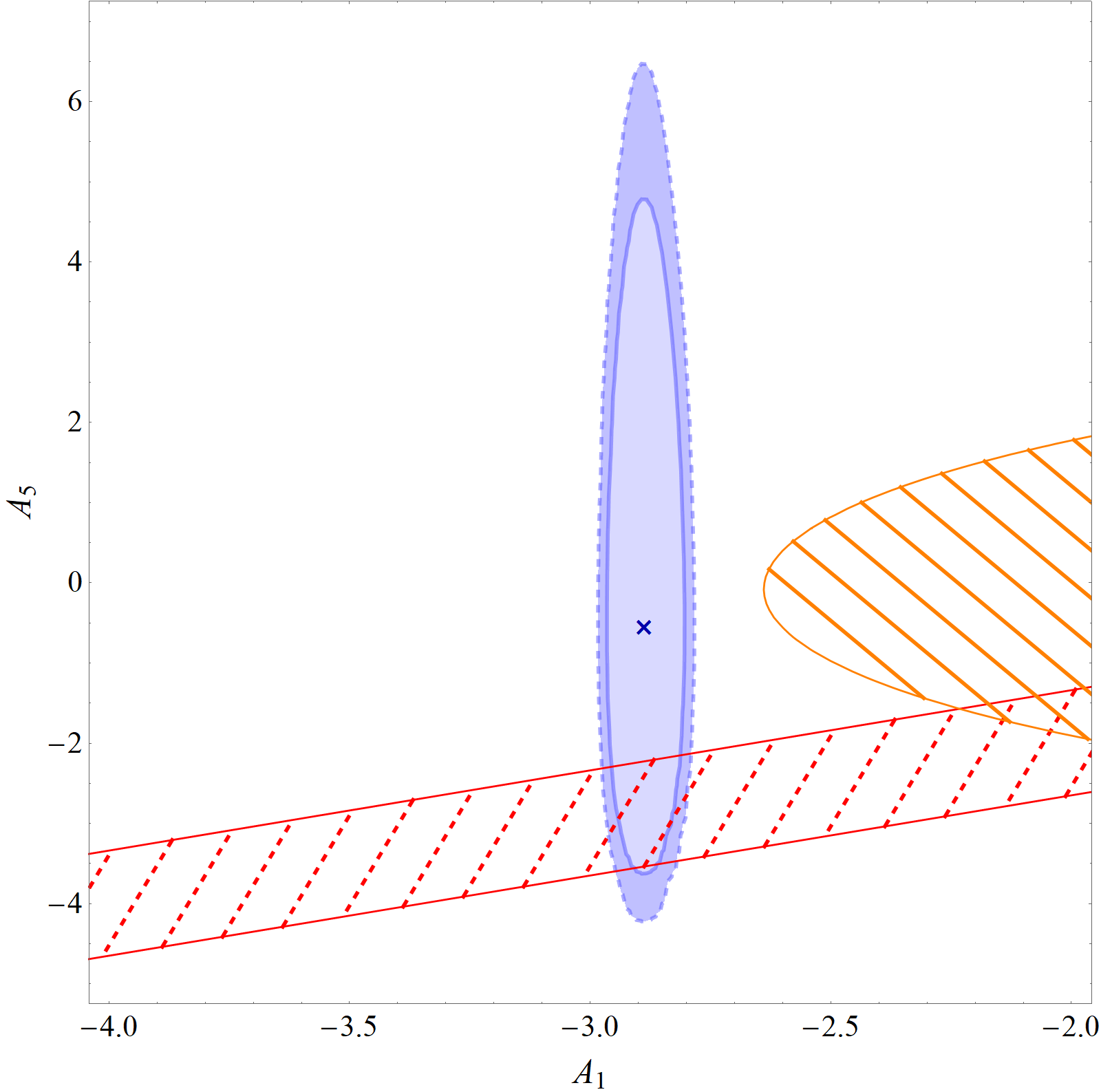}\label{fig:fewM5a1a5}}\\
	\subfloat[Mod V: $A_1$ vs. $A_5$ (C1)]{\includegraphics[height=3.7cm]{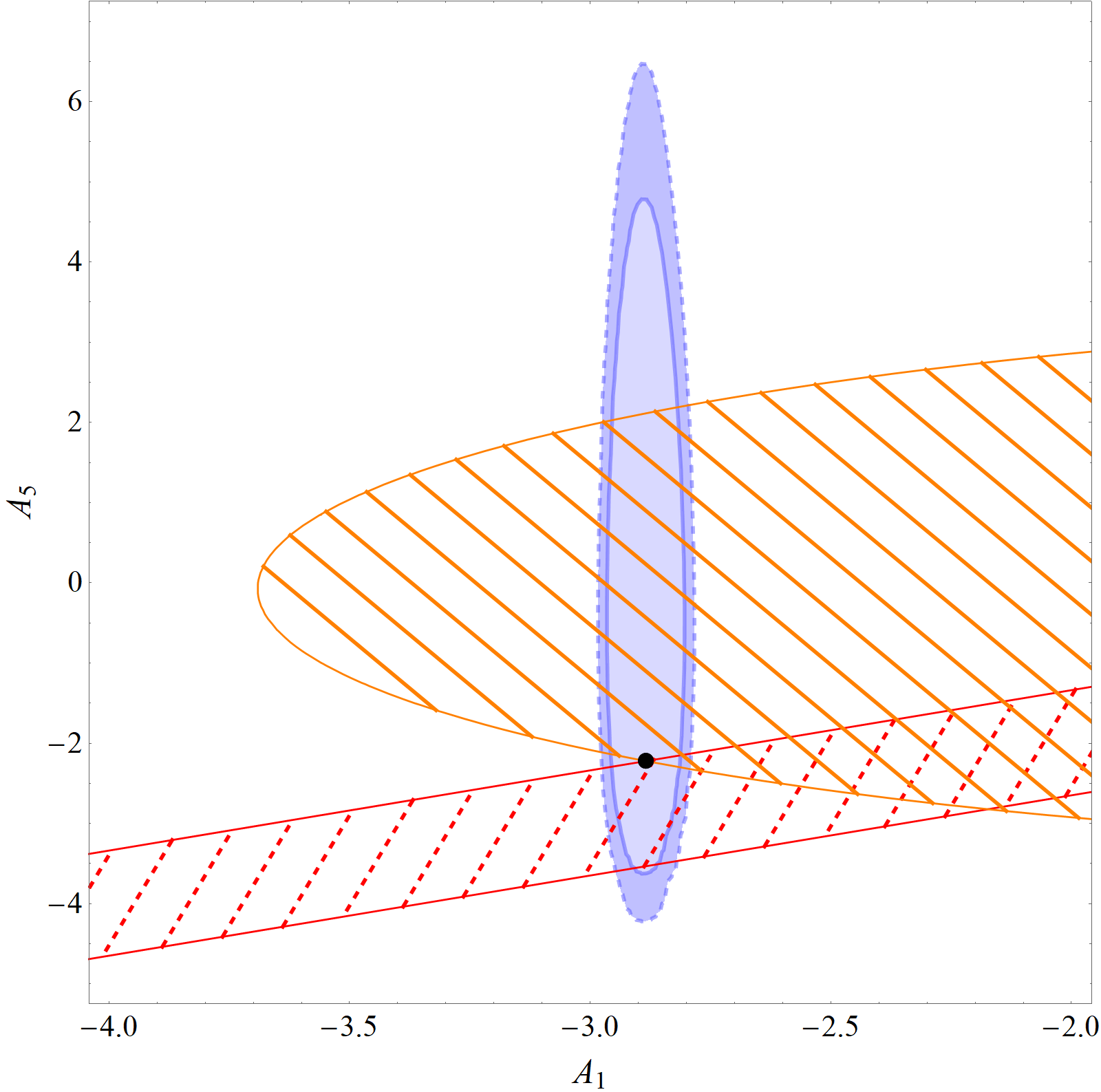}\label{fig:fewM5a1a5C}}~~
	\subfloat[Mod V: $A_5$ vs. $\sin\theta$ (C1)]{\includegraphics[height=3.7cm]{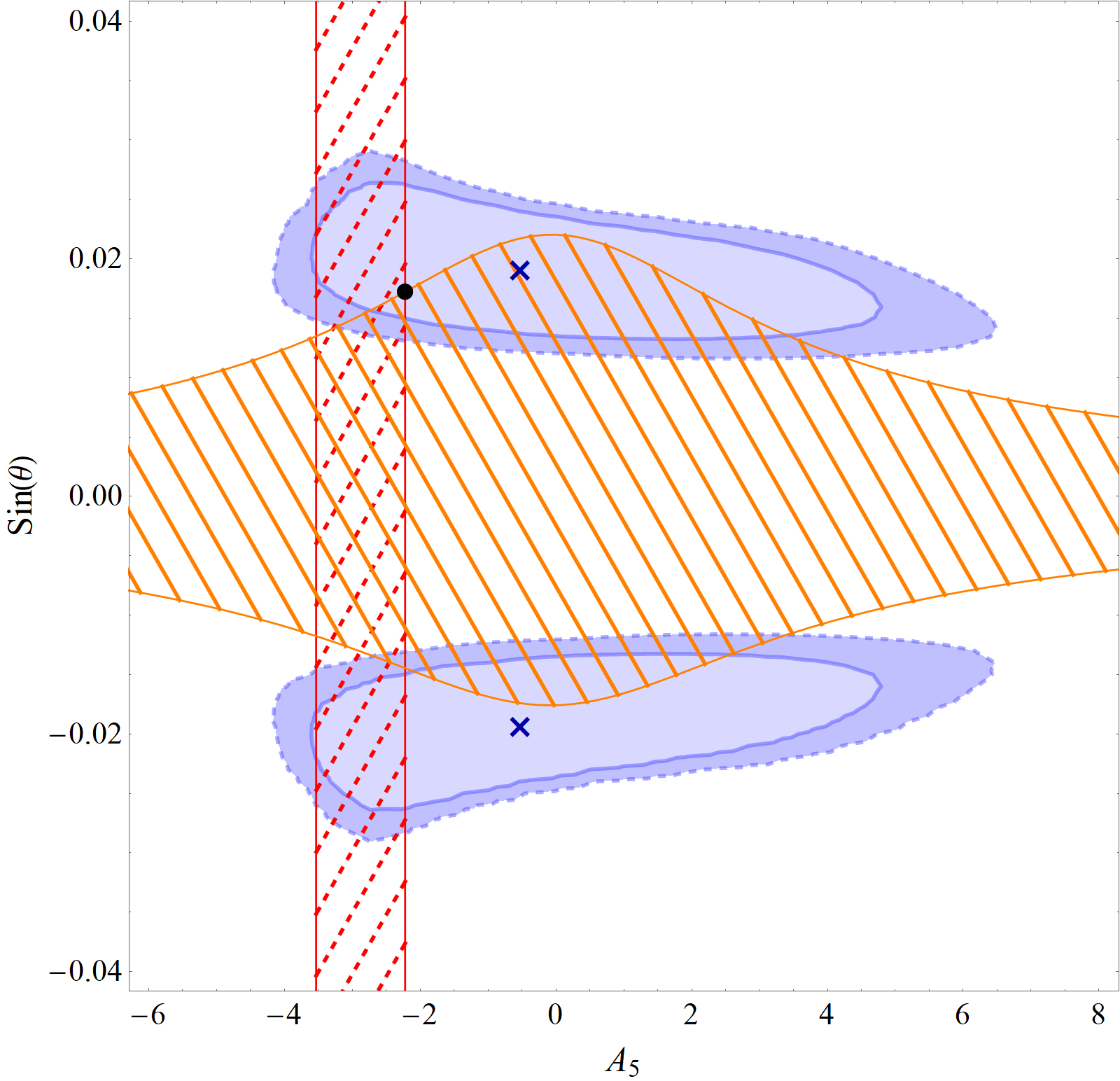}\label{fig:fewM5a5stC}}~~
	\subfloat[Legends]{\includegraphics[height=3.5cm]{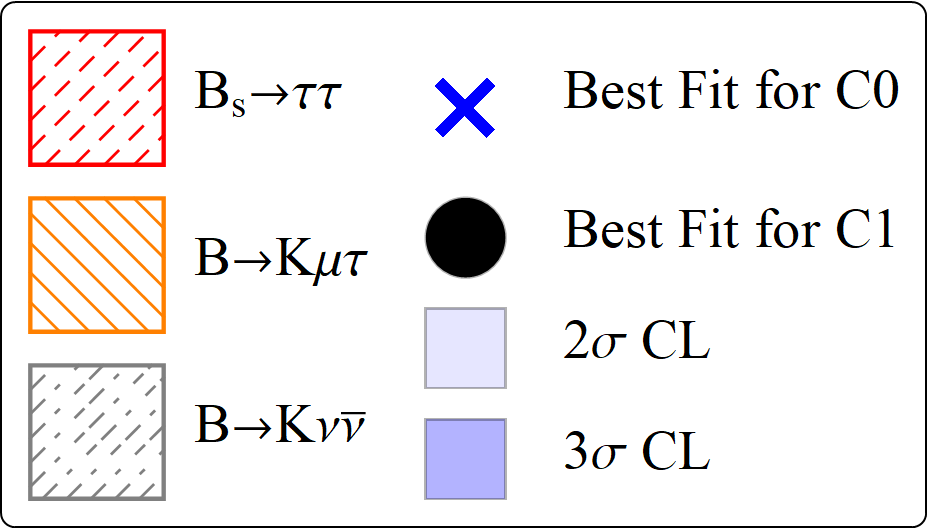}\label{fig:legend2D}}
	\caption{Two dimensional parameter confidence regions for all models for the `small dataset' (see section \ref{sec:anafew}). For most models, both fits C0 and C1 are shown. Only when the constraint regions are almost identical, we just show the result for one of them, with different markers for the best-fit points. The accompanying legend \ref{fig:legend2D} shows different markers to read the plots.}
	\label{fig:fewdat2D1}
\end{figure*}

\subsection{Motivation}\label{sec:motivation}
There are several reasons to take a look at these models again:
\paragraph{Data} Though the main motivation for the conception of these models is to explain the persistent deviations present in the LFUV observables, passing these through the grind of the bulk of observations available for $b\to s\ell\ell$ transitions, e.g. angular observables for the $B\to K^*\ell\ell$ decay, 
is never a bad idea. Though these observables are less precise than the `clean' ones mainly used in the small dataset, their inclusion ensures that the statistical inference we are seeking for remains unbiased, i.e. the data are not cherry-picked. In the worst case, the process would prove to be an overkill, keeping the inferences intact, or, in the best case, would point to new information/inference, such as new deviations to rule out a model or finding a different parameter space for a previously disallowed model. We need to mention here, that with the larger uncertainties, any new (normalized) deviation for these additional data are even more significant.

In addition, there has been a new measurement of $D^*$ polarization ($F_L(D^*)$) in the decay mode $B\to D^* \tau \nu$ since the proposition of these models \cite{Adamczyk:2019wyt}. Though still imprecise, it is consistent with SM within $2\sigma$. 
\paragraph{Theory} SM calculations of observables in $B\to D^{(*)} \tau \nu$ mode have earlier been taken from older works (see section \ref{sec:dataset}) and new results slightly reduce the deviation. We want to consistently follow \cite{Jaiswal:2017rve}, which will also help us keep track of the theoretical correlations. Though the effects of these in the final results are small, their effects need a thorough examination in presence of a larger number of observables.

Also, in this work we provide the most precise SM calculation of $R(J/\Psi)$ yet, and we need to check the effect of that in the analysis too.  
\paragraph{Analysis} Our most important motivation for this analysis is the actual way models were classified as `realistic' and `semi-realistic' in ref. \cite{Choudhury:2017ijp} based on whether the best-fits were satisfying all constraints from all important limits. For model I, first the goodness of fit was assessed using reduced $\chi^2$, then it was argued that the best-fit violates the $B_s\to \tau^+\tau^-$ limit. But, the same is true for models IV and V, where the actual global minimum does not satisfy both limits from $B_s\to \tau^+\tau^-$ and $B\to K\mu\tau$. Here, a process equivalent to constrained minimization was performed to obtain the new minimum satisfying the constraints, which is quite a good fit too. It is not clear whether this same procedure has been followed in the case of Models I-III as well, before putting them aside as `semi-realistic'. It had also been argued against model I that to tweak the WCs to satisfy the $B_s\to \tau^+\tau^-$ limit, the best fit for $R(K^*)$ (explanation of which is one of the main motivations) moves nearly $2\sigma$ away from the global average. It is apparent from the results of models IV and V that this argument is valid for them as well. Thus, we felt the need to look into whether models I-III are really unworthy, and if models IV and V really work that well.

Another small, but persistent concern is the effect correlations have on the fit. 
Admittedly, for small number of data, the experimental correlations (which only are available for some of the observables) will not play a significant role, and the same is true for the theoretical ones (the uncertainties themselves are very small compared to the experimental ones). On the other hand, for any analysis involving a large number of mostly correlated data, those correlations may play a significant role to change the quality of the fits.

\section{Analysis}
\subsection{Methodology}\label{sec:method}
\subsubsection{Types of Fits and Datasets}\label{sec:dataset}
We have done our analysis in two parts. First, we exactly recreate the analyses of ref.s \cite{Choudhury:2017qyt,Choudhury:2017ijp} to be sure of the nature of the parameter space and their relation to the constraints. The eight observables they considered are two global averaged $R_{D^{*}}$, three $R_{K^(*)}$, $R(J/\psi)$, the $q^2\in[1,6]$ GeV$^2$ bin for $B_s\to\phi\mu\mu$ and the $B_s\to\mu\mu$ branching ratio. Data for these fits are taken from those papers. We will call it the `small dataset' from now on. We redo the analysis in the second part with a total of 170 observables listed below, which will be mentioned as the `large dataset' from now on: 
\paragraph{$\bm{b\to c\tau\nu}$:} We have not only added the newest $R(J/\Psi)$ \cite{Aaij:2017tyk} and the $F_L(D^*)$ \cite{Adamczyk:2019wyt} data, but have also included individual measurements of $R(D^{(*)})$ and $P_{\tau}(D^*)$ as separate entries \cite{Lees:2012xj,Lees:2013uzd,Huschle:2015rga,Sato:2016svk,Aaij:2015yra,Hirose:2017dxl,Hirose:2016wfn,Aaij:2017uff,Aaij:2017deq}. This ensures the correct treatment of experimental correlations and increases the degrees of freedom for the fit. There are thus 11 data-points in this segment. In addition to the widely used last generation of SM results for $R(D)$ \cite{Lattice:2015rga,Na:2015kha,Aoki:2016frl,Bigi:2016mdz} and $R(D^*)$ \cite{Fajfer:2012vx}, several calculations have become available since 2017 \cite{Bernlochner:2017jka,Bigi:2017jbd,Jaiswal:2017rve}. Of these, we have extensively followed ref. \cite{Jaiswal:2017rve} and the results therein, to not only calculate $R(D^{(*)})$, but other observables for the $B\to D^{(*)}\tau\nu$ as well. This gives us the opportunity to follow the theoretical correlations among them correctly. Theoretical estimate of $R(J/\Psi)$ used in this analysis is obtained by following the perturbative QCD (PQCD) parametrization of the form factors \cite{Wen-Fei:2013uea}. For this parametrization, the form factor parameters appear in the exponent, and the estimate of uncertainty generally do not take the correlations between the numerator and the denominator into account. For our analysis, we have calculated the uncertainty by generating a Monte-Carlo (MC) dataset and obtaining the result numerically. This result is checked with varying number of data-points and the MC uncertainty is orders of magnitude smaller than the quoted one, i.e. the result is stable. This is the most precise calculation of $R(J/\Psi)$ till date as per our knowledge. We are aware of the fact that the SM calculation of this observable depends heavily on the form factor parametrization and to compare with a competing parametrization, we showcase another result using the light-front covariant quark (LFCQ) model and the obtained smaller uncertainties increase the tension between them by many folds:
\begin{align}
	\nn R(J/\Psi)_{SM(PQCD)} &= 0.289\pm 0.005\\
	R(J/\Psi)_{SM(LFCQ)} &= 0.249\pm 0.006
\end{align} 
\paragraph{$\bm{ b\to s \ell \ell}$:} We have considered 159 observables in this sector. These include angular observables ($F_L$, $S_3$, $S_4$, $A_5$, $A_6$, $S_7$, $A_8$, $A_9$)obtained from the unbinned maximum likelihood fit and branching ratios in three $q^2$ bins\footnote{All $q^2$ bins have been provided in units of GeV$^2$} ($0.1\to  2$, $2\to 5$, and $1\to 6$) for $B_s\to \phi\mu\mu$ channel \cite{Aaij:2015esa}, $B_s\to \mu^+\mu^-$ branching ratio \cite{CMS:2014xfa}, and $B^+\to K^{*+}\mu^+\mu^-$ branching ratios in four $q^2$ bins ($0.1\to 2$, $2\to 4$, $4\to 6$, and $1.1\to 6$) \cite{Aaij:2014pli}. Next, for the $ B^+\to K^+\mu^+\mu^-$ channel, we have branching ratios in seven $q^2$ bins  ($0.1\to 0.98$, $1.1\to 2$, $2\to 3$, $3\to 4$, $4\to 5$, $5\to 6$, and $1.1\to 6$) \cite{Aaij:2014pli} and $R(K)$ \cite{Aaij:2014ora}. For the $B^0\to K^{*0}\mu^+\mu^-$ channel, we have added angular observables ($F_L$, $S_3$, $S_4$, $S_5$, $A_{FB}$, $S_7$, $S_8$, $S_9$, $A_3$, $A_4$, $A_5$, $A_{6s}$, $A_7$, $A_8$, $A_9$, $P_1$, $P_2$, $P_3$) evaluated using the method of moments in six $q^2$ bins ($0.1\to 0.98$, $1.1\to 2$, $2\to 3$, $3\to 4$, $4\to 5$, $5\to 6$) \cite{Aaij:2015oid}, branching ratios in five $q^2$ bins ($0.1\to 0.98$, $1.1\to 2.5$, $2.5\to 4$, $4\to 6$, $1.1\to 6$) \cite{Aaij:2016flj}, $R_{K^*}^{central}$ and $R_{K^*}^{low}$ \cite{Aaij:2017vbb}. Finally, we have  $B^0\to K^0\mu^+\mu^-$ branching ratios in four $q^2$ bins ($0.1\to 2$, $2\to 4$, $4\to 6$, $1.1\to 6$)	\cite{Aaij:2014pli}. For the $B\to K$ form factors, a combined fit to the recent lattice computation as well as LCSR predictions at $q^2$ = 0, as given in \cite{Altmannshofer:2014rta} was used and the SM predictions for the observables in $ B^+\to K^+\mu^+\mu^-$ and $B^0\to K^0\mu^+\mu^-$ have been made. For the observables in $B^+\to K^{*+}\mu^+\mu^-$ , $B^0\to K^{*0}\mu^+\mu^-$ and $B_s\to \phi\mu\mu$ channels, LCSR form factor parametrization \cite{Straub:2015ica} was used in the analysis and the SM estimates have been made. The definition of the observables in the $B\to V \mu^+ \mu^-$ modes have been given in ref.s \cite{Descotes-Genon:2013vna,Altmannshofer:2008dz}. The SM prediction for $ B_s\to \mu^+\mu^-$ branching ratio has been given in ref. \cite{Becirevic:2012fy}.

All optimizations in this paper are done using \textit{Mathematica}\textsuperscript{\textcopyright} in the form of a package \cite{OptEx}. As we use the assumption that the optimizing statistic follows $\chi^2$ distribution (equivalent to the `PROB' method in literature \cite{cern1996cernlib,Aaij:2016kjh}), the parameter confidence levels (CLs) are obtained without any external constraints. Constraints such as \bstt, \bKmt, and \bKnn~ can only be applied for the best fit points, after the CLs are obtained\footnote{We have found that compared to the size of the parameter CLs, the constraint regions for \bKnn~ and \bKstnn~ are almost of the same order and thus we will only mention \bKnn~ from hereon, when we mean both of these decays}. We have found that almost all of the parameter spaces relevant to our analysis is conveniently allowed by the recently widely discussed $\sim 30\%$ theoretical constraint for $\mathcal{B}(B_c\to \tau\nu)$ \cite{Alonso:2016oyd}, and so we are not going to mention it explicitly from hereon.  As experimental constraints from \bstt~ and \bKmt~ decays were found to be the determining factors to discard models in ref. \cite{Choudhury:2017ijp}, we have obtained, for both small and large datasets, three types of best fit results: 
\begin{enumerate*}
	\item one without any constraints, (called \textbf{C0} from hereon)
	\item one with only $B_s\to\tau\tau$ and $B\to K\mu\tau$ decays (\textbf{C1}), and finally,
	\item one with all constraints (\textbf{C2}). 
\end{enumerate*}

\begin{figure*}
	\centering
	\subfloat[Mod IV: $A_1$ (C0)]{\includegraphics[height=5cm]{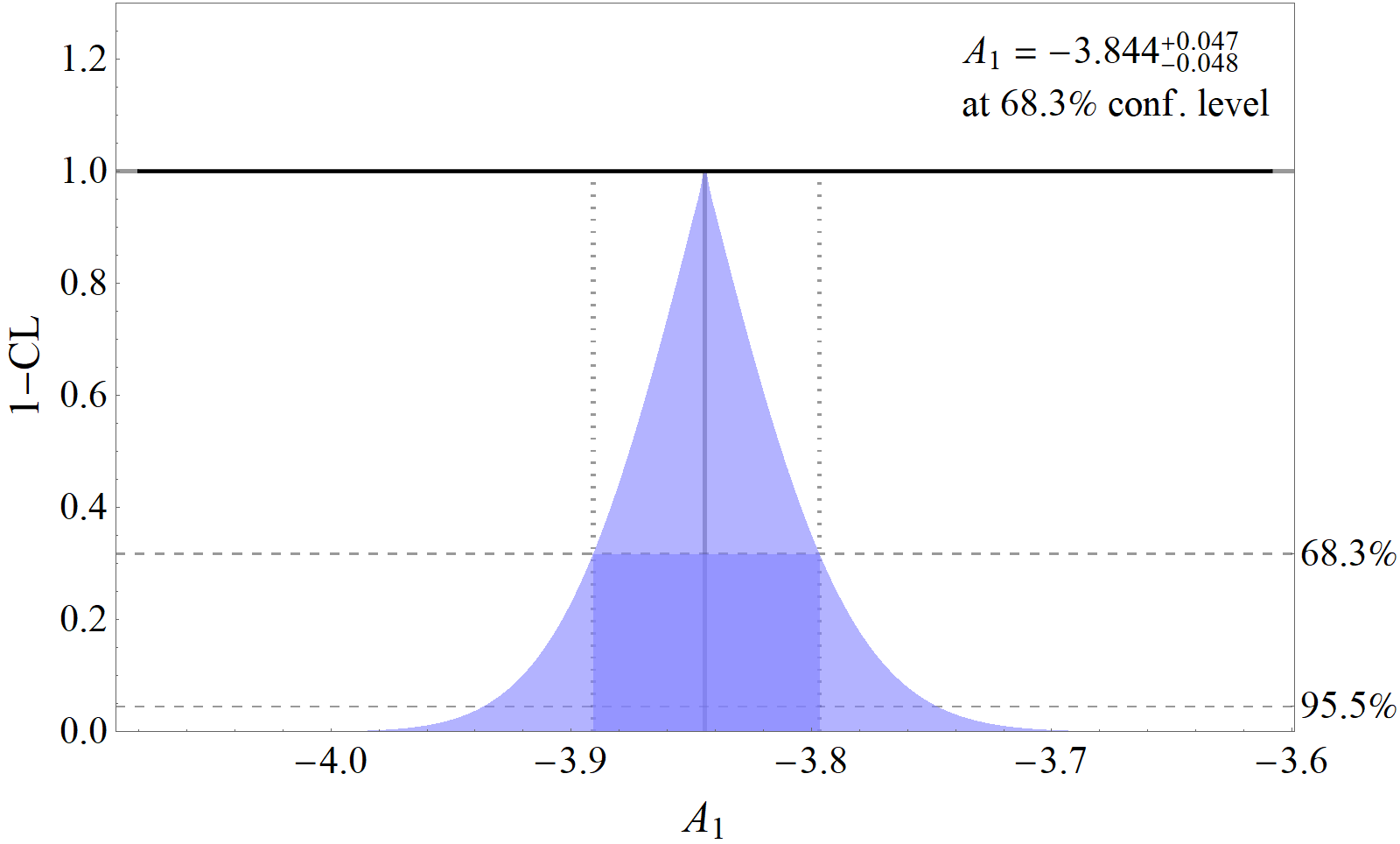}\label{fig:fewM4a11D}}~~
	\subfloat[Mod IV: $A_5$ (C0)]{\includegraphics[height=5cm]{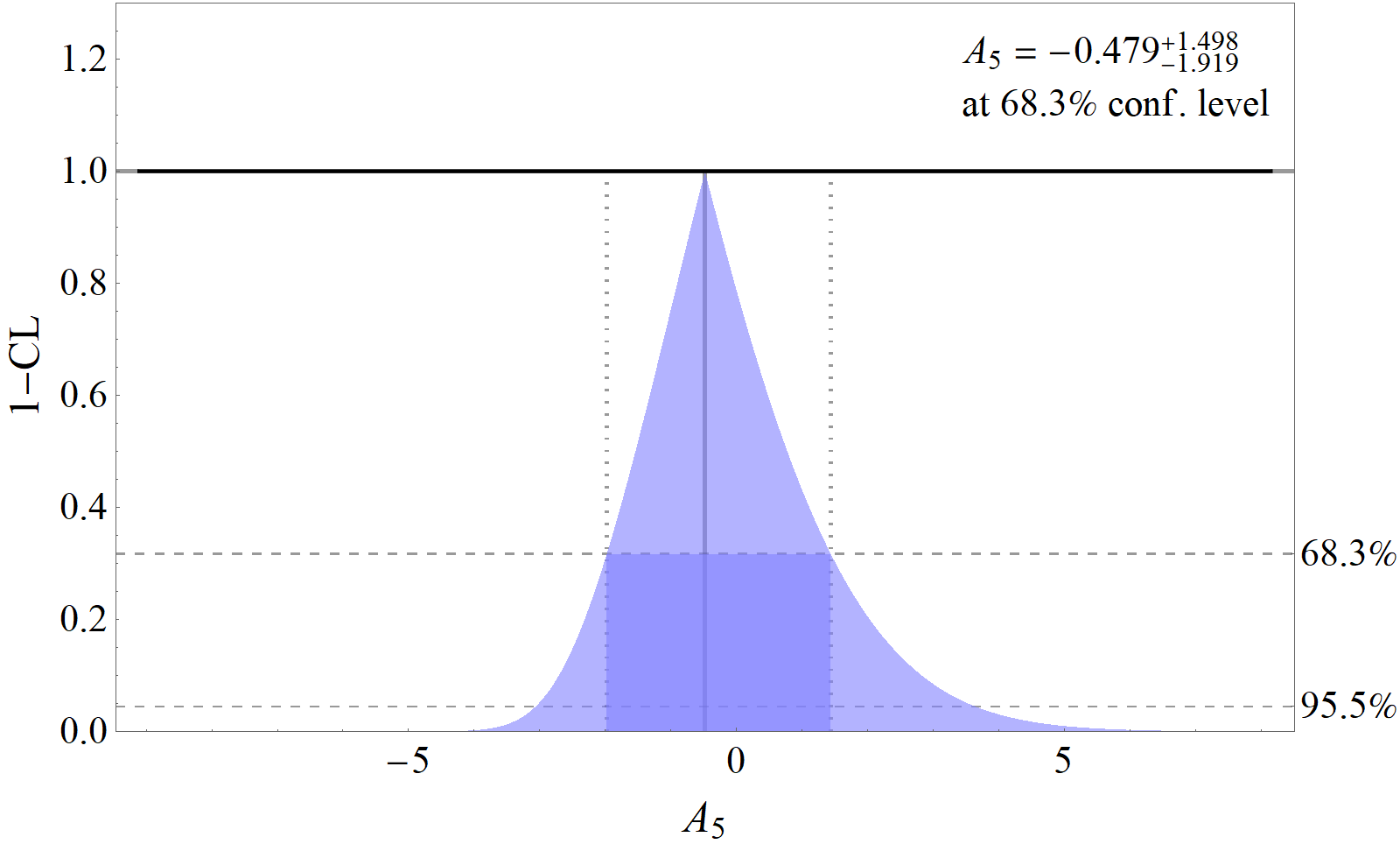}\label{fig:fewM4a51D}}\\
	\subfloat[Mod IV: $\sin\theta$ (C0)]{\includegraphics[height=5cm]{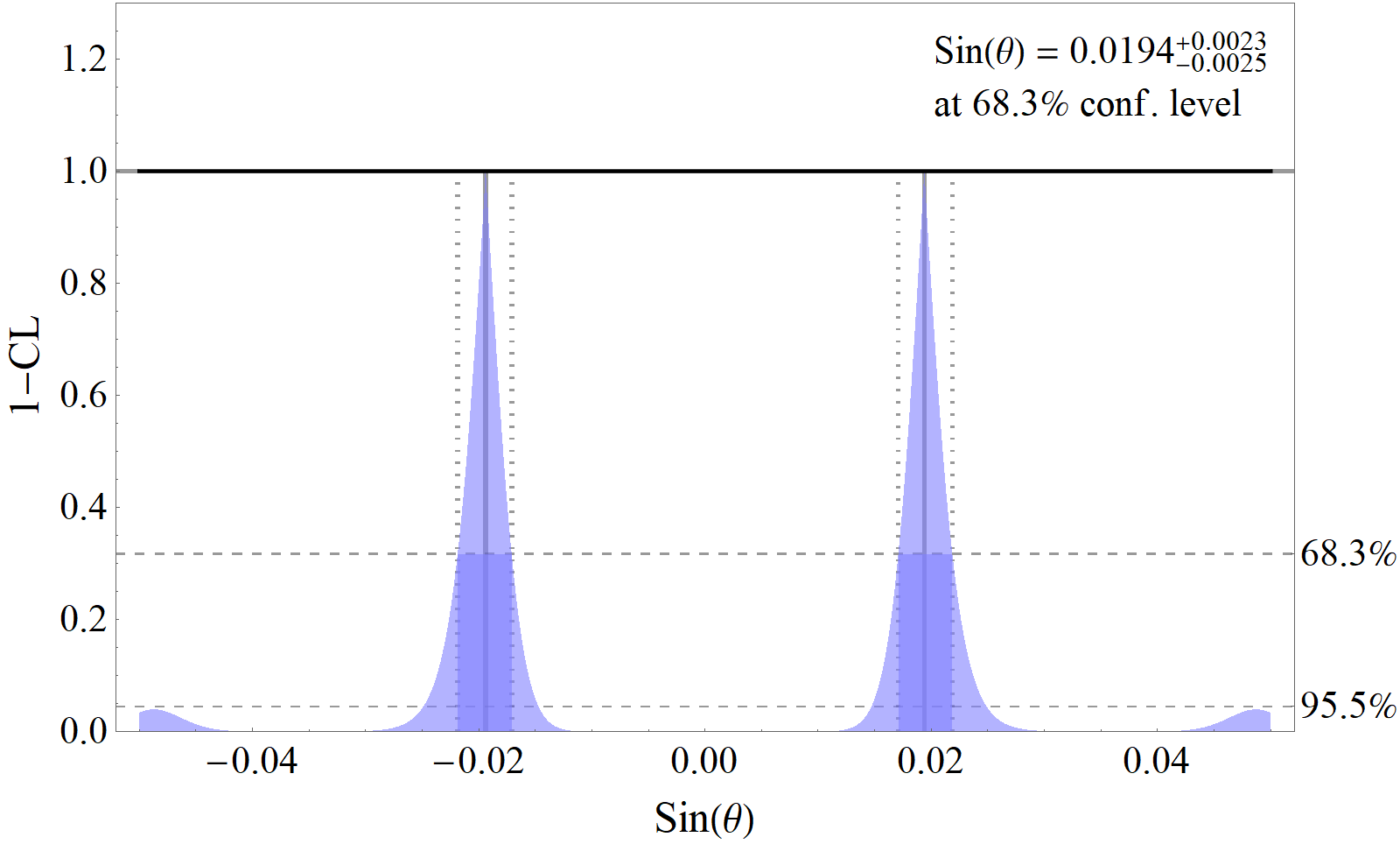}\label{fig:fewM4st1D}}~~
	\subfloat[Mod V: $A_1$ (C0)]{\includegraphics[height=5cm]{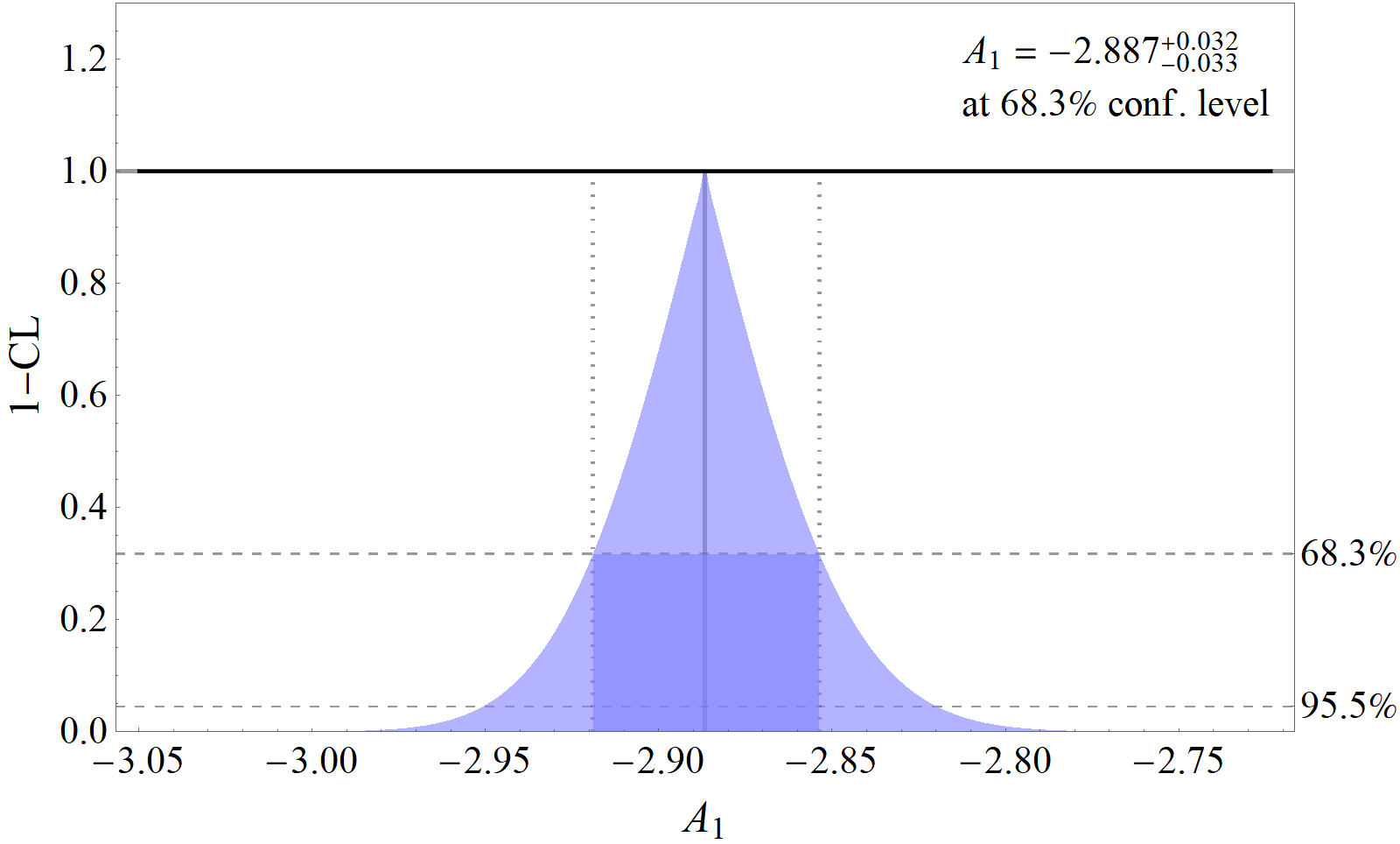}\label{fig:fewM5a11D}}\\
	\subfloat[Mod V: $A_5$ (C0)]{\includegraphics[height=5cm]{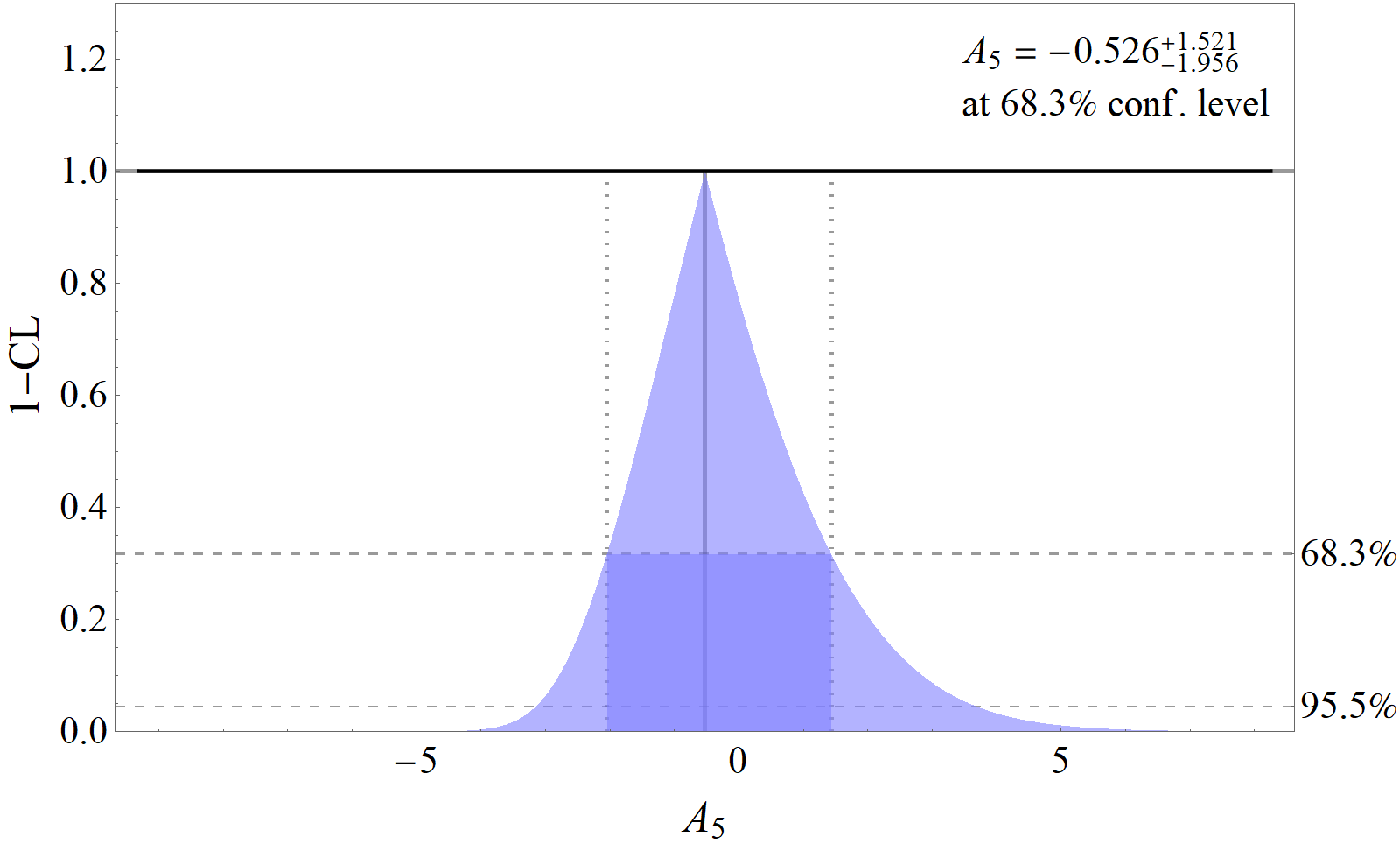}\label{fig:fewM5a51D}}~~
	\subfloat[Mod V: $\sin\theta$ (C0)]{\includegraphics[height=5cm]{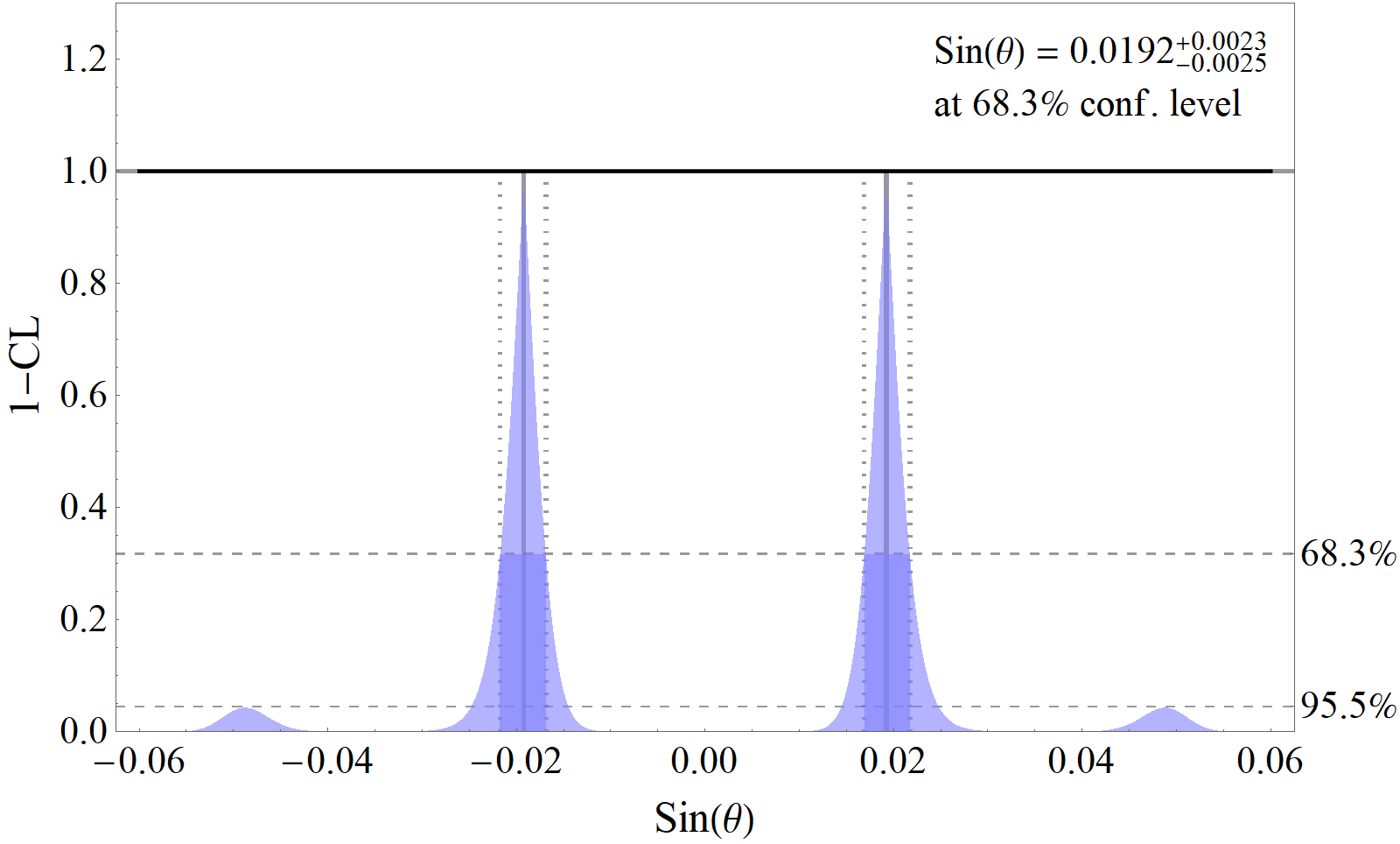}\label{fig:fewM5st1D}}
	\caption{One dimensional profile likelihoods, $1$ and $2\sigma$ CLs of the parameters for models IV and V (fit C0) for the small dataset (see sec. \ref{sec:anafew}). Each figure contains the $1\sigma$ results for the parameters in the top panel. For $\sin\theta$ plots, we quote the positive result, as it is preferred by the constraints.}
	\label{fig:fewdat1D1}
\end{figure*}

\begin{figure*}
	\centering
	\subfloat[Fit C0]{\includegraphics[height=5cm]{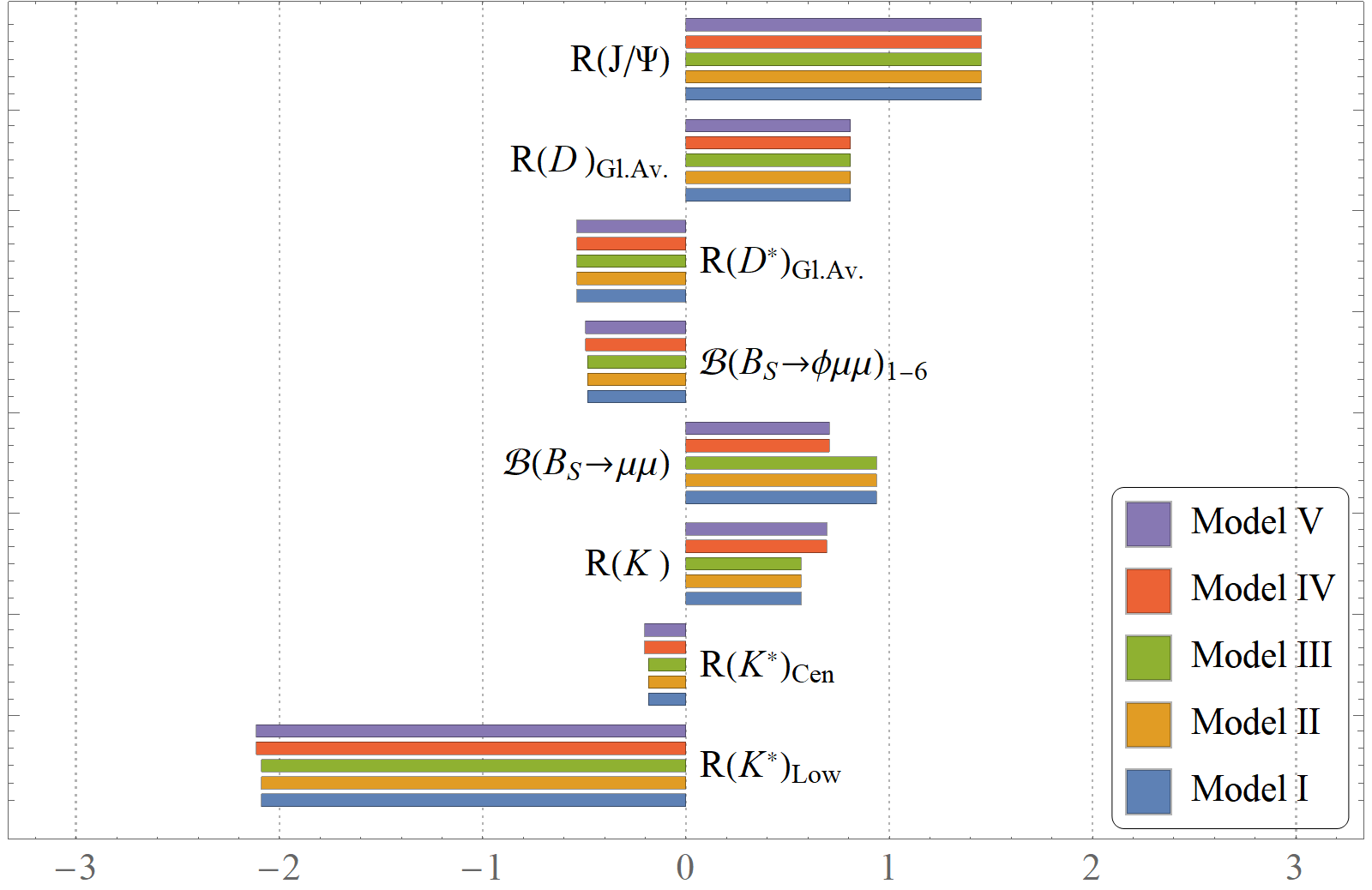}\label{fig:fewpullsC0}}~~
	\subfloat[Fit C1]{\includegraphics[height=5cm]{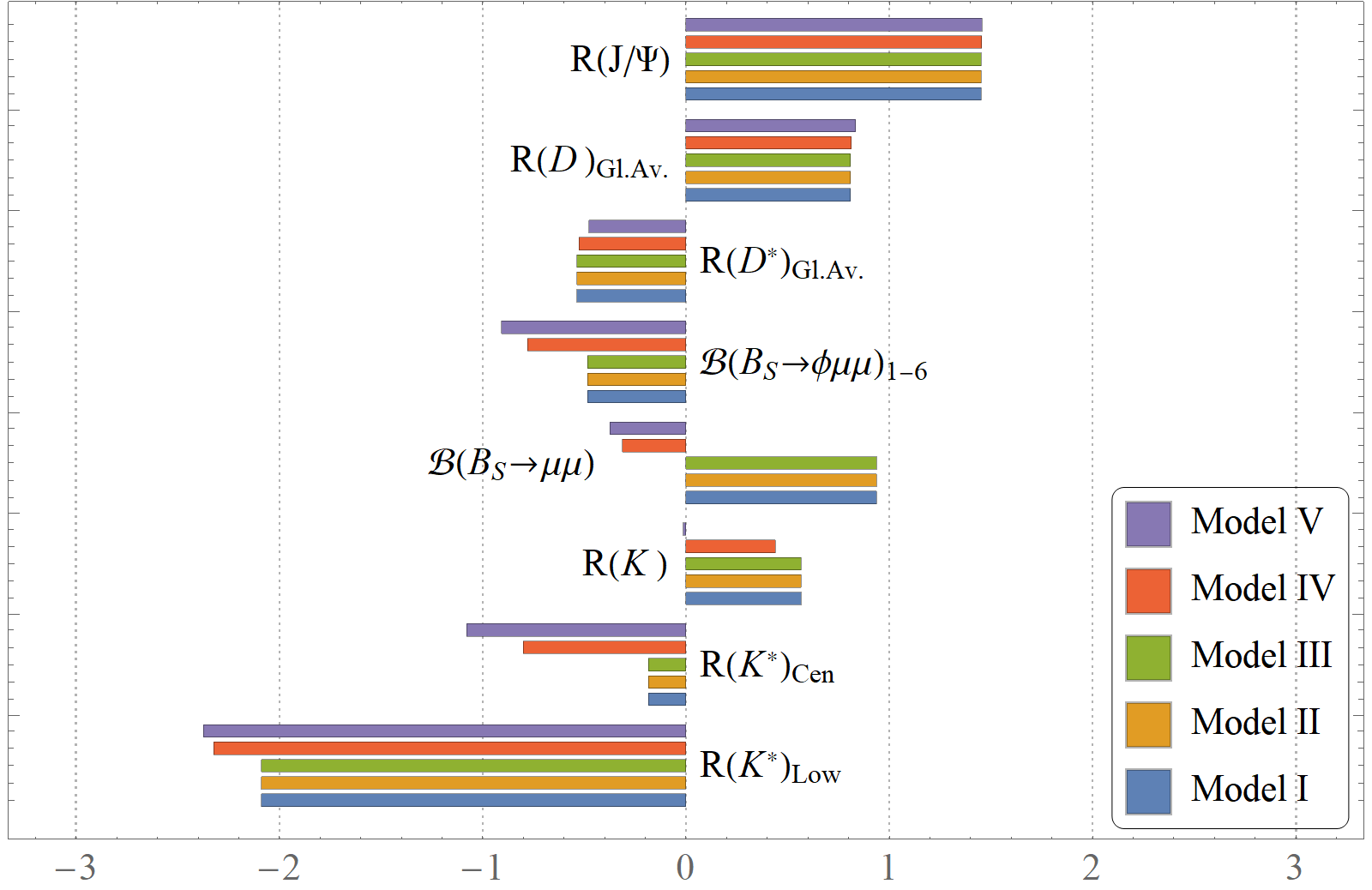}\label{fig:fewpullsC1}}
	\caption{Pull results for all models for the small dataset (see sec. \ref{sec:anafew}).}
	\label{fig:fewdatpull}
\end{figure*}

\begin{figure*}
	\centering
	\subfloat[Fit C0]{\includegraphics[height=5cm]{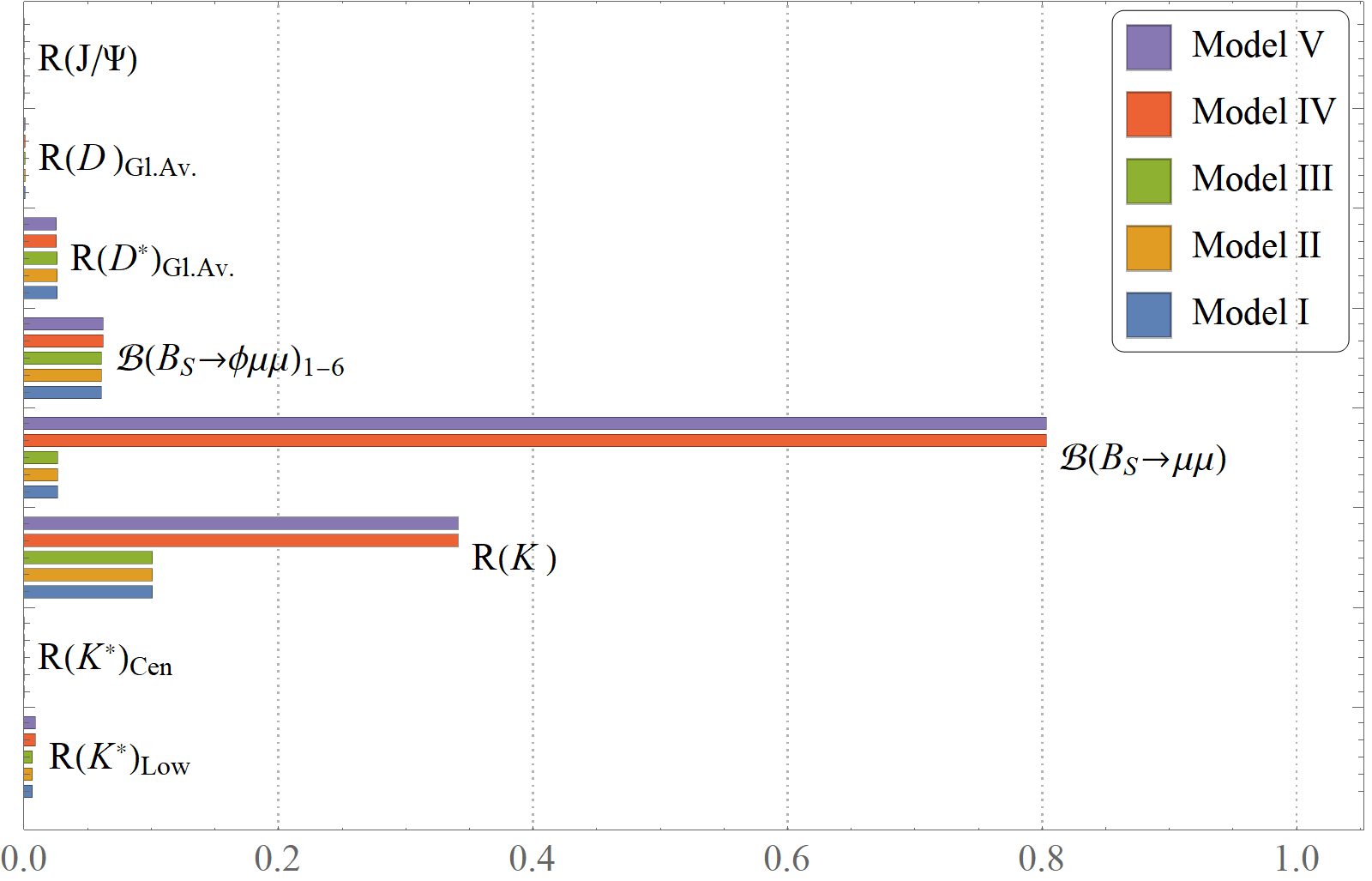}\label{fig:fewCDC0}}~~
	\subfloat[Fit C1]{\includegraphics[height=5cm]{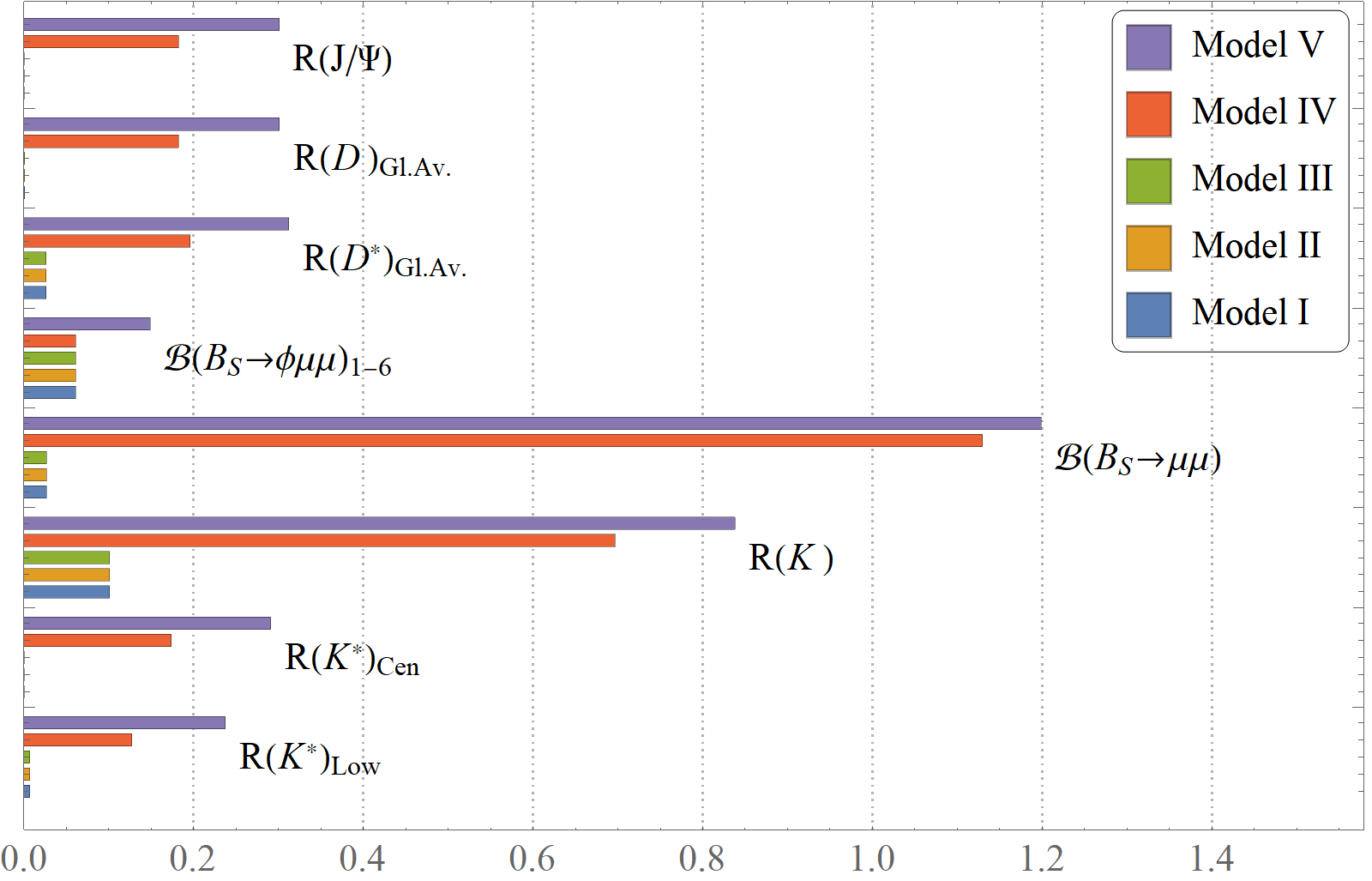}\label{fig:fewCDC1}}
	\caption{Cook's distances for the observables for all models for the small dataset (see sec. \ref{sec:anafew}).}
	\label{fig:fewdatCD}
\end{figure*}

\subsubsection{Fits and Confidence Levels}
In our analysis the $\chi^2$ is constructed in the following way: 
\begin{align}\label{chidef}
\nn \chi^2 = &\sum^{{\rm data}}_{i,j = 1} \left(O^{{exp}}_i - O^{th}_i\right) \left(V^{stat} + V^{syst} + V^{theo} \right)^{-1}_{i j} \nn \\
&\left(O^{{exp}}_j - O^{th}_j\right) \,.
\end{align}
Here, $O^{th}_p$ is the theoretical expression and $O^{{exp}}_p$ the central value of 
the experimental result of the $p^\text{th}$ observable used in the analysis. $V^{type}$ is the  covariance matrix, where $type$ refers to either the statistical, systematic or the theoretical. The $V^{theo}$ is created by propagating the uncertainties of the SM expressions of the observables\footnote{The dependence of $V^{theo}$ on the (small) NP parameters are at a higher order and extremely small and hence neglected.}. $O^{th}_p$ are functions of the model parameters.

Parameter uncertainties are obtained in two ways,
\begin{enumerate*}
	\item when the profile likelihoods of the parameters are approximately Gaussian, it is possible to obtain the `HESSE' errors \cite{James:1975dr}, which are, obviously, symmetric;
	\item when possible, we attempt to find the uncertainties directly from the $1\sigma$ CL of the profile likelihoods of the said parameter. One and two dimensional profile likelihoods are depicted in this analysis as 1-CL plots, closely following the PROB method followed in ref. \cite{Aaij:2016kjh}.
\end{enumerate*}
As we will see in sections below, for most cases in our analysis, we face scenarios where the $\chi^2$ minima are contours instead of isolated points. In those cases, we rely on two-dimensional parameter confidence plots and show the position of the numerically obtained minimum whenever possible.

\subsubsection{Pulls and Cook's Distances}
In general, `pull's in particle physics are defined in parameter estimation problems where several experimental results are combined. Pulls are usually defined for nuisance parameters (parameters which have a measured value) of a fit \cite{CYRSP303}. There are many alternative ways to define a pull depending on the nature of the fit \cite{Demortier:2008}. The simplest definition is
\begin{align}\label{pulldef}
	pull(\theta) = \frac{\theta^{exp} - \theta^{fit}}{\sigma^{exp}}\,,
\end{align}
where $\theta$ is the nuisance parameter in question, $\theta^{exp}$ is its experimentally measured value, $\theta^{fit}$ is the fitted value, and $\sigma^{exp}$ is the uncertainty of the experimental result. Defined this way, they are distributed as a unit Gaussian distribution for a healthy fit. In that way, they are closely related to the Studentized Residuals in linear regression. As these appear in the fit statistic in the same way as the other experimental inputs, it is a standard practice to define similar quantities for all observables of the fit. A prominent example is the global electroweak fit. In this way, in addition to quantifying the quality of a fit, they show the deviation of the fitted observables with respect to their measured values. We will use this definition of pulls in the rest of the paper.

It is not enough to know whether a datum is an outlier; we need to know the effect/influence a specific observable has on a fit. A measure of that can be achieved by using some `deletion statistic', which will determine the amount of change in the fitted results in absence of that datum. Cook's distance \cite{CookDist1,CookDist2} is one of many such quantities for doing this. Observables with a large Cook's distance merit closer examination in the analysis. Cook's distance of the $i$-th observable in a fit can be defined as
\begin{align}\label{cookdef}
	CD_{i} = \frac{\sum_{j=1}^{data} (\hat{y} - \hat{y}_{j(i)})^2}{p~MSE}\,,
\end{align}
where $\hat{y}$ is the fitted value of the $j$-th observable, $\hat{y}_{j(i)}$ is the fitted value of the $j$-th observable when the $i$-th observable is excluded from the fit and $MSE$ is the mean squared error for the fitted model, defined as
\begin{align}\label{msedef}
	MSE = \frac{\sum_{j=1}^{data} \sigma^2_{y_j}~pull(y_j)^2}{n-p}\,,
\end{align}
with $n$ and $p$ the number of observables and number of parameters respectively. Defined in this way, Cook's distance takes into account both the `pull' and `impact' \cite{CYRSP303} of an observable. To detect a highly influential observable, we need to put a cutoff value on the Cook's distances ($Cut_{CD}$). As these follow a Fisher-Snedecor distribution ($F(p,n-p)$), the median point $F_{0.5}(p,n-p)$ is generally chosen as a cutoff \cite{CookCutoff}. 

\subsection{Results}\label{sec:results}
\begin{table}
	\centering
	\begin{tabular}{|c|c|c|c|c|c|c|}
		\hline
		\text{Few Obs.} & \multicolumn{3}{|c|}{ \text{C0}}  & \multicolumn{3}{|c|}{\text{C1}}   \\
		\hline
		\text{Model} & \text{$p$-value(\%)} &\multicolumn{2}{|c|}{ \text{Best-fit value} } & \text{$p$-value(\%)} & \multicolumn{2}{|c|}{ \text{Best-fit value}}   \\
		\hline
		I & 9.88 & \text{$A_1$} & -3.244 & 9.88 & \text{$A_1$} & -0.657 \\
		&  & \text{$A_2$} & -1.484 &  & \text{$A_2$} & -0.190 \\
		&  & \text{$S_\theta $} & -0.018 &  & \text{$S_\theta $} & 0.044 \\
		\hline
		II & 9.88 & \text{$A_1$} & -2.923 & 9.88 & \text{$A_1$} & -2.107 \\
		&  & \text{$A_2$} & -6.123 &  & \text{$A_2$} & 1.560 \\
		&  & \text{$S_\theta $} & -0.015 &  & \text{$S_\theta $} & 0.061 \\
		\hline
		III & 9.88 & \text{$A_1$} & 0.139 & 9.88 & \text{$A_1$} & -2.887 \\
		&  & \text{$A_3$} & -5.855 &  & \text{$A_3$} & 0.903 \\
		&  & \text{$S_\theta $} & 0.013 &  & \text{$S_\theta $} & -0.044 \\
		\hline
		IV & 10.19 & \text{$A_1$} & -3.844 & 2.51 & \text{$A_1$} & -3.840 \\
		&  & \text{$A_5$} & -0.479 &  & \text{$A_5$} & -2.218 \\
		&  & \text{$S_\theta $} & 0.019 &  & \text{$S_\theta $} & 0.017 \\
		\hline
		V & 10.19 & \text{$A_1$} & -2.887 & 3.15 & \text{$A_1$} & -2.884 \\
		&  & \text{$A_5$} & -0.526 &  & \text{$A_5$} & -2.222 \\
		&  & \text{$S_\theta $} & -0.019 &  & \text{$S_\theta $} & 0.017 \\
		\hline
	\end{tabular}
	\caption{Fit results for small dataset with fits C0 and C1. There are no physically allowed best fit results for the fit C2 for small dataset. Parameter uncertainties are not quoted because of reasons stated in sec. \ref{sec:anafew}.}
	\label{table:FC01}
\end{table}

\begin{figure*}
	\centering
	\subfloat[Mod I: $A_1$ vs. $A_2$ (C1)]{\includegraphics[height=3.8cm]{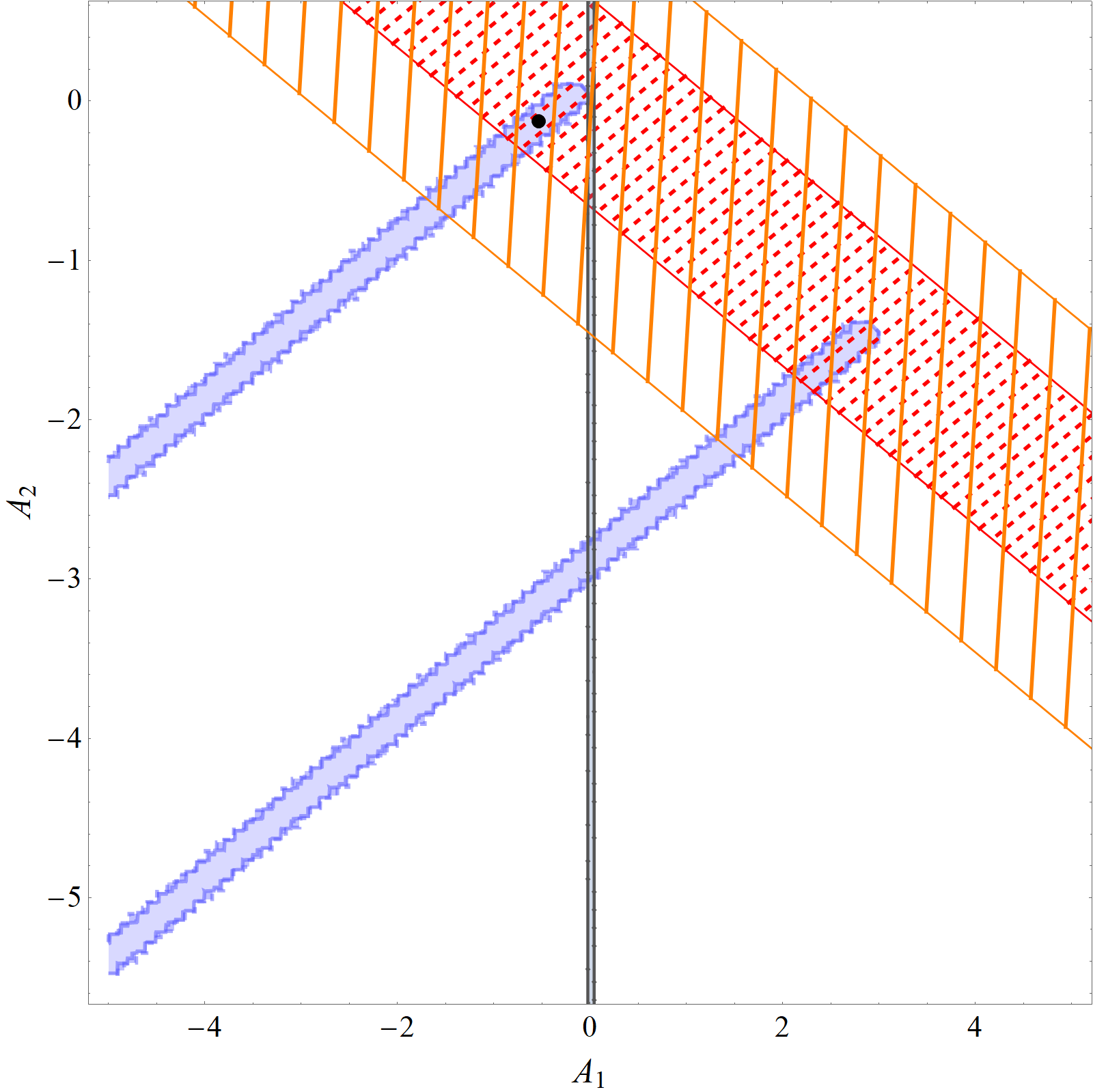}\label{fig:allM1a1a2C}}~~
	\subfloat[Mod I: $A_2$ vs. $\sin\theta$ (C1)]{\includegraphics[height=3.8cm]{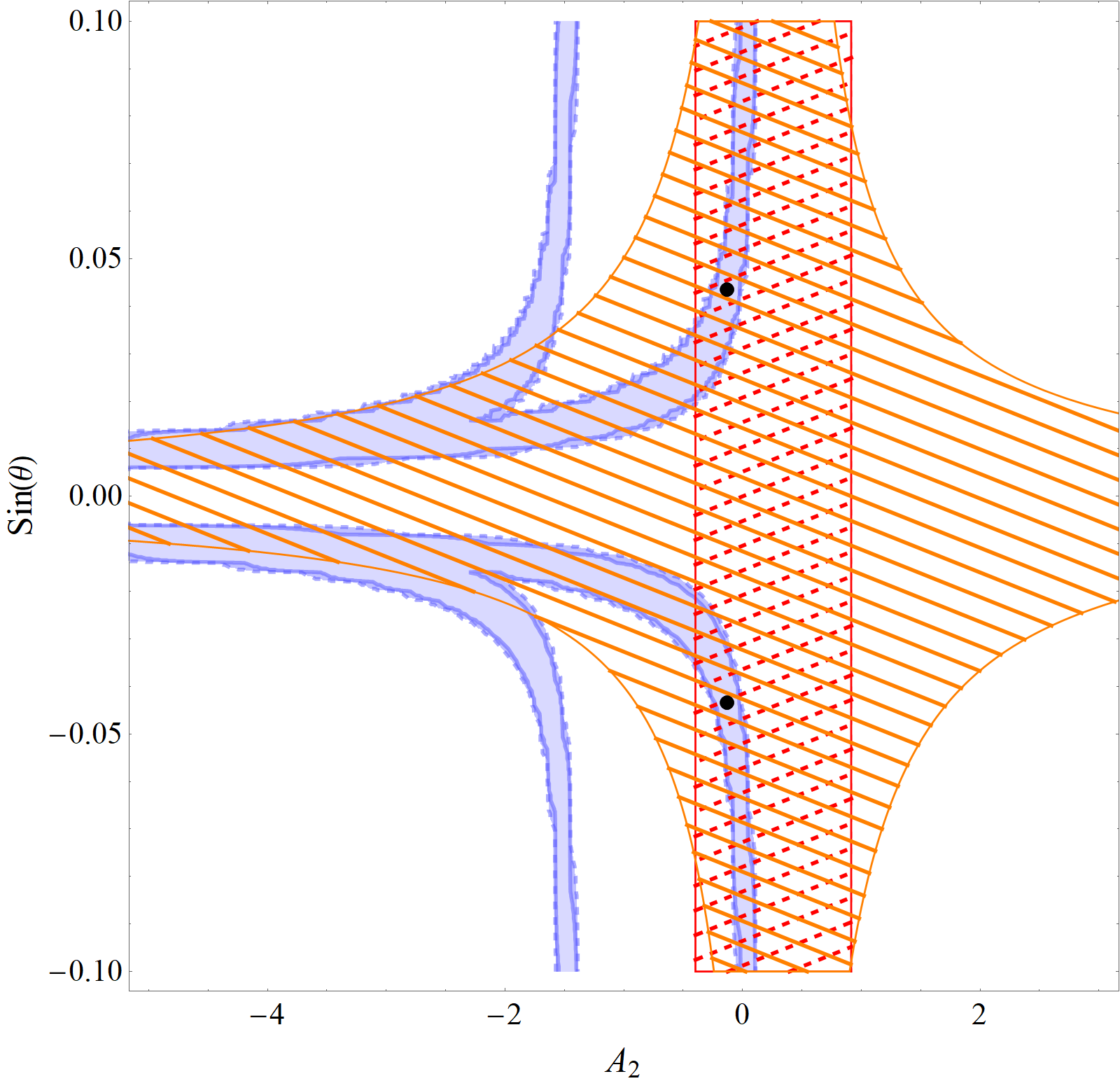}\label{fig:allM1a2stC}}~~
	\subfloat[Mod II: $A_1$ vs. $A_2$ (C1)]{\includegraphics[height=3.8cm]{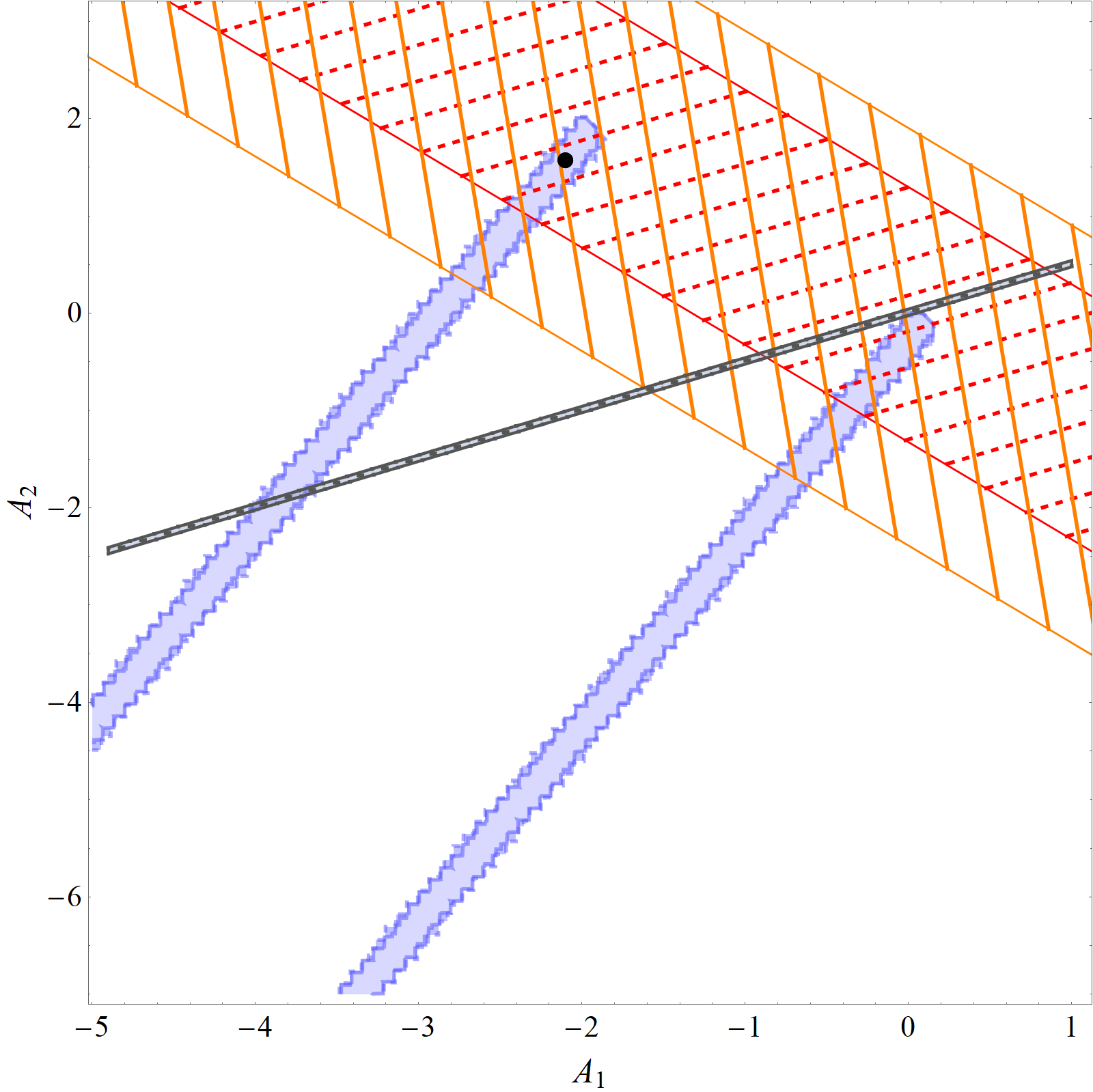}\label{fig:allM2a1a2C}}~~
	\subfloat[Mod II: $A_2$ vs. $\sin\theta$ (C1)]{\includegraphics[height=3.8cm]{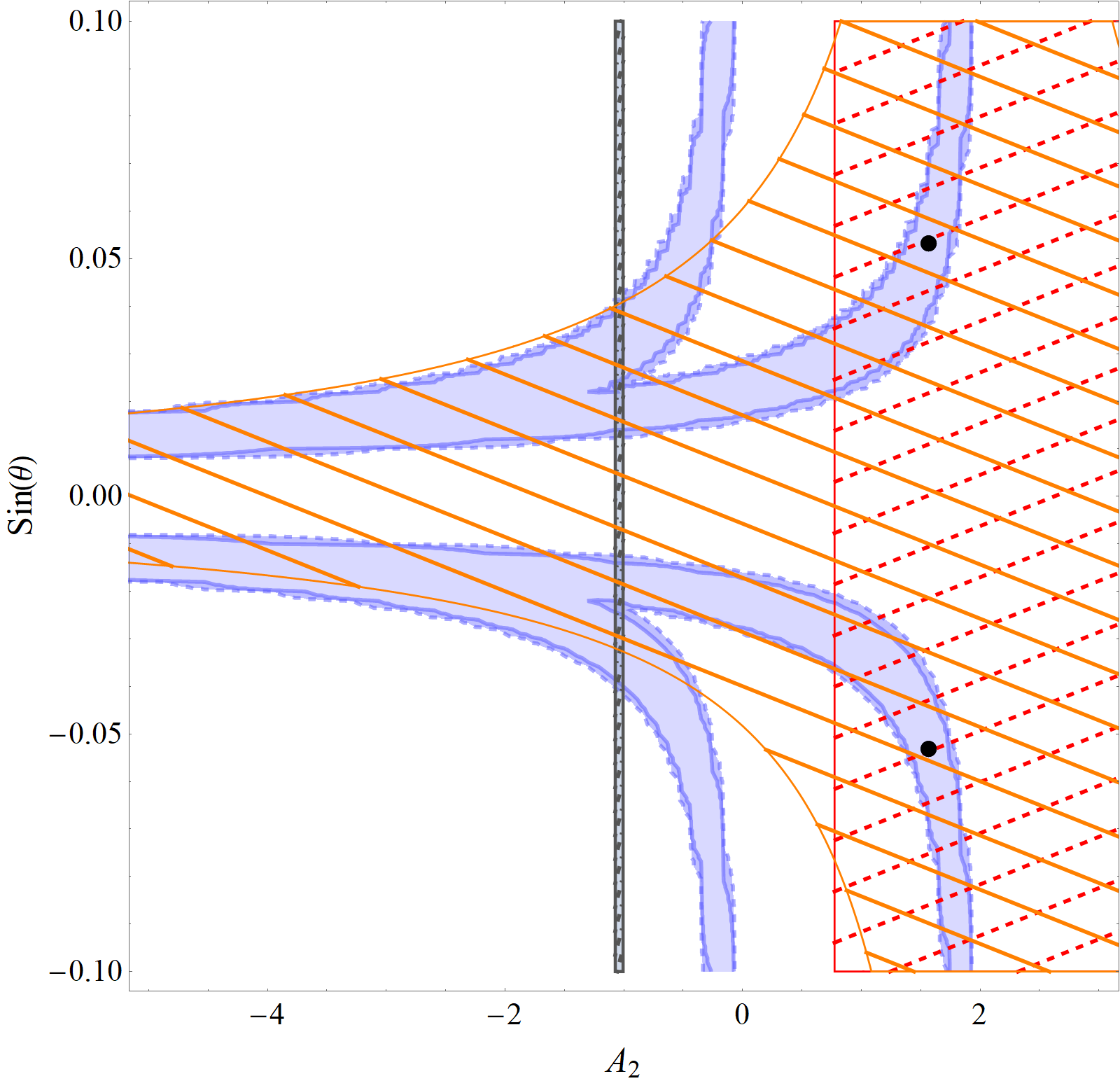}\label{fig:allM2a2stC}}\\
	\subfloat[Mod III: $A_1$ vs. $A_3$ (C1)]{\includegraphics[height=3.8cm]{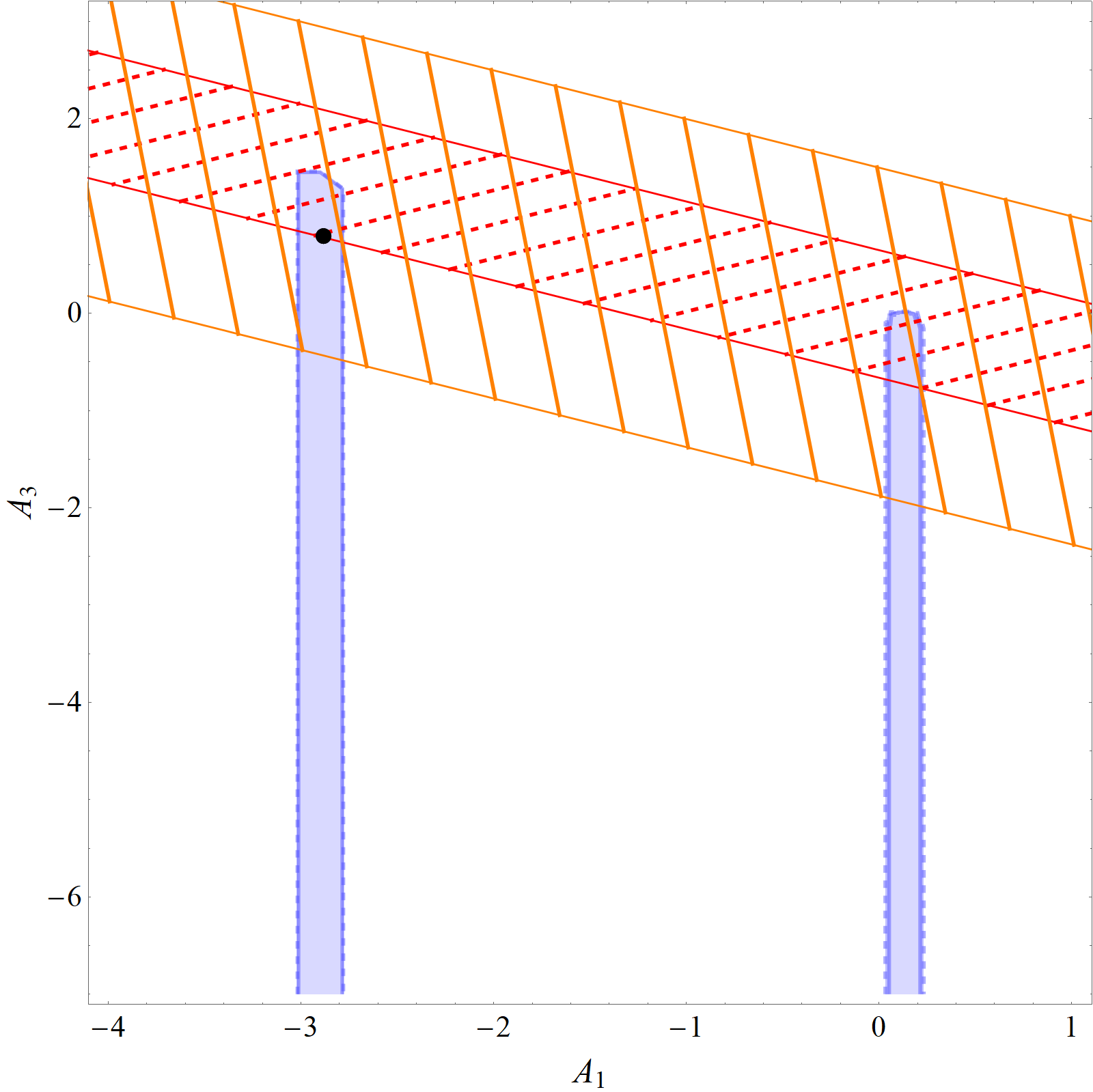}\label{fig:allM3a1a3C}}~~
	\subfloat[Mod III: $A_3$ vs. $\sin\theta$ (C1)]{\includegraphics[height=3.8cm]{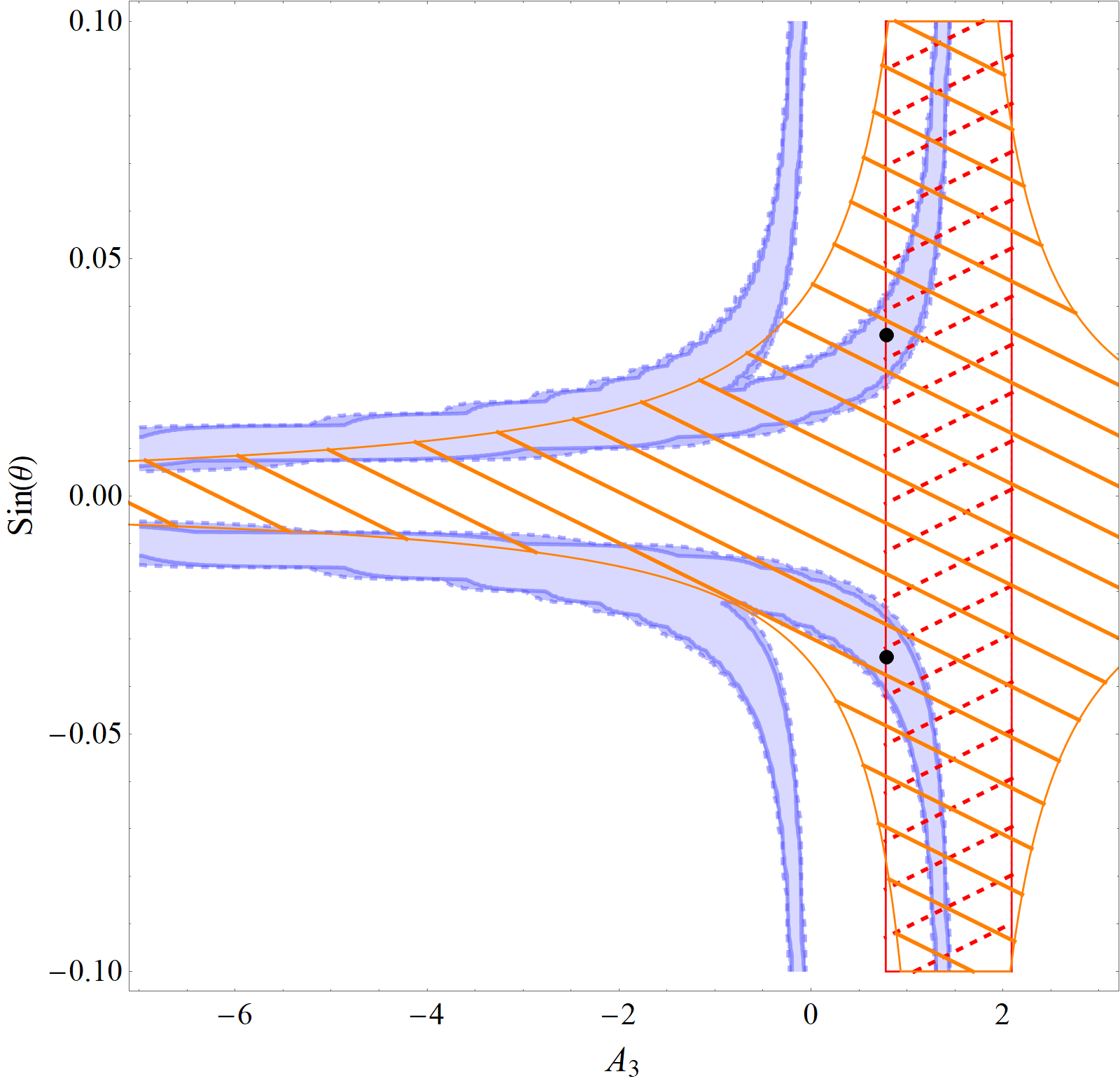}\label{fig:allM3a3stC}}~~
	\subfloat[Mod IV: $A_1$ vs. $A_5$ (C0)]{\includegraphics[height=3.8cm]{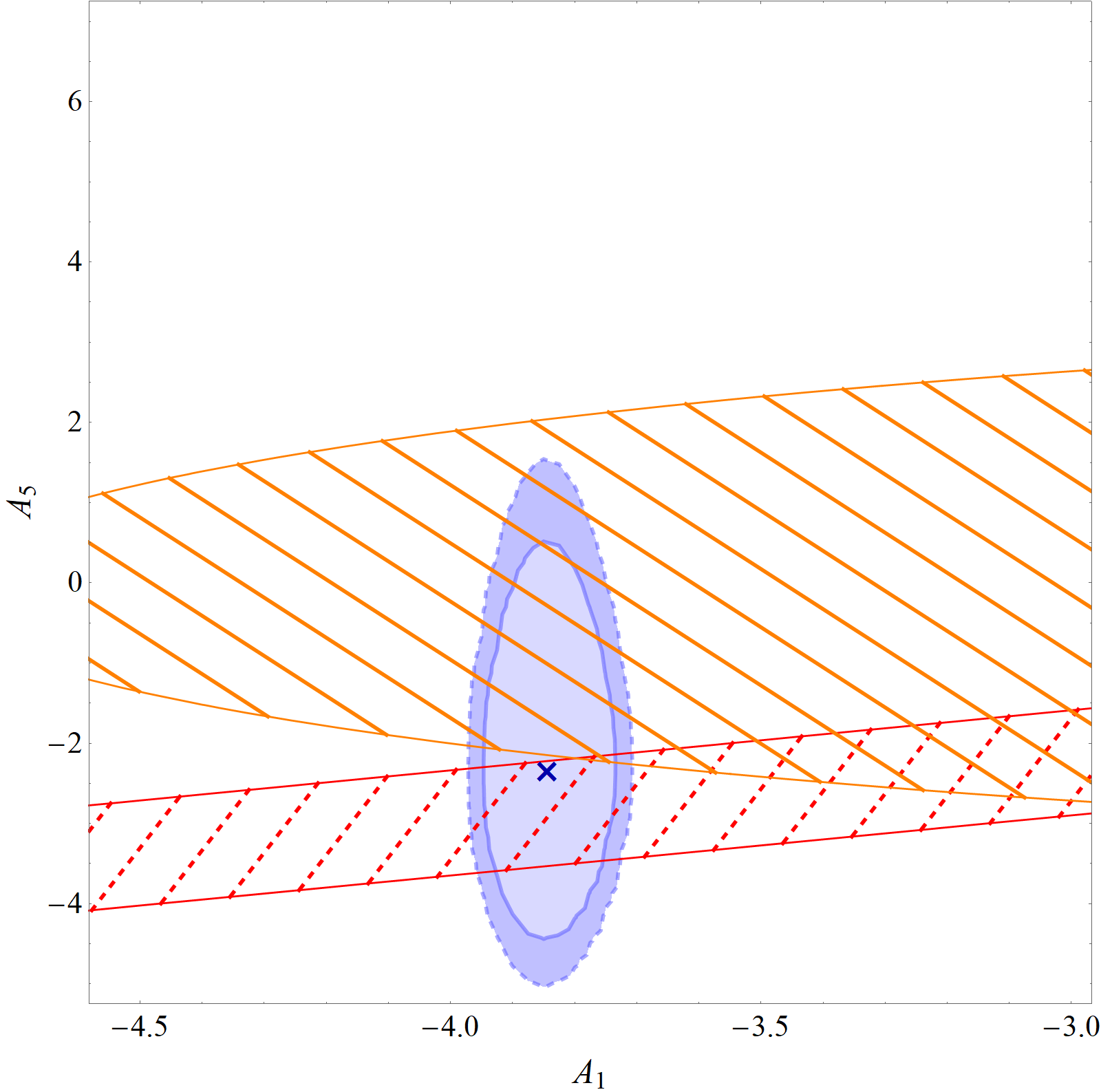}\label{fig:allM4a1a5}}~~
	\subfloat[Mod IV: $A_1$ vs. $A_5$ (C1)]{\includegraphics[height=3.8cm]{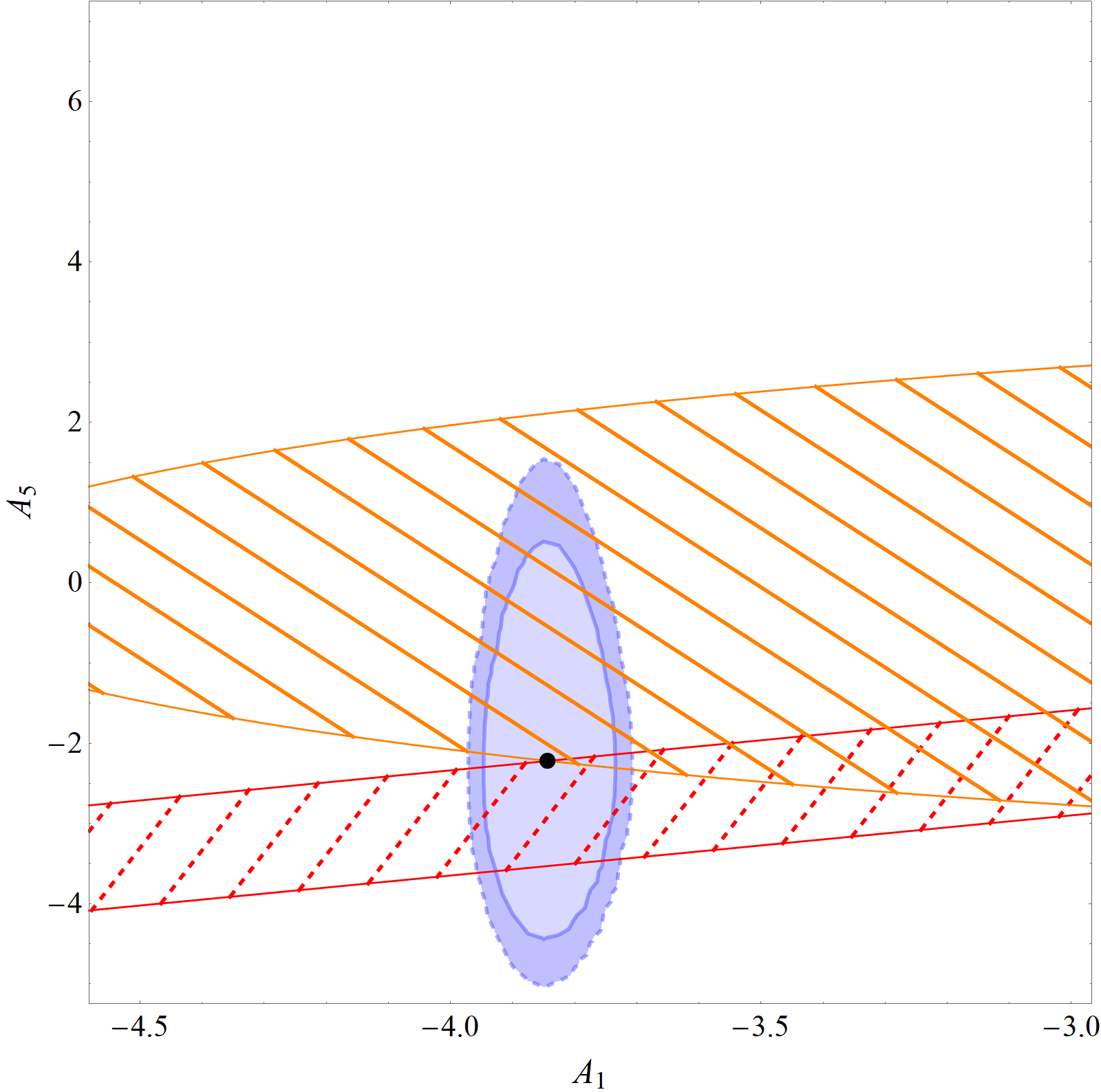}\label{fig:allM4a1a5C}}\\
	\subfloat[Mod IV: $A_5$ vs. $\sin\theta$ (C1)]{\includegraphics[height=3.8cm]{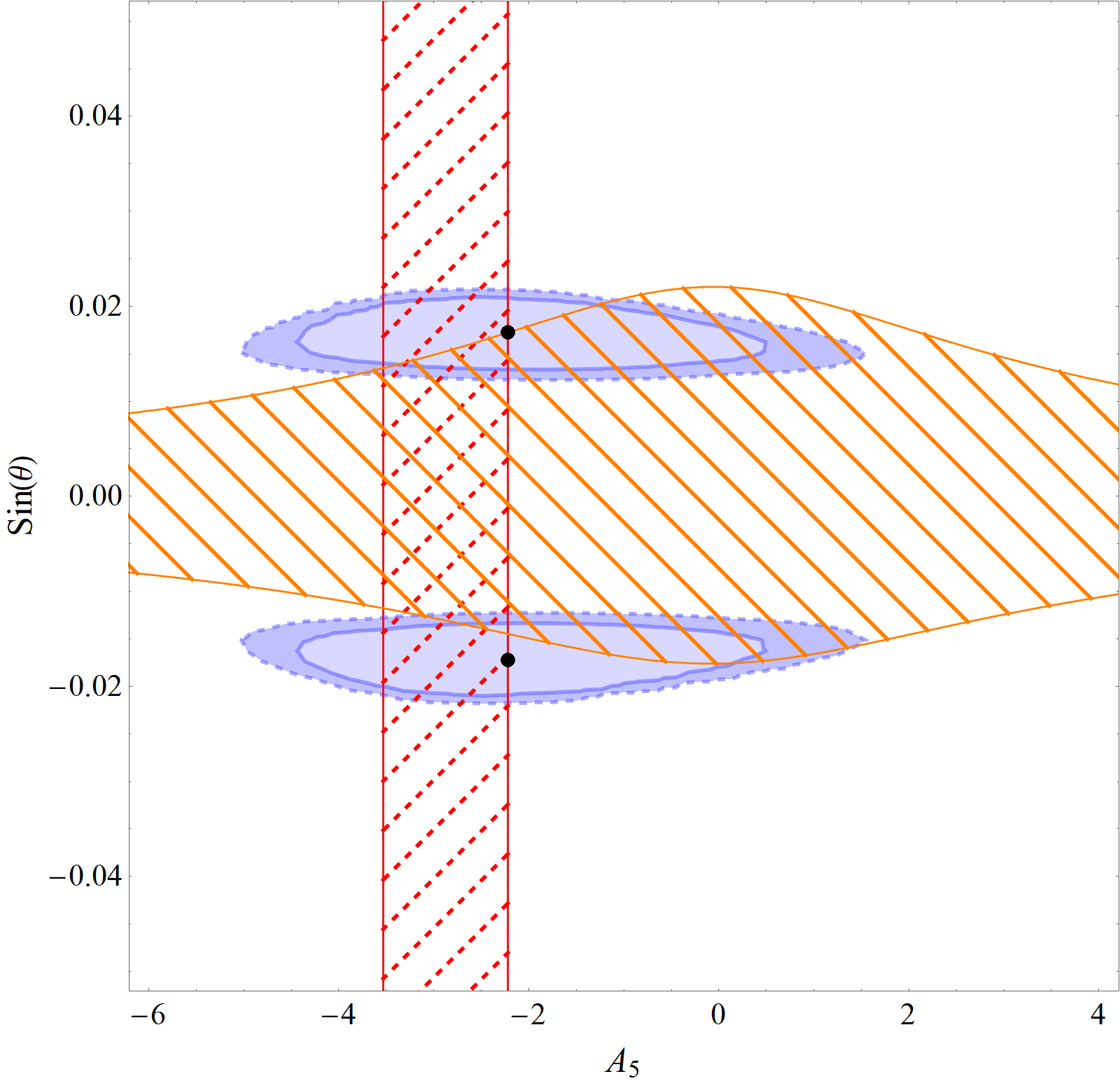}\label{fig:allM4a5stC}}~~
	\subfloat[Mod V: $A_1$ vs. $A_5$ (C0)]{\includegraphics[height=3.8cm]{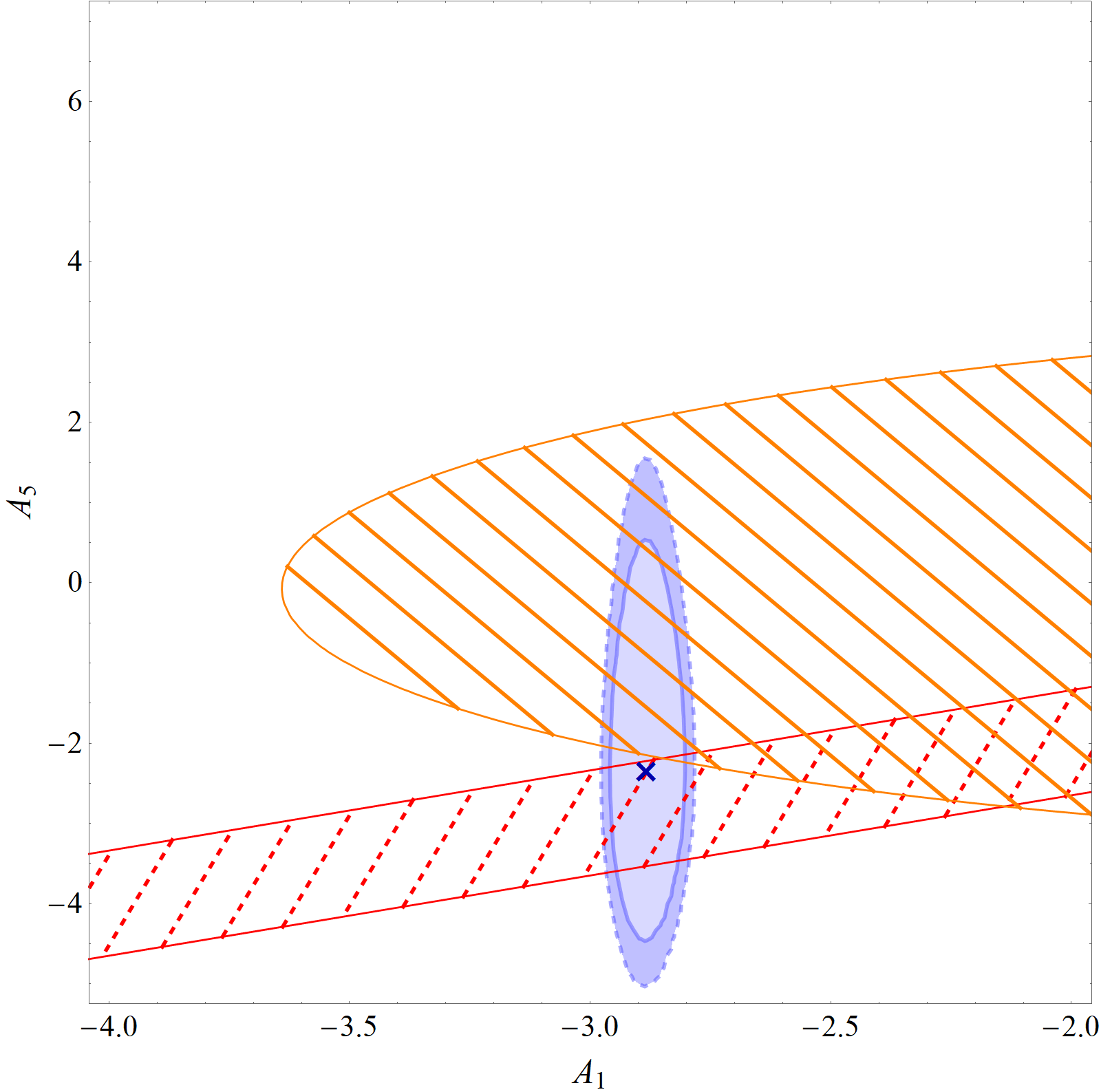}\label{fig:allM5a1a5}}~~
	\subfloat[Mod V: $A_1$ vs. $A_5$ (C1)]{\includegraphics[height=3.8cm]{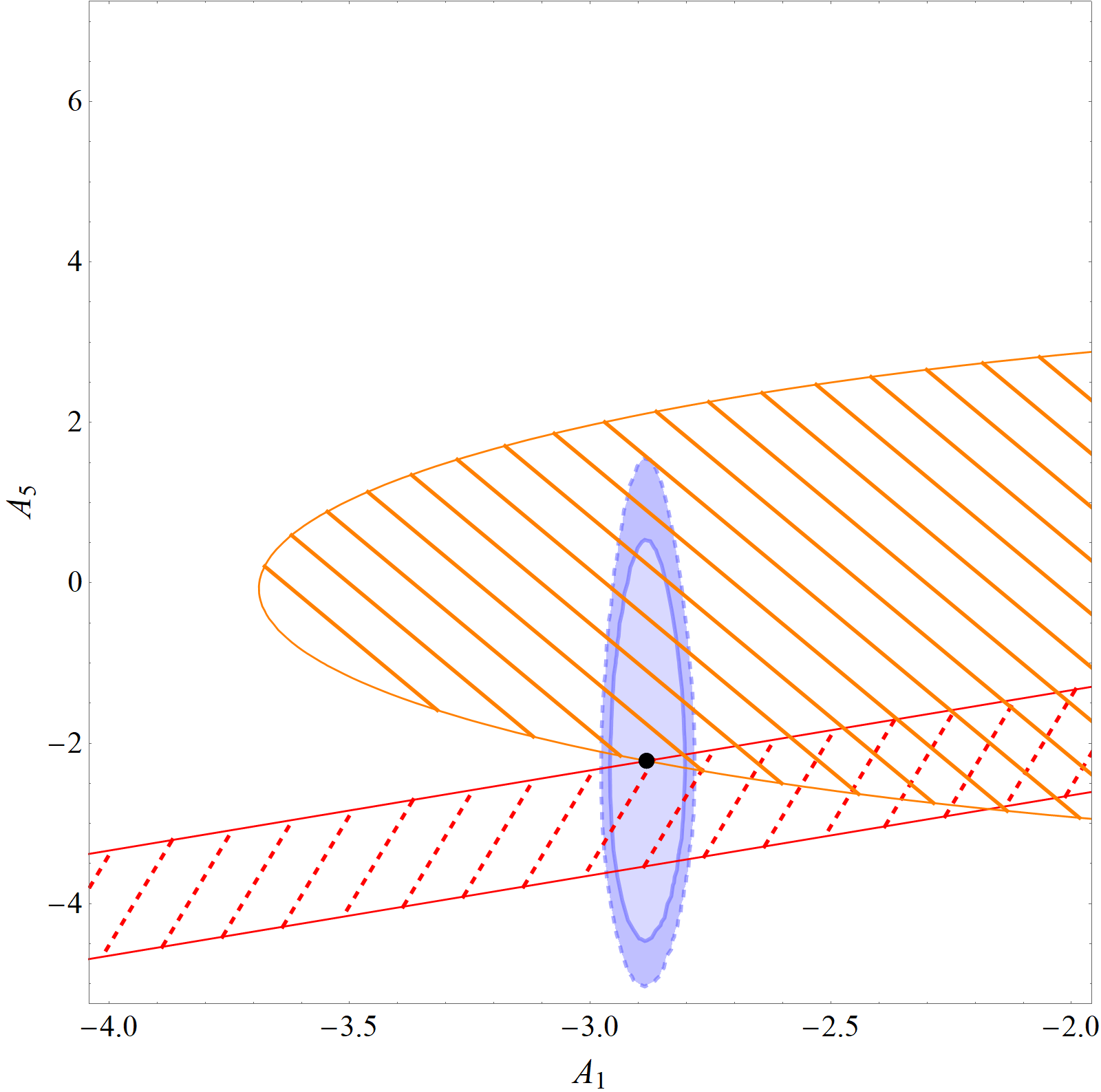}\label{fig:allM5a1a5C}}~~
	\subfloat[Mod V: $A_5$ vs. $\sin\theta$ (C1)]{\includegraphics[height=3.8cm]{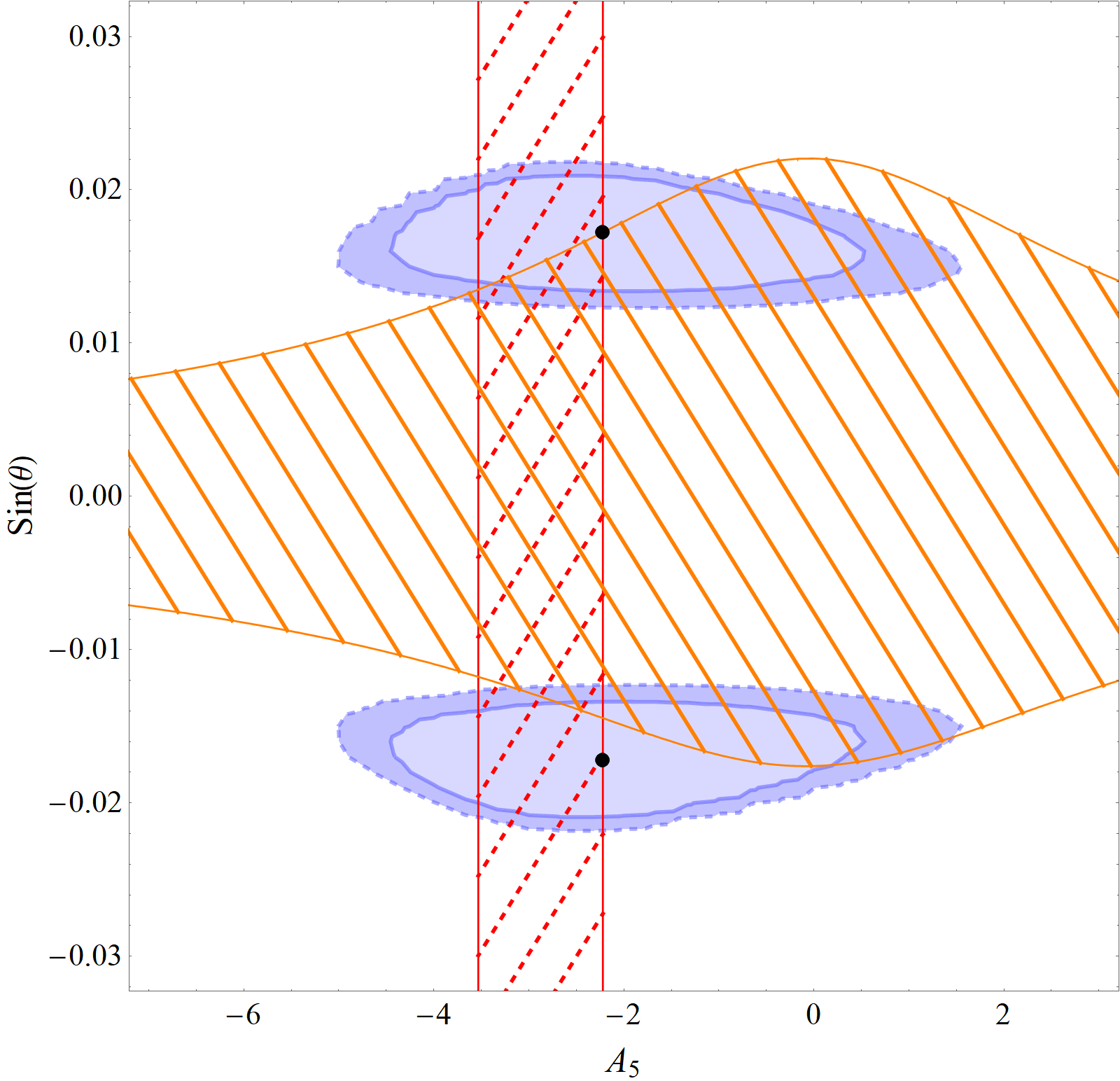}\label{fig:allM5a5stC}}
	\caption{Two dimensional parameter confidence regions for all models for the `large dataset' (see section \ref{sec:anaall}). For most models, both fits C0 and C1 are shown. Only when the constraint regions are almost identical, we just show the result for one of them, with different markers for the best-fit points. In fig. \ref{fig:allM3a1a3C}, the constraint region for \bKnn~lies outside the shown region, and there are no simultaneous overlaps. The legend \ref{fig:legend2D} in figure \ref{fig:fewdat2D1} shows different markers to read the plots.}
	\label{fig:alldat2D1}
\end{figure*}

\begin{figure*}
	\centering
	\subfloat[Mod IV: $A_1$ (C0)]{\includegraphics[height=5cm]{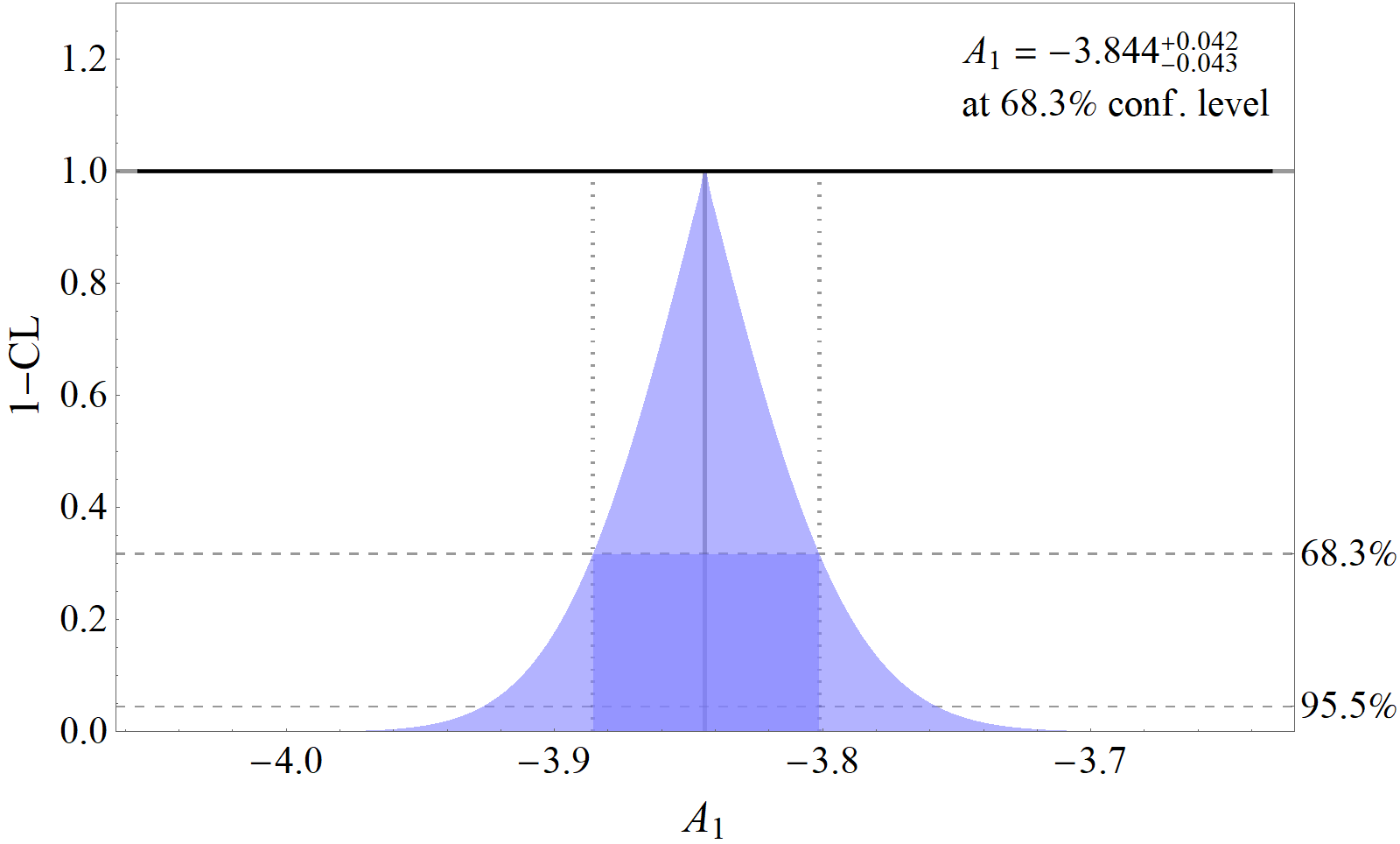}\label{fig:allM4a11D}}~~
	\subfloat[Mod IV: $A_5$ (C0)]{\includegraphics[height=5cm]{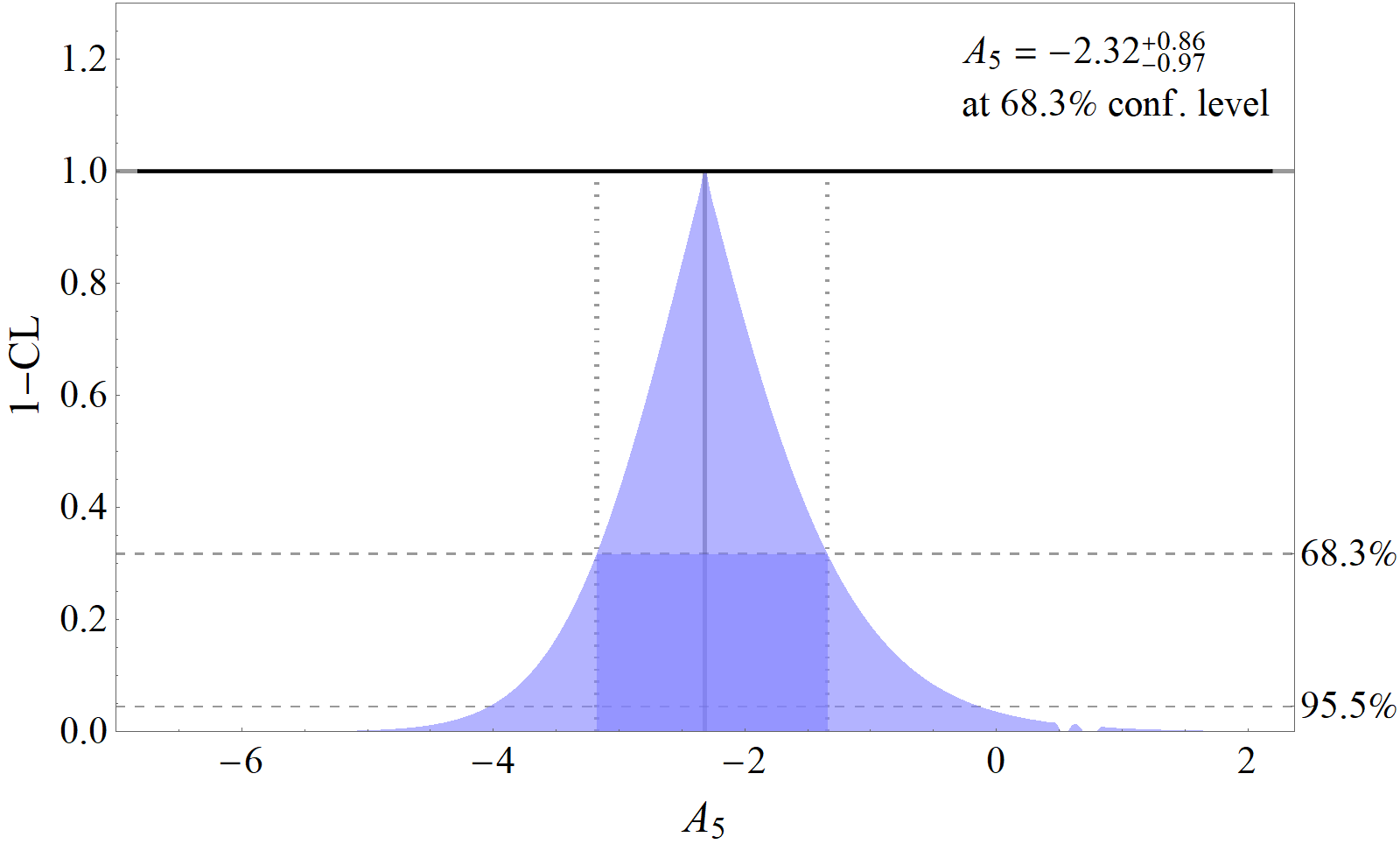}\label{fig:allM4a51D}}\\
	\subfloat[Mod IV: $\sin\theta$ (C0)]{\includegraphics[height=5cm]{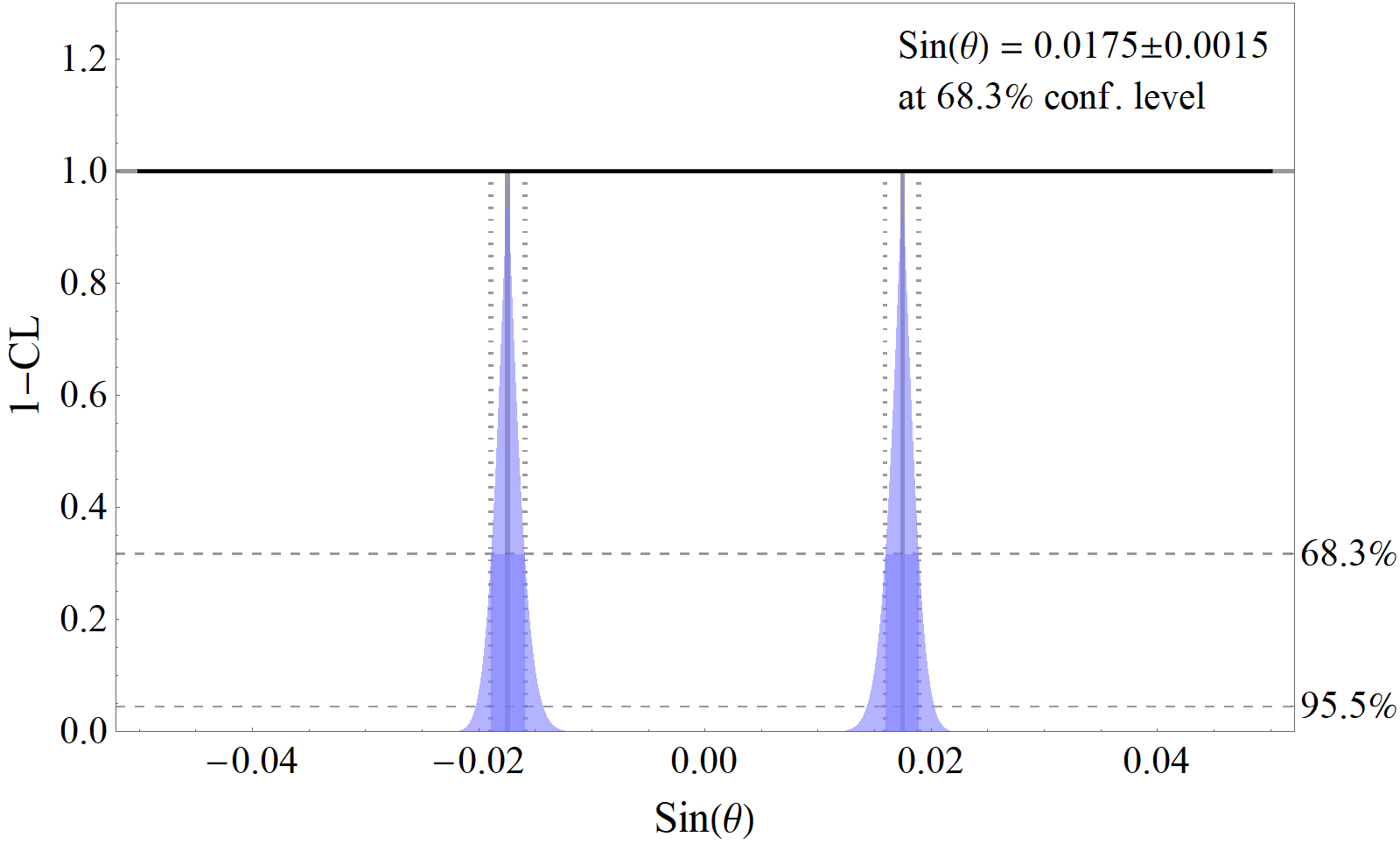}\label{fig:allM4st1D}}~~
	\subfloat[Mod V: $A_1$ (C0)]{\includegraphics[height=5cm]{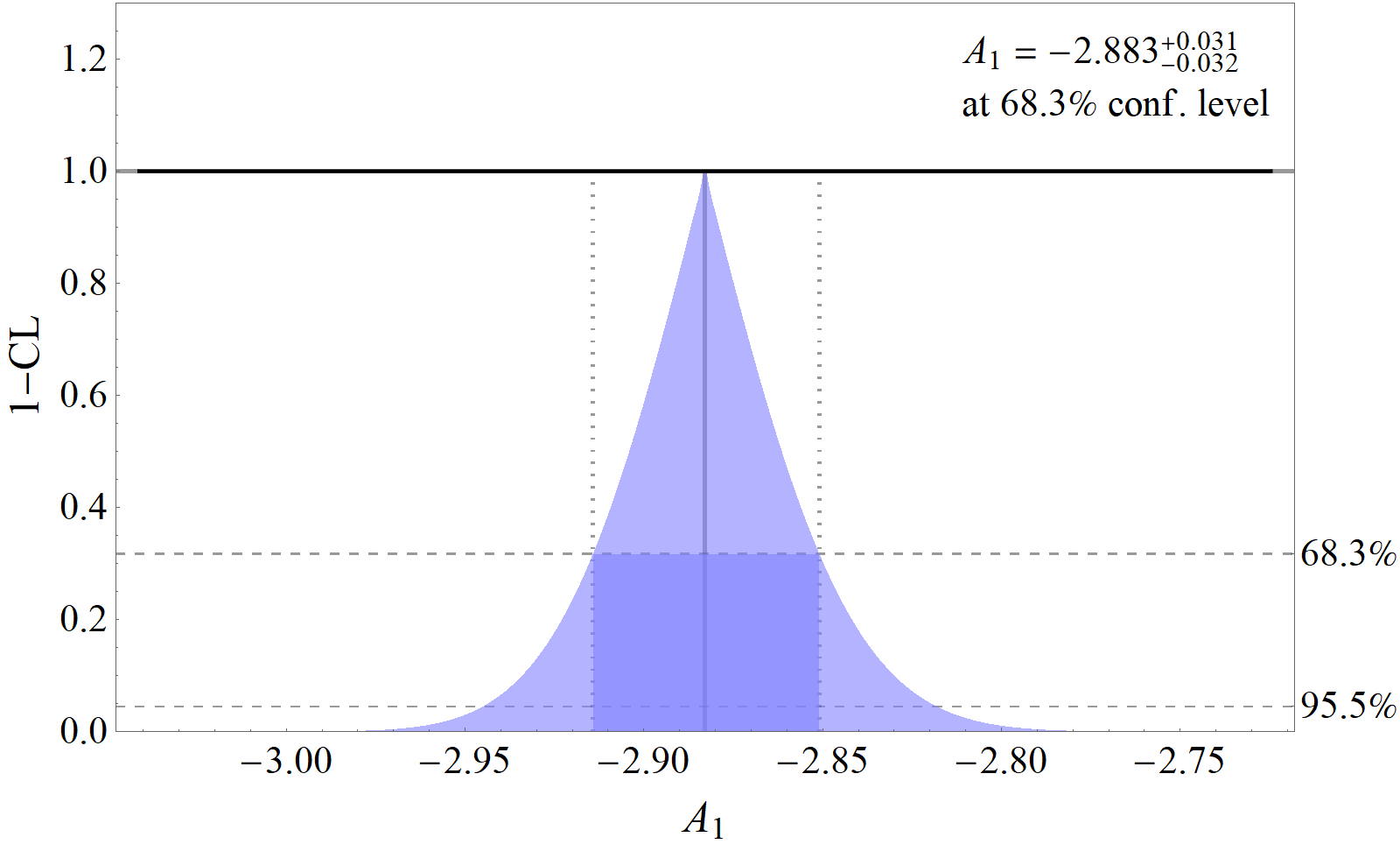}\label{fig:allM5a11D}}\\
	\subfloat[Mod V: $A_5$ (C0)]{\includegraphics[height=5cm]{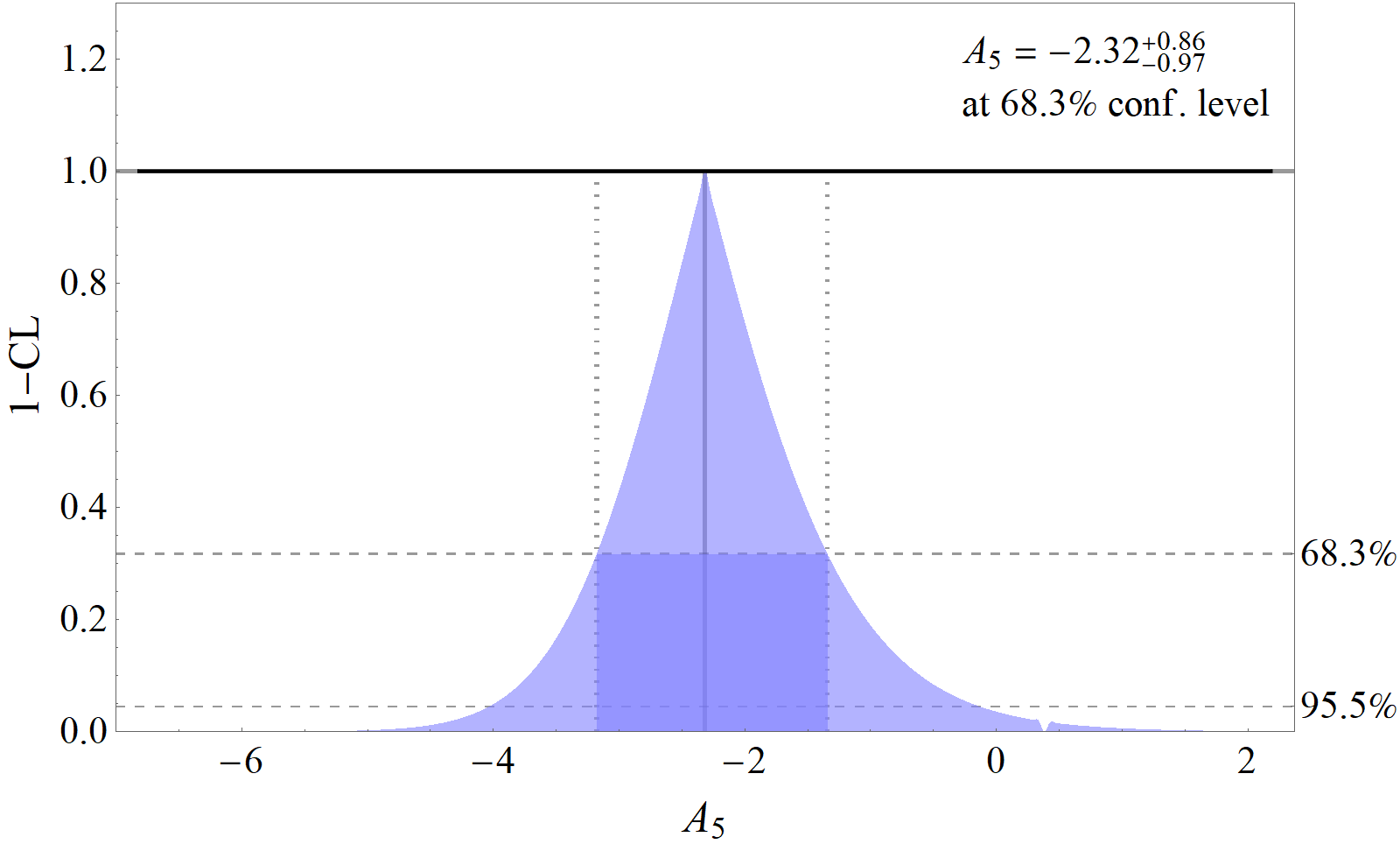}\label{fig:allM5a51D}}~~
	\subfloat[Mod V: $\sin\theta$ (C0)]{\includegraphics[height=5cm]{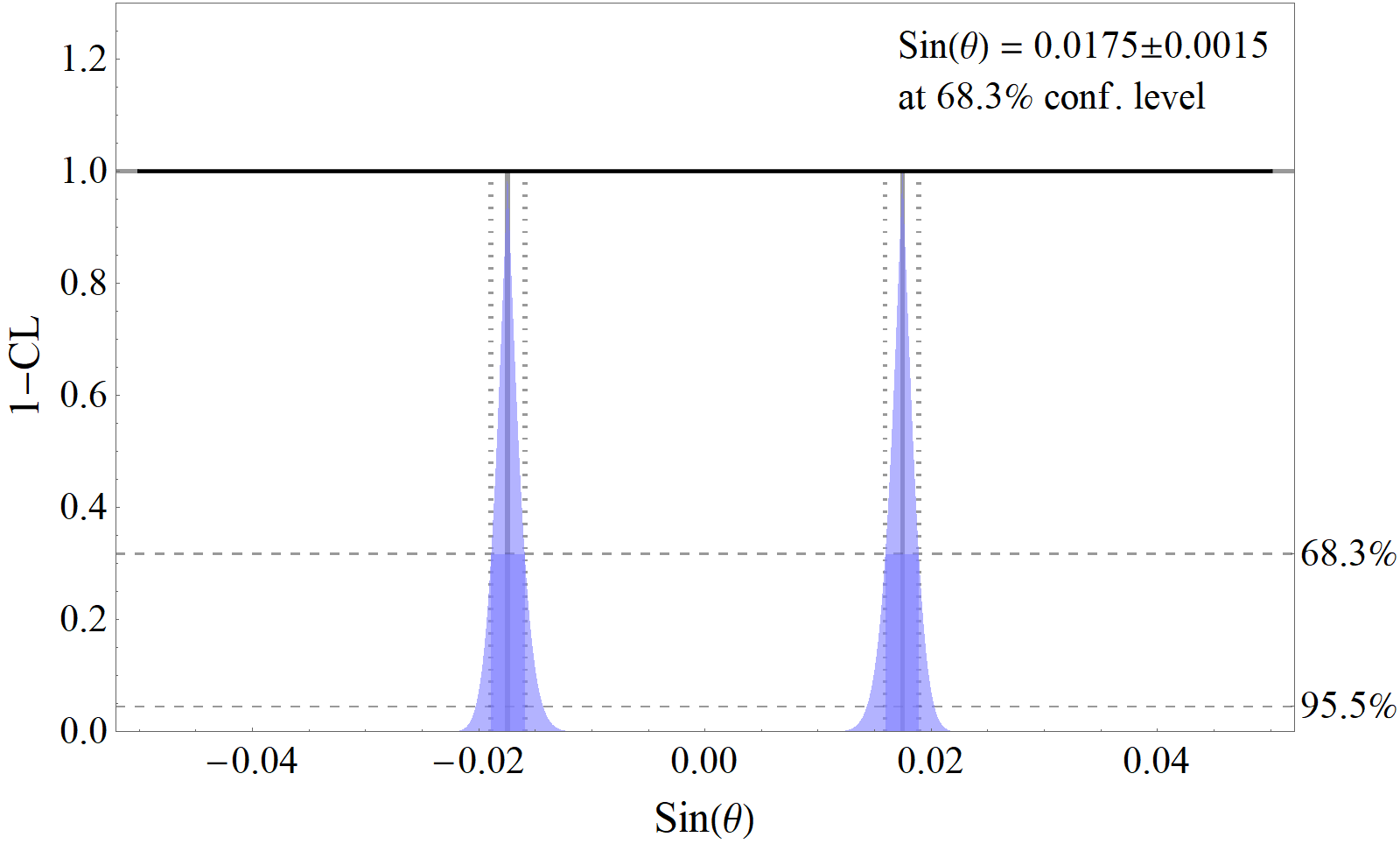}\label{fig:allM5st1D}}
	\caption{One dimensional profile likelihoods, $1$ and $2\sigma$ CLs of the parameters for models IV and V (fit C0) for the large dataset (see sec. \ref{sec:anaall}). Each figure contains the $1\sigma$ results for the parameters in the top panel. For $\sin\theta$ plots, we quote the positive result, as it is preferred by the constraints.}
	\label{fig:alldat1D1}
\end{figure*}

\begin{turnpage}
\begin{figure*}
	\centering
	\includegraphics[height=14cm]{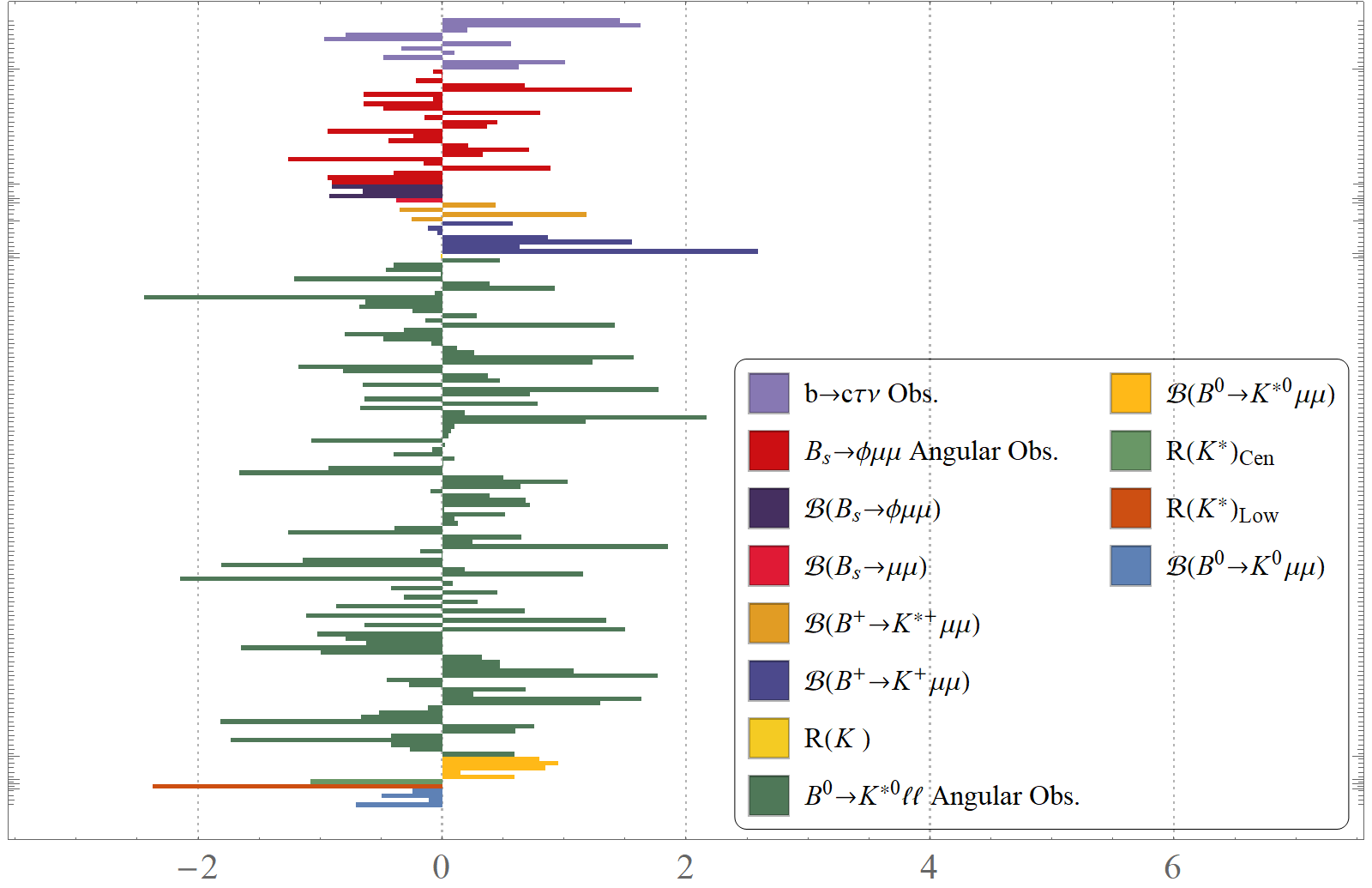}
	\caption{Pull results for Model V (fit C1) for the large dataset (see sec. \ref{sec:anaall}). Results are color coded and organized serially. For wrapped columns, first go from top-to-bottom in the left column, then go to the right one similarly.}
	\label{fig:pullAllmod5}
\end{figure*}
\end{turnpage}

\begin{figure*}
	\centering
	\subfloat[Rest of the obs.]{\includegraphics[height=5.5cm]{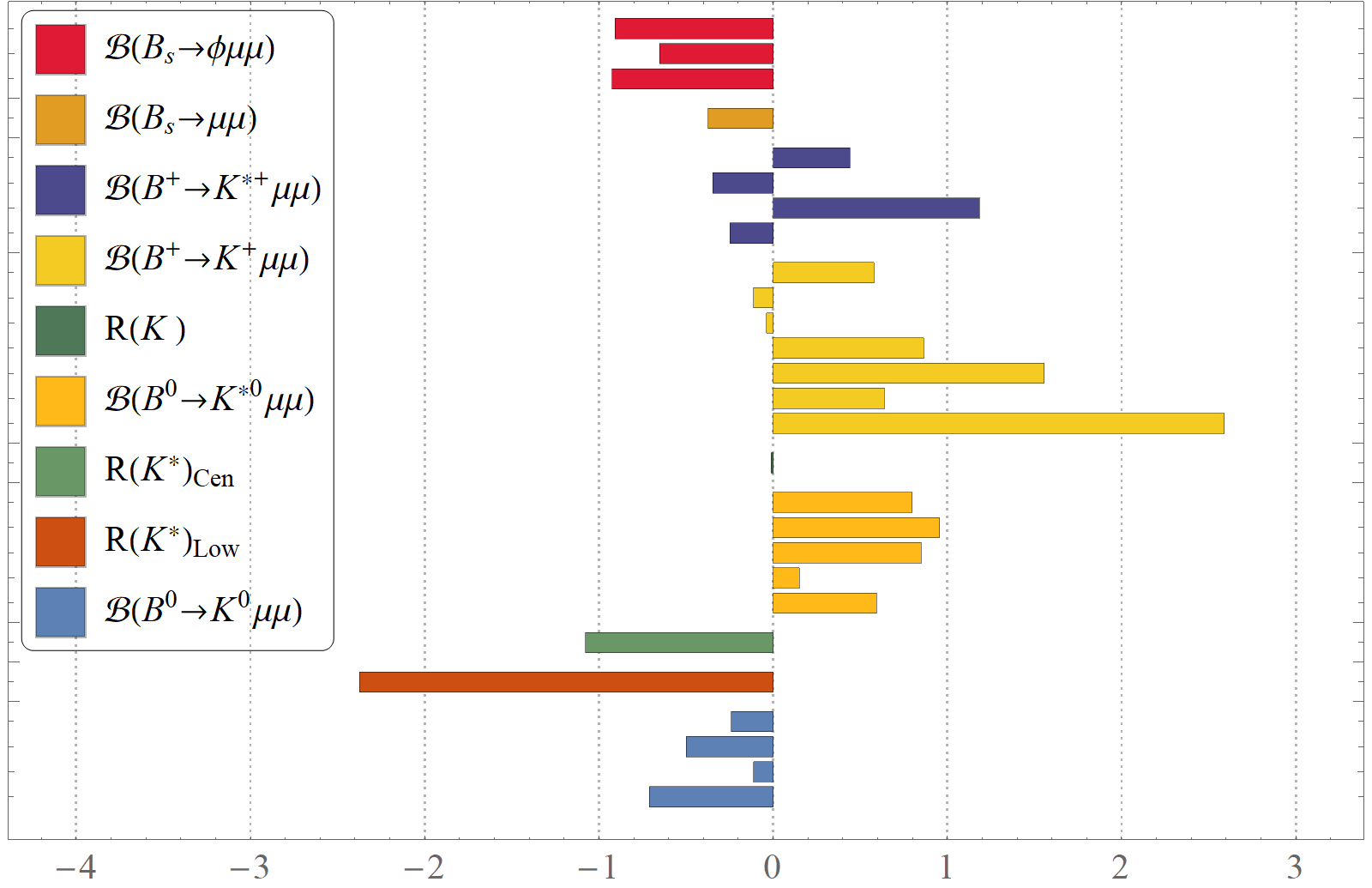}\label{fig:pullAllmod5gr1C}}~~
	\subfloat[$b\to c\tau\nu$ observables]{\includegraphics[height=5.5cm]{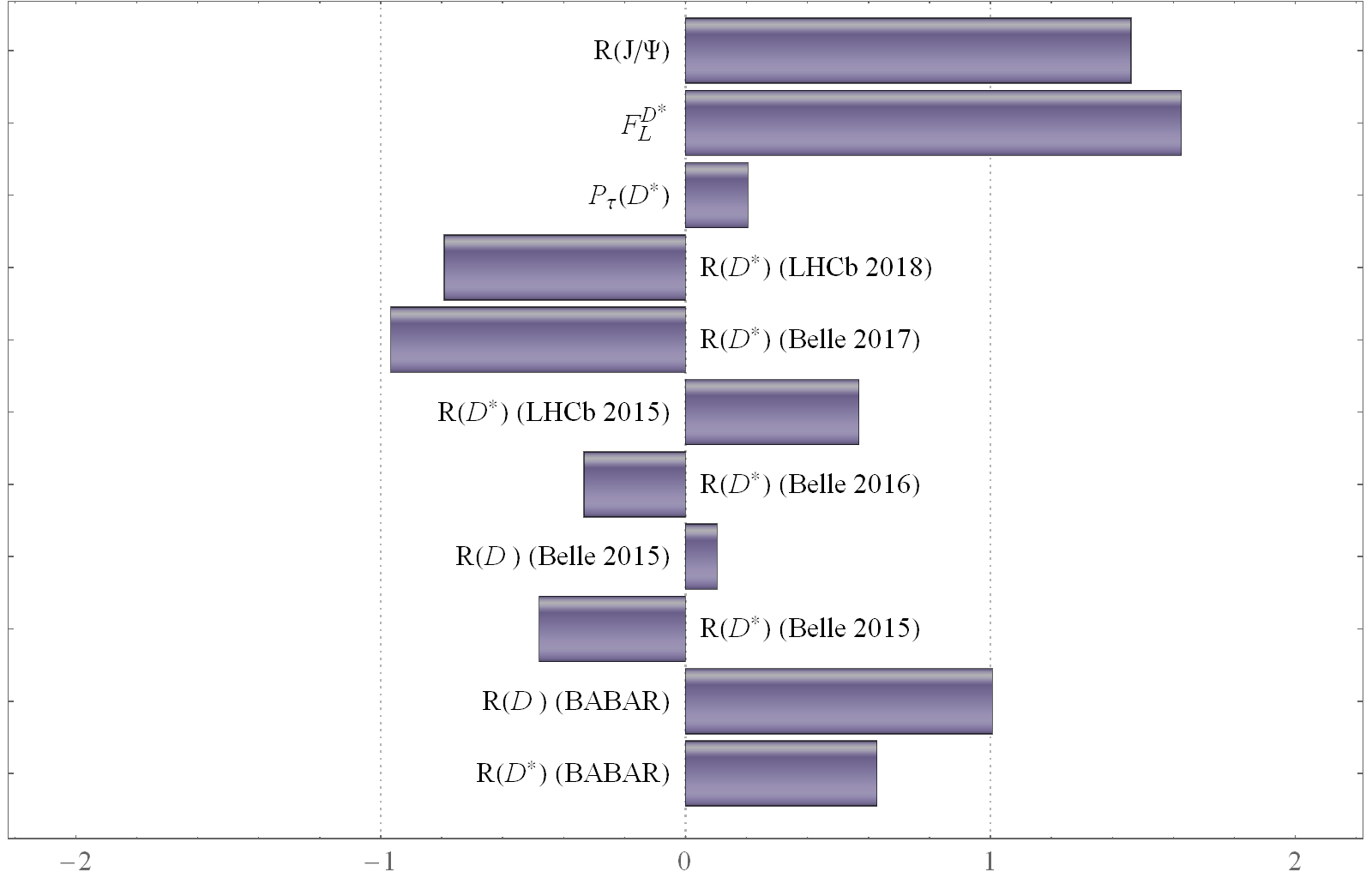}\label{fig:pullAllmod5b2ctnC}}\\
	\subfloat[$B\to K^*\ell\ell$ angular obs.]{\includegraphics[height=11cm]{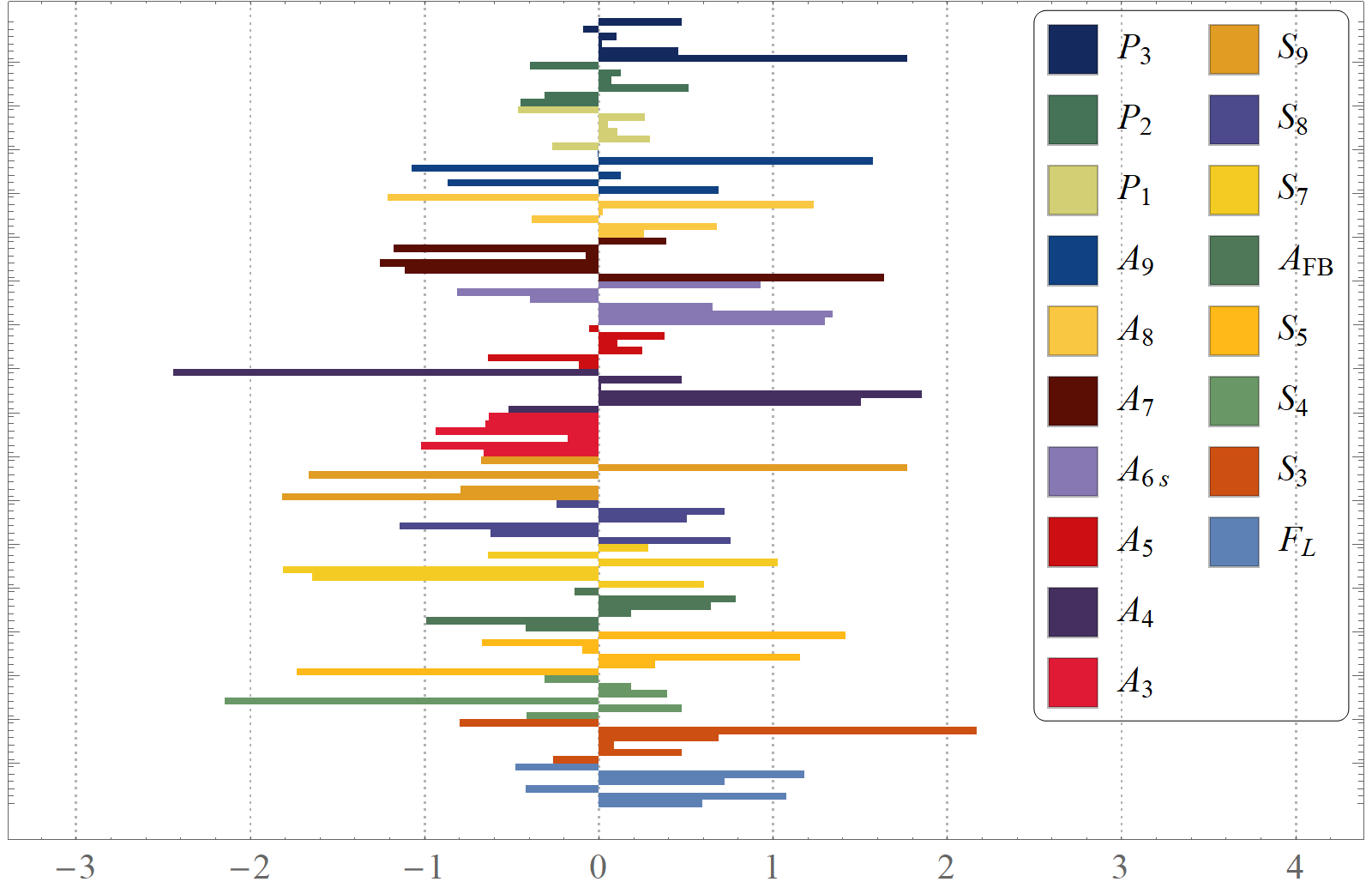}\label{fig:pullAllmod5angobsC}}~~
	\caption{Pull results for Model V (fit C1) for the large dataset (see sec. \ref{sec:anaall}). Results are color coded and organized serially. For wrapped columns, first go from top-to-bottom in the left column, then go to the right one similarly.}
	\label{fig:pullpartmod5}
\end{figure*}

\subsubsection{Small dataset:}\label{sec:anafew}
Following the discussion in the previous section, we have done two sets of fits here, one with all $V$s diagonal, i.e. without considering any correlation, and the other with appropriate correlations. We have found that the effect of the correlations on the fits with fewer observables is really small and main inferences do not change considerably. Thus, it will be implied that all shown results are obtained with correlations, if not mentioned otherwise.

Table \ref{table:FC01} lists the $p$-values and best fit results for all models with the small dataset, with both unconstrained fit (C0) and with constraints from $B_s\to\tau\tau$ and $B\to K\mu\tau$ decays (C1). We see that even in the absence of any constraints (C0), the fits are of poor quality, with $p$-values $\sim 10\%$ and with the fit C1, models IV and V are essentially discarded while the fit qualities of the rest of the models are same as the C0 fits. We have also found that there is no minimum in the physically allowed region for the fit with all constraints turned on. 
Let us mention here that $\chi^2_{SM} \sim 54$, corresponding to a $p$-value $\sim 10^{-8}\%$ (for 5 degrees of freedom). 

As can be seen from figures \ref{fig:fewM1a1a2} and \ref{fig:fewM1a1a2C} (depicting the $A_1$ - $A_2$ plane in the parameter space for Model I), the minima are straight line contours and thus all points on those lines are equally good, in terms of the fit. Constraints from \bstt~ and \bKnn~ are independent of $\sin\theta$ and allow only part of the $3\sigma$ CLs and there is no common parameter space between them within that CL. The allowed region for the constraint from \bKmt~ depends heavily on the value of $\sin\theta$ and thus varies a lot, but its effect is independent of the tension between \bstt~ and \bKnn. There is only one region where these two constraints overlap very close to the margin of the $3\sigma$ CL, and this tells us Model I will only be allowed if future experimental results shift the CLs in that direction.

We can see a different perspective of the same picture in the $A_2$ - $\sin\theta$ parameter space (figures \ref{fig:fewM1a2st} and \ref{fig:fewM1a2stC}). Here, \bKnn~ does not depend on any of the parameters for this model but the other two constraints do. The new result from here is the dual CL regions due to sign ambiguity of $\sin\theta$ for this analysis. Analyzing these results clearly shows us why Model I is indeed discarded 
but not due to the tension between \bstt~ and the $R(K^*)_{Low}$, 
but rather because of that between \bstt~ and \bKnn. This same thing happens for models II (figures \ref{fig:fewM2a1a2} - \ref{fig:fewM2a2stC}) and III (figures \ref{fig:fewM3a1a3} - \ref{fig:fewM3a3stC}). Thus, these models are also ruled out in view of the existing experimental results by just more than $3\sigma$.

Models IV (figures \ref{fig:fewM4a1a5} - \ref{fig:fewM4a5stC}) and V (figures \ref{fig:fewM5a1a5} - \ref{fig:fewM5a5stC}) do not affect \bKnn~ decays, and thus the only important constraints are from \bstt~ and \bKmt. Here too we see that the best-fit of the unconstrained fit (C0) is allowed by \bKmt, but not by \bstt. More importantly, of the two symmetric fit points for $\sin\theta$ (with opposite signs), the positive ones are preferred by the \bKmt~ constraint, while the negative ones are not. As an example, the \bKmt~region in figure \ref{fig:fewM4a1a5} is drawn for the positive $\sin\theta$ value, whereas that of figure \ref{fig:fewM5a1a5} is drawn for the negative one. Though the $p$-values for C1 fits show that these fits are statistically disallowed, the best fit points lie within $2\sigma$ CLs. This indicates that if the overall statistical significance of the fits increase with a bigger set of observables, then the C1 fits will also become significant. 

As we have got closed confidence levels for these last two models, we can obtain the 1-dimensional marginal confidence levels for the parameters in these. Figure \ref{fig:fewdat1D1} shows the 1-dimensional CLs for the parameters of these models for the unconstrained fits.

To know the importance of the tension between $R(K^*)_{Low}$ and \bstt~ and other observables in allowing or discarding the models, and to get a quantitative estimate of the deviation between the fitted and experimental results, we need to check the `pulls' for these fits. Figure \ref{fig:fewdatpull} showcases the pulls of fit C0 and C1. Figure \ref{fig:fewpullsC0} shows us that the pulls do not change considerably for the different models (the only slight change happens for models IV and V for $B_s\to\mu\mu$ and $R(K)$). It shows us that for all the models, the unconstrained fits do not explain $R(K^*)_{Low}$ adequately (consistent deviation of more than $2\sigma$) and it has an opposite pull to $R(J/\Psi)$, the next most important deviation. In fit C1 (figure \ref{fig:fewpullsC1}), on incorporating \bstt~ and \bKmt, we see that models IV and V actually have larger deviations for both bins of $R(K^*)$ and $B_s\to\phi\mu\mu$ branching fractions ($B_s\to\mu\mu$ is actually more consistent with the fit result for these models). This tells us that the reason for models I-III not working is not actually $R(K^*)_{Low}$ being in tension with one of the constraints, but, as explained earlier, the incompatibility between the constraints themselves.

Figure \ref{fig:fewdatCD} portrays the Cook's distances for the observables for fits C0 and C1. We can clearly see that the impact of $\mathcal{B}(B_s\to\mu\mu)$ is largest for models IV and V. As the cutoff value $Cut_{CD} \sim 0.9$ for all five models, this observable essentially becomes a valid outlier for the C1 fit of models IV and V. Surprisingly, $R(K^*)_{Low}$ has quite low $CD$, indicating a small impact on the fits. This also brings into question the impact these observables would have in presence of all other observables in the $b\to s\ell\ell$, which we can only answer with the fits for the full dataset.

\begin{figure*}
	\centering
	\subfloat[All Observables]{\includegraphics[height=5cm]{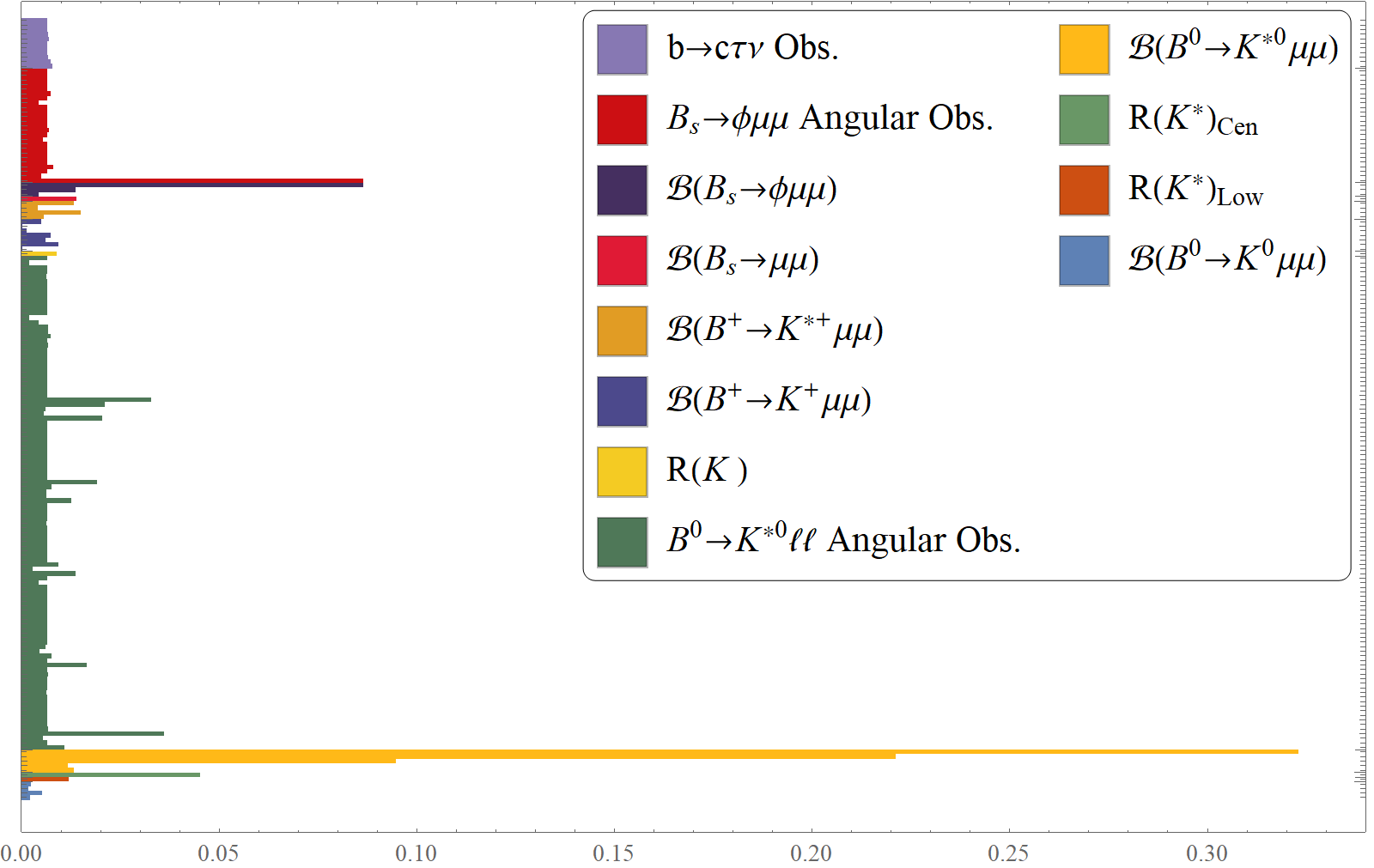}\label{fig:CDAllmod5AllC}}~~
	\subfloat[Obs. other than $b\to c\tau\nu$ and Angular obs.]{\includegraphics[height=5cm]{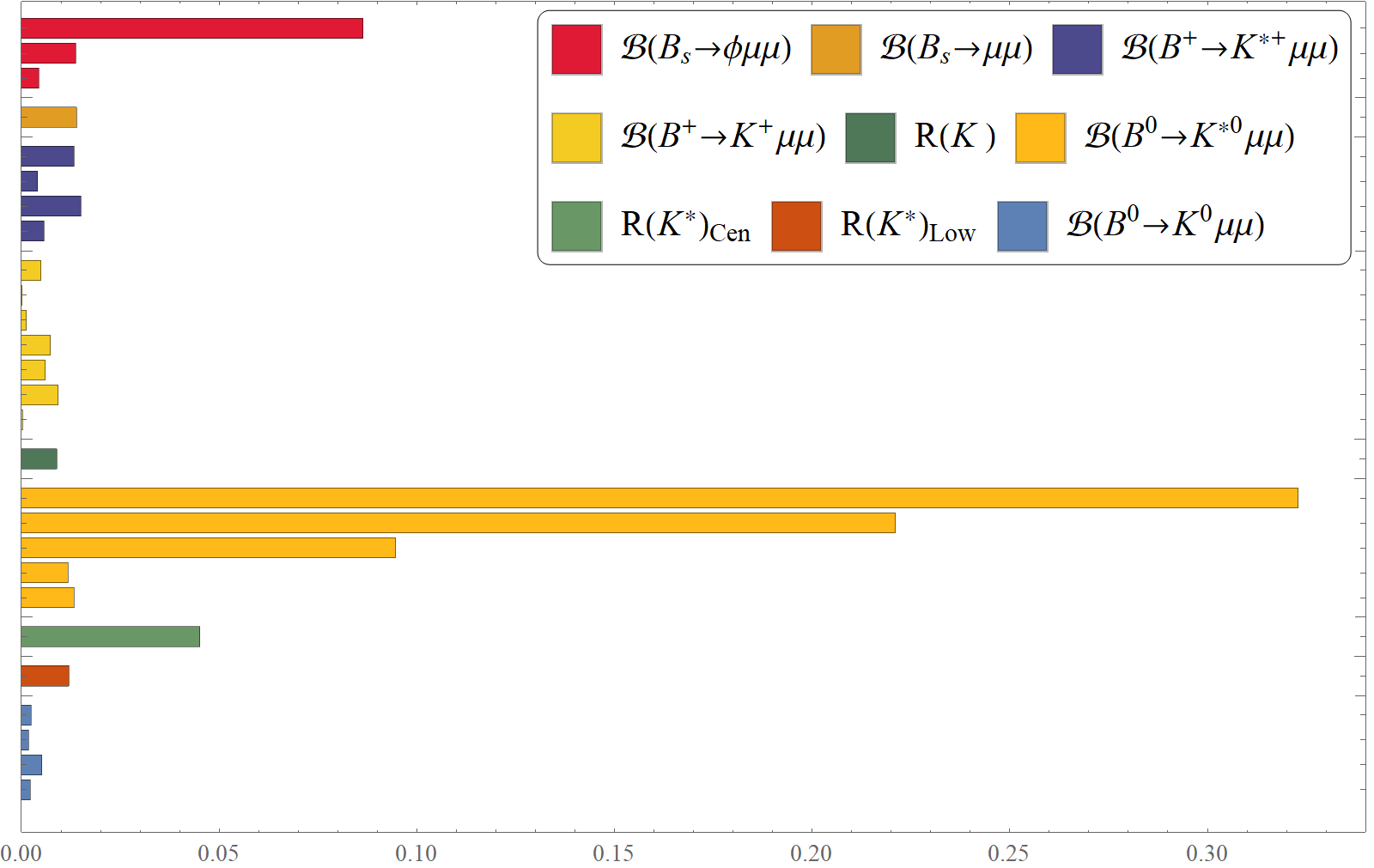}\label{fig:CDAllmod5gr1C}}
	\caption{Cook's distances for Model V (fit C1) for the large dataset (see sec. \ref{sec:anaall}). Results are color coded and organized serially. For figure \ref{fig:CDAllmod5AllC}, first go from top-to-bottom in the left column, then go to the right one similarly. For figure \ref{fig:CDAllmod5gr1C}, go from left-to-right from top, then do the same in the next row.}
	\label{fig:CDAllmod5}
\end{figure*}

\subsubsection{Large dataset:}\label{sec:anaall}
\begin{table}
	\centering
	\begin{tabular}{|c|c|c|c|c|c|c|}
		\hline
		\text{All Obs.} & \multicolumn{3}{|c|}{ \text{C0}}  & \multicolumn{3}{|c|}{\text{C1}}   \\
		\hline
		\text{Model} & \text{$p$-value(\%)} &\multicolumn{2}{|c|}{ \text{Best-fit value} } & \text{$p$-value(\%)} & \multicolumn{2}{|c|}{ \text{Best-fit value}}   \\
		\hline
		I & 65.48 & \text{$A_1$} & -2.689 & 65.48 & \text{$A_1$} & -0.531 \\
		&  & \text{$A_2$} & -1.210 &  & \text{$A_2$} & -0.130 \\
		&  & \text{$S_\theta $} & -0.017 &  & \text{$S_\theta $} & 0.044 \\
		\hline
		II & 65.48 & \text{$A_1$} & -1.653 & 65.48 & \text{$A_1$} & -2.098 \\
		&  & \text{$A_2$} & -3.575 &  & \text{$A_2$} & 1.569 \\
		&  & \text{$S_\theta $} & 0.017 &  & \text{$S_\theta $} & 0.053 \\
		\hline
		III & 65.48 & \text{$A_1$} & -2.883 & 65.48 & \text{$A_1$} & -2.883 \\
		&  & \text{$A_3$} & -1.167 &  & \text{$A_3$} & 0.787 \\
		&  & \text{$S_\theta $} & -0.017 &  & \text{$S_\theta $} & 0.034 \\
		\hline
		IV & 74.20 & \text{$A_1$} & -3.844 & 74.14 & \text{$A_1$} & -3.843 \\
		&  & \text{$A_5$} & -2.317 &  & \text{$A_5$} & -2.220 \\
		&  & \text{$S_\theta $} & 0.017 &  & \text{$S_\theta $} & 0.017 \\
		\hline
		V & 74.20 & \text{$A_1$} & -2.883 & 74.14 & \text{$A_1$} & -2.882 \\
		&  & \text{$A_5$} & -2.317 &  & \text{$A_5$} & -2.220 \\
		&  & \text{$S_\theta $} & 0.017 &  & \text{$S_\theta $} & 0.017 \\
		\hline
	\end{tabular}
	\caption{Fit results for large dataset with fits C0 and C1. There are no physically allowed best fit results for the fit C2 for the large dataset. Parameter uncertainties are not quoted because of reasons stated in sec. \ref{sec:anaall}.}
	\label{table:AC01}
\end{table}

Table \ref{table:AC01} lists the $p$-values and best fit results for all models with the large dataset, with fits C0 and C1. Here too we find that there is no minimum in the physically allowed region for the fit C2. The reason for that, similar to the fits with the small dataset, is that constraints from \bstt~ and \bKnn~ allow only part of the $3\sigma$ CLs and there is no common parameter space between them within that CL. There is still only one region where these two constraints overlap very close to the margin of the $3\sigma$ CL (figures \ref{fig:allM1a1a2C} - \ref{fig:allM5a5stC}). Actually, the nature of the interplay of CLs and constraints remain quite the same as that for the small dataset for models I-III and they are still ruled out with all the constraints in place. Due to the similarity in the CLs of these fits with the small dataset, we only show parameter spaces for the C1 fit for models I-III in figure \ref{fig:alldat2D1}. The major change for these three models with a larger dataset is a large increase in the statistical significance of the fits ($p$-values $\sim 65\%$ for all of them), which is expected. Let us mention here that $\chi^2_{SM} \sim 214$, corresponding to a $p$-value $\sim 0.78\%$ (167 degrees of freedom).

With the large dataset, the most important and noticeable change comes for models IV and V. While the C1 fits for these models were statistically ruled out with a significance $< 5\%$, with $p$-values $\sim 74\%$ for both C0 and C1 fits, they are not only allowed, but are more significant than models I-III. Also, the best-fits for the unconstrained fits (C0) are actually almost same as those for the constrained ones. More importantly, the parameter CLs are considerably reduced in size, making the models not only more statistically significant, but also more precise. Figure \ref{fig:alldat1D1} shows the 1-dimensional CLs for the parameters of these models for the unconstrained fits.

At first glance, this increase in statistical significance of even the constrained fit (C1) is surprising. To understand why this is happening, we need to examine the pull distributions and the Cook's distances. Figures \ref{fig:pullAllmod5} and \ref{fig:pullpartmod5} show the pull values of the observables for the fit C1. Pulls for the fit C0 do not have qualitatively different results. At first glance, the pull distribution of all the observables (figure \ref{fig:pullAllmod5}) shows us that though $R(K^*)_{Low}$ still has an absolute pull $>2$, it no longer is the only one (figure \ref{fig:pullAllmod5gr1C}). One bin in $\mathcal{B}(B^+\to K^+\mu\mu$ results and at least three $B^0\to K^{*0}\ell\ell$ angular observables join it (figure \ref{fig:pullAllmod5angobsC}). In addition, many of the angular observables have quite large pulls. Their final effect is ultimately canceled due to the almost equal number of pulls on both sides of zero. We have to remember that these are quite imprecise observables and more precise measurements in the future may completely change the nature of the fits, as there are so many of them. Thus, $R(K^*)_{Low}$ still remains a point of tension. On the other hand, the other observable with a large pull for the small dataset, $R(J/\Psi)$, has now decreased a lot. In fact, the whole $b\to c\tau\nu$ sector has quite well behaved pulls, including the new addition of $F_L(D^*)$ (figure \ref{fig:pullAllmod5b2ctnC}).

As before, we check the impact of these observables on the fit with the help of Cook's distances and figure \ref{fig:CDAllmod5} frames these results. $Cut_{CD}\sim 0.8$ for these fits. In contrast to those for the small dataset, all observables have quite small Cook's distances. This makes the fits quite stable and that explains the higher significance of the fits. $\mathcal{B}(B_s\to\mu\mu)$, which was an observable with the most impact earlier, now sits well with other data. 

\section{Conclusion}\label{sec:conclusions}
In the scope of our present article, we revisit the proposition put forward by the authors of refs.~\cite{Choudhury:2017qyt,Choudhury:2017ijp} where they introduce a class of models with the potential for explaining the observed deviation in the charged ($b\to c$) and neutral ($b\to s$) current anomalies subject to the same source, and with a minimal number of NP parameters. The `models' basically correspond to Lagrangians involving diquark-dilepton operators, written in the flavor basis. These involve the second and third generation quarks but only third generation leptons. The mass eigenstates corresponding to the $\mu$ and $\tau$ leptons are generated by rotating the charged lepton states.

We perform our analysis in two parts. The first part our analysis is essentially an exact recreation of that in the aforementioned articles, with even the data obtained from their paper (eight data-points). The second part of our analysis takes into consideration most of the available experimental data corresponding to the $b\to c$ and $b\to s$ transitions, which amount to a total of 170 data points. This results in a considerable increase in the degrees of freedom for the fit. In both these parts, we obtain three types of best fit results: (a) without constraints (\textbf{C0}), (b) with constraints from $B_s\to\tau\tau$ and $B\to K\mu\tau$ only~(\textbf{C1}) and (c) with constraints from $B_s\to\tau\tau$, $B\to K\mu\tau$ and $B\to K^{(*)}\nu\bar\nu$~(\textbf{C2}). We also take care of all possible correlations (statistical, systematic and theoretical) among the datasets, even though we find that the correlations have but a tiny effect on our inferences regarding the small dataset. In the course of this analysis, we present the most precise theoretical estimate for the $R(J/\psi)$ observable till date to the best of our knowledge.

We observe that models I-III are `semi realistic', in a sense that they are unable to satisfy all the constraints simultaneously. 
By examining the two-dimensional profile-likelihoods of the parameters for these models, we find that instead of best-fit points, we have best fit contours of the parameters, some points of which are allowed by one constraint, some by another. Though there are parameter CLs allowed by both \bstt~ and \bKmt~ modes, the small region allowed simultaneously by \bstt, \bKmt, and $B\to K^{(*)}\nu\bar\nu$ is away by just more than $3\sigma$ from the nearest point on the best-fit contour. On top of this, we examine the `pulls' of the observables for all models to ascertain the level of agreement between the fitted results and data, and find that the fitted $R(K^*)_{Low}$ is always away by more than $2\sigma$, irrespective of the constraints. Thus, these models are disallowed not because of the tension between the $B_s\to\tau\tau$ and the $R(K^*)_{Low}$ data, but rather because of that between \bstt~(\bKmt) and $B\to K^{(*)}\nu\bar\nu$. 

In the case of models IV and V, though there are parameter CLs allowed by the constraints, the models are ruled out from the point of statistical significance ($p$-value $<5\%$; see Tab.~\ref{table:FC01}). Actually, the fits are of poor statistical significance in general for any model. 
We also try to find any observable with unnaturally large impact in our analysis using Cook's distances, and find that the data-point for $\mathcal{B}(B_s\to \mu\mu)$ is a one and a possible outlier. 

The immediate effect that the `large dataset' results in is a huge enhancement in the statistical quality of the fits for the NP parameters. 
Models I-III are still ruled out when one incorporates all the constraints. However, the most important changes are noticed in models IV and V. These are now not only statistically allowed, but also more significant than models I-III, statistically. This increase in the statistical significance can be attributed to the equal number of pulls on either side of zero due to the addition of the large number of angular observables to the data set. $R(K^*)_{Low}$, with a large pull, still remains `problematic', but its effect on the overall fit is now canceled by large pulls of some of the other observables. The $B_s\to\mu\mu$ mode that was an outlier for the fits corresponding to the `small dataset', now fit into the scheme of things quite well, making the fit results quite robust.

We thus conclude that the models IV and V, while still incapable of a satisfactory explanation of $R(K^*)_{Low}$, is not only allowed by the plethora of flavor observables,  but also is of high statistical significance, with quite a precise prediction for effective Wilson coefficients, providing a possible pathway for future model builders. High precision experiments in the coming days will illuminate the map even further.

\acknowledgments 
The authors thank Anirban Kundu for numerous discussions and clarifications essential for this work. S.K.P. is supported by the grants IFA12-PH-34 and SERB/PHY/2016348. 

\appendix
\section{Transitions in Effective Theory}\label{app:effHamil}
\subsection{$b\to s \mu^{+}\mu^{-}$}
The effective Hamiltonian for this transition is:
\begin{align}
{\cal H}_{\rm eff} = -\frac{4G_F}{\sqrt{2}} V_{tb} V^*_{ts} \frac{\alpha_{em}}{4\pi} \sum_{i} C_i(\mu) O_i(\mu)\,,
\label{h-eff1}
\end{align}
where the relevant operators for our analysis are:
\begin{align}
O_{7} &= \frac {m_b}{e} (\bar s \sigma_{\mu \nu} P_R b) F^{\mu \nu}\,,
\nonumber\\
O_{9} &= (\bar s \gamma_\mu P_L b)(\bar\mu \gamma^\mu \mu)\,,
\nonumber\\
O_{10} &= (\bar s \gamma_\mu P_L b)(\bar\mu \gamma^\mu \gamma_5 \mu)\,.
\label{ops}
\end{align}
If the NP operators are of only (V-A)(V-A) type, then one can denote the modified WCs as:
\begin{align}
C_9 \to C_9^{SM} + C_9^{NP}\,,
C_{10} \to C_{10}^{SM} + C_{10}^{NP}\,.
\end{align}

\subsection{$b\to c \tau \bar\nu $}

Considering the NP operators of only (V-A)(V-A) structure, one can write the effective Hamiltonian for this process as:

\begin{align}
{\cal H}_{\rm eff} = \frac{4G_F}{\sqrt{2}} V_{cb} (1+C^{NP}) (\bar c_L \gamma_{\mu} b_L)(\bar\tau_L \gamma^{\mu} \nu_{\tau L})
\label{h-eff2}
\end{align}

\subsection{$b\to s \nu \bar\nu $}

Again here, considering only the (V-A)(V-A) operator structure, the effective Hamiltonian for this transition can be written as:

\begin{align}
\nn {\cal H}_{\rm eff} &= -\frac{4G_F}{\sqrt{2}} V_{tb} V^*_{ts} \frac{\alpha_{em}}{4\pi} C_L^{SM} (1+C_{\nu}^{NP})\\
& \quad\times 2\ (\bar s_L \gamma_{\mu} b_L)(\bar \nu_L \gamma^{\mu} \nu_L) \,
\label{h-eff3}
\end{align}

where, $C_L^{SM}$ = $-X_t/s^2_w$.

\section{Details on Models}\label{app:models}
\subsection{Models I-III}
In terms of the component field, Eq.~\eqref{eq:op1} reduces to:
\begin{align}\label{eq:Ia_comp}
\nn {\cal O}_{\rm I} &= A_1 (s,b)(\nu_{\tau},\nu_{\tau})+ A_1 (c,t)(\tau,\tau)\\
&+\left(\frac{1}{2}A_1+A_2\right)(c,t)(\nu_{\tau},\nu_{\tau}) + \left(A_2-\frac{1}{2}A_1\right)(c,b)(\tau,\nu_{\tau})\\ &+\left(A_2-\frac{1}{2}A_1\right)(s,t)(\nu_{\tau},\tau)+ 
\left(\frac{1}{2}A_1+A_2\right)(s,b)(\tau,\tau).
\end{align}
The term relevant for the $b\to s \mu \mu$ transition can be generated from the $(s,b)(\tau,\tau)$ term in the above equation by rotating the $\tau$ as in eq.~\eqref{eq:rot}. To find the relation between ($C_9^{NP}$,$C_{10}^{NP}$) and the model parameters we decompose the corresponding NP operator and compare the same to the operator structure of $O_{9}$ and $O_{10}$. For example, the relevant operator for $b\to s \mu \mu$ channel
is given as:
\begin{align}
\nn&\left(\frac{1}{2} A_1 +A_2\right)\left[\frac{1}{2}(\bar s \gamma_\mu P_L b)(\bar\mu \gamma^\mu \mu)\sin ^2\theta \right.\\
&\qquad\left. -\frac{1}{2}(\bar s \gamma_\mu P_L b)(\bar\mu \gamma^\mu \gamma_5 \mu)\sin ^2\theta\right]\,.
\end{align}
Thus, one can write the effective Hamiltonian for this Model as:
\begin{align}
{\cal H}_{\rm eff} &= -\frac{4G_F}{\sqrt{2}} V_{tb} V^*_{ts} \frac{\alpha_{em}}{4\pi}
\left( C_7^{SM} O_7+ C_9^{SM} O_9 + C_{10}^{SM} O_{10}\right)
\nonumber\\
& + \frac{1}{2}\left(\frac{1}{2} A_1 +A_2\right)\sin ^2\theta\, O_9
-\frac{1}{2}\left(\frac{1}{2} A_1 +A_2\right)\sin ^2\theta\, O_{10} \,.
\label{h-effmod1b2s}
\end{align}
which can be equivalently written as:
\begin{align}
{\cal H}_{\rm eff} &= -\frac{4G_F}{\sqrt{2}} V_{tb} V^*_{ts} \frac{\alpha_{em}}{4\pi}[
\left( C_7^{SM} O_7+ C_9^{SM} O_9 + C_{10}^{SM} O_{10}\right)
\nonumber\\
& + \frac{ \frac{1}{2}\left(\frac{1}{2} A_1 +A_2\right)\sin ^2\theta}
{-\frac{4G_F}{\sqrt{2}} V_{tb} V^*_{ts} \frac{\alpha_{em}}{4\pi}}\, O_9 
-\frac{\frac{1}{2}\left(\frac{1}{2} A_1 +A_2\right)\sin ^2\theta}
{-\frac{4G_F}{\sqrt{2}} V_{tb} V^*_{ts} \frac{\alpha_{em}}{4\pi}}\, O_{10}].
\label{h-effmod11b2s}
\end{align}
Hence, from Eq.~\eqref{h-eff1} and Eq.~\eqref{h-effmod11b2s}, we get:
\begin{align}
C_9^{NP} = -C_{10}^{NP}=\frac{ \frac{1}{2}\left(\frac{1}{2} A_1 +A_2\right)\sin ^2\theta}
{-\frac{4G_F}{\sqrt{2}} V_{tb} V^*_{ts} \frac{\alpha_{em}}{4\pi}}\,
\end{align}
Next, the term relevant for $b\to c \tau \bar\nu $ process from Eq.~\eqref{eq:Ia_comp} is :
$ \left(A_2-\frac{1}{2}A_1\right)(c,b)(\tau,\nu_{\tau})$.

Thus the effective Hamiltonian for this channel can be written as:
\begin{align}
{\cal H}_{\rm eff} = \frac{4G_F}{\sqrt{2}} V_{cb}\,(c,b)(\tau,\nu_{\tau})+\left(A_2-\frac{1}{2}A_1\right)(c,b)(\tau,\nu_{\tau})
\label{h-effmod1b2c}
\end{align}
which can be equivalently written as:
\begin{align}
{\cal H}_{\rm eff} = \frac{4G_F}{\sqrt{2}} V_{cb}\,\left[1 +
\frac{\left(A_2-\frac{1}{2}A_1\right)}{\frac{4G_F}{\sqrt{2}} V_{cb}}\right](c,b)(\tau,\nu_{\tau})
\label{h-effmod11b2c}
\end{align}
Thus, from Eq.~\eqref{h-eff2} and Eq.~\eqref{h-effmod11b2c}, we get:
\begin{align}
C^{NP} = \frac{\left(A_2-\frac{1}{2}A_1\right)}{\frac{4G_F}{\sqrt{2}} V_{cb}}
\end{align}
Proceeding in a similar fashion as above, we can find the relation between the WCs and the model parameters for the rest of the models.
In terms of the component fields, eq. \ref{eq:op2} reduces to:
\begin{align}\label{eq:IIa_comp}
{\cal O}_{\rm II} &= \left(A_2-\frac{1}{2} A_1\right)(s,b)(\nu_{\tau},\nu_{\tau})+ \left(A_2-\frac{1}{2}A_1\right) (c,t)(\tau,\tau)\nn \\
&+\frac{1}{2}\left(A_1+A_2\right)(c,t)(\nu_{\tau},\nu_{\tau}) + \left(A_1-\frac{1}{2}A_2\right)(c,b)(\tau,\nu_{\tau})\nn \\ &+\left(A_1-\frac{1}{2}A_2\right)(s,t)(\nu_{\tau},\tau)+ 
\frac{1}{2}\left(A_1+A_2\right)(s,b)(\tau,\tau)\,.
\end{align}
We hence obtain
\begin{align}
C_9^{NP} = -C_{10}^{NP}=\frac{ \frac{1}{4}\left( A_1 +A_2\right)\sin ^2\theta}
{-\frac{4G_F}{\sqrt{2}} V_{tb} V^*_{ts} \frac{\alpha_{em}}{4\pi}}\,
\end{align}
\begin{align}
C^{NP} = \frac{\left(A_1-\frac{1}{2}A_2\right)}{\frac{4G_F}{\sqrt{2}} V_{cb}}
\end{align}
In terms of the component fields, eq. \ref{eq:op3} gives:
\begin{align}\label{eq:IIIa_comp}
{\cal O}_{\rm III} &= \left(A_3-\frac{1}{2} A_1\right)(s,b)(\nu_{\tau},\nu_{\tau})+ \left(A_3-\frac{1}{2}A_1\right) (c,t)(\tau,\tau)\nn \\
&+\left(\frac{1}{2}A_1+A_3\right)(c,t)(\nu_{\tau},\nu_{\tau}) + A_1 (c,b)(\tau,\nu_{\tau})\nn \\
&+ A_1(s,t)(\nu_{\tau},\tau)+ \left(\frac{1}{2}A_1+A_3\right)(s,b)(\tau,\tau)\,,
\end{align}
providing us with
\begin{align}
C_9^{NP} = -C_{10}^{NP}=\frac{ \frac{1}{2}\left(\frac{1}{2} A_1 +A_3\right)\sin ^2\theta}
{-\frac{4G_F}{\sqrt{2}} V_{tb} V^*_{ts} \frac{\alpha_{em}}{4\pi}}\,
\end{align}
\begin{align}
C^{NP} = \frac{A_1}{\frac{4G_F}{\sqrt{2}} V_{cb}}\,.
\end{align}

\subsection{Models IV and V}
In terms of component fields, eq. \ref{eq:op4} gives:
\begin{align}\label{eq:IVa_comp}
{\cal O}_{\rm IV} & = \frac{3 A_1}{4} (c, b) (\tau, \nu_\tau)
+ \frac{3 A_1 }{4} (s, b) (\tau, \tau) + A_5 (s, b) \{\tau, \tau\} 
\nonumber\\
& + \frac{3 A_1}{4} (s, t) (\nu_{\tau}, \tau) 
+ A_5 (c, t) \{\tau, \tau\}
+ \frac{3 A_1 }{4}(c, t) (\nu_\tau, \nu_\tau).
\end{align}
The corresponding Wilson coefficients are
\begin{align}
C_9^{NP} &= \frac{ \frac{1}{2}\left(\frac{3}{4} A_1 +A_5\right)\sin ^2\theta}
{-\frac{4G_F}{\sqrt{2}} V_{tb} V^*_{ts} \frac{\alpha_{em}}{4\pi}}\,
\nonumber\\
C_{10}^{NP} &= \frac{ \frac{1}{2}\left(A_5-\frac{3}{4} A_1\right)\sin ^2\theta}
{-\frac{4G_F}{\sqrt{2}} V_{tb} V^*_{ts} \frac{\alpha_{em}}{4\pi}}\nn\\
C^{NP} &= \frac{\frac{3}{4}A_1}{\frac{4G_F}{\sqrt{2}} V_{cb}}
\end{align}
In terms of the component fields, eq. \ref{eq:op5} reduces to:
\begin{align}\label{eq:Va_comp}
{\cal O}_{\rm V} & = A_1 (c, b) (\tau, \nu_\tau) + A_1 (s, b) (\tau, \tau)
+ A_5 (s, b) \{\tau, \tau\} \nonumber\\
&+ A_1 (s, t) (\nu_{\tau}, \tau) + A_1 (c, t) (\nu_\tau, \nu_\tau)
+ A_5 (c, t) \{\tau, \tau\}.
\end{align}
Therefore, we have
\begin{align}
C_9^{NP} &= \frac{ \frac{1}{2}\left( A_1 +A_5\right)\sin ^2\theta}
{-\frac{4G_F}{\sqrt{2}} V_{tb} V^*_{ts} \frac{\alpha_{em}}{4\pi}}\,
\nonumber\\
C_{10}^{NP} &= \frac{ \frac{1}{2}\left(A_5- A_1\right)\sin ^2\theta}
{-\frac{4G_F}{\sqrt{2}} V_{tb} V^*_{ts} \frac{\alpha_{em}}{4\pi}}\nn\\
C^{NP} &= \frac{A_1}{\frac{4G_F}{\sqrt{2}} V_{cb}}\,.
\end{align}

\bibliography{b_s_global_AK}

\end{document}